\pdfoutput=1
\documentclass[12pt,a4paper]{article}
\usepackage[T1]{fontenc}

\usepackage{ifthen} 
\usepackage[compat=1.1.0]{tikz-feynman}
\tikzfeynmanset{warn luatex=false}
\usepackage{subcaption}
\usepackage{placeins}
\newboolean{pdflatex}
\setboolean{pdflatex}{true} 

\newboolean{articletitles}
\setboolean{articletitles}{true} 

\newboolean{uprightparticles}
\setboolean{uprightparticles}{false} 

\def\paperauthors{LHCb collaboration} 
\def\paperasciititle{Search for $B$ meson decays to multimuon final states} 
\def\papertitle{Search for $B$ meson decays to multimuon final states} 
\def\paperkeywords{{High Energy Physics}, {LHCb}} 
\def\papercopyright{\the\year\ CERN for the benefit of the LHCb collaboration} 
\def\paperlicence{CC BY 4.0 licence}
\def\paperlicenceurl{https://creativecommons.org/licenses/by/4.0/}

\newif\ifEnableSectionTOCLinks
\EnableSectionTOCLinksfalse 
\usepackage{booktabs}

\usepackage[top=1in, bottom=1.25in, left=1in, right=1in]{geometry}

\usepackage{microtype}
\usepackage{lineno}  
\usepackage{xspace} 
\usepackage{caption} 

\usepackage{graphicx}  
\usepackage{color}
\usepackage{colortbl}
\graphicspath{{./figs/}} 

\usepackage{amsmath} 
\usepackage{amssymb}
\usepackage{amsfonts}
\usepackage{upgreek} 

\newcommand*\patchAmsMathEnvironmentForLineno[1]{%
\expandafter\let\csname old#1\expandafter\endcsname\csname #1\endcsname
\expandafter\let\csname oldend#1\expandafter\endcsname\csname
end#1\endcsname
 \renewenvironment{#1}%
   {\linenomath\csname old#1\endcsname}%
   {\csname oldend#1\endcsname\endlinenomath}%
}
\newcommand*\patchBothAmsMathEnvironmentsForLineno[1]{%
  \patchAmsMathEnvironmentForLineno{#1}%
  \patchAmsMathEnvironmentForLineno{#1*}%
}
\AtBeginDocument{%
\patchBothAmsMathEnvironmentsForLineno{equation}%
\patchBothAmsMathEnvironmentsForLineno{align}%
\patchBothAmsMathEnvironmentsForLineno{flalign}%
\patchBothAmsMathEnvironmentsForLineno{alignat}%
\patchBothAmsMathEnvironmentsForLineno{gather}%
\patchBothAmsMathEnvironmentsForLineno{multline}%
\patchBothAmsMathEnvironmentsForLineno{eqnarray}%
}

\usepackage[pdftex,
            pdfauthor={\paperauthors},
            pdftitle={\paperasciititle},
            pdfkeywords={\paperkeywords}]{hyperref}
\usepackage{hyperxmp}
\hypersetup{
    pdfcopyright={Copyright (C) \papercopyright},
    pdflicenseurl={\paperlicenceurl}
}

\usepackage[colorinlistoftodos,textsize=scriptsize]{todonotes}

\usepackage[bottom,flushmargin,hang,multiple]{footmisc}

\usepackage[all]{hypcap} 

\usepackage{xspace} 
\usepackage{upgreek}

\def\lhcb   {\mbox{LHCb}\xspace}

\def\MagUp {\mbox{\em Mag\kern -0.05em Up}\xspace}

\ifthenelse{\boolean{uprightparticles}}%
{

 \def\Pmu         {\ensuremath{\upmu}\xspace}

 \def\Ppsi        {\ensuremath{\uppsi}\xspace}

 \def\PDelta      {\ensuremath{\Delta}\xspace}                 
 \def\PXi         {\ensuremath{\Xi}\xspace}                 
 \def\PLambda     {\ensuremath{\Lambda}\xspace}                 
 \def\PSigma      {\ensuremath{\Sigma}\xspace}                 
 \def\POmega      {\ensuremath{\Omega}\xspace}                 
 \def\PUpsilon    {\ensuremath{\Upsilon}\xspace}
 \let\oldPi\Pi
 \def\PPi         {\ensuremath{\oldPi}\xspace}

 \def\PB      {\ensuremath{\mathrm{B}}\xspace}                 
 \def\PD      {\ensuremath{\mathrm{D}}\xspace}                 
 \def\PJ      {\ensuremath{\mathrm{J}}\xspace}                 
 \def\PK      {\ensuremath{\mathrm{K}}\xspace}                 
 \def\Pb      {\ensuremath{\mathrm{b}}\xspace}                 
 \def\Pc      {\ensuremath{\mathrm{c}}\xspace}

 \def\Pp      {\ensuremath{\mathrm{p}}\xspace}                 

 \def\Ps      {\ensuremath{\mathrm{s}}\xspace}

 \def\thebaroffset{0.0em}
}
{

 \def\Pmu         {\ensuremath{\mu}\xspace}

 \def\Ppsi        {\ensuremath{\psi}\xspace}                 
                  
 \mathchardef\PDelta="7101
 \mathchardef\PXi="7104
 \mathchardef\PLambda="7103
 \mathchardef\PSigma="7106
 \mathchardef\POmega="710A
 \mathchardef\PUpsilon="7107
 \mathchardef\PPi="7105
 \def\PB      {\ensuremath{B}\xspace}                 
 \def\PD      {\ensuremath{D}\xspace}                 
 \def\PJ      {\ensuremath{J}\xspace}                 
 \def\PK      {\ensuremath{K}\xspace}                 
 \def\Pb      {\ensuremath{b}\xspace}                 
 \def\Pc      {\ensuremath{c}\xspace}

 \def\Pp      {\ensuremath{p}\xspace}                 

 \def\Ps      {\ensuremath{s}\xspace}

 \def\thebaroffset{0.18em}
}
\newcommand{\offsetoverline}[2][\thebaroffset]{\kern #1\overline{\kern -#1 #2}}%

\makeatletter
\ifcase \@ptsize \relax
  \newcommand{\miniscule}{\@setfontsize\miniscule{4}{5}}
\or
  \newcommand{\miniscule}{\@setfontsize\miniscule{5}{6}}
\or
  \newcommand{\miniscule}{\@setfontsize\miniscule{5}{6}}
\fi
\makeatother

\DeclareRobustCommand{\optbar}[1]{\shortstack{{\miniscule (\rule[.5ex]{1.25em}{.18mm})}
  \\ [-.7ex] $#1$}}

\def\mumu       {{\ensuremath{\Pmu^+\Pmu^-}}\xspace}

\def\squark    {{\ensuremath{\Ps}}\xspace}

\def\cquark    {{\ensuremath{\Pc}}\xspace}

\def\bquark    {{\ensuremath{\Pb}}\xspace}

\def\kaon    {{\ensuremath{\PK}}\xspace}

\def\KorKbar {\kern \thebaroffset\optbar{\kern -\thebaroffset \PK}{}\xspace}

\def\Kp      {{\ensuremath{\kaon^+}}\xspace}

\def\KS      {{\ensuremath{\kaon^0_{\mathrm{S}}}}\xspace}

\def\D       {{\ensuremath{\PD}}\xspace}

\def\DorDbar {\kern \thebaroffset\optbar{\kern -\thebaroffset \PD}\xspace}

\def\Dp      {{\ensuremath{\D^+}}\xspace}
\def\Dm      {{\ensuremath{\D^-}}\xspace}

\def\DpDm    {\ensuremath{\Dp {\kern -0.16em \Dm}}\xspace}

\def\B       {{\ensuremath{\PB}}\xspace}

\def\BorBbar {\kern \thebaroffset\optbar{\kern -\thebaroffset \PB}\xspace}
\def\Bz      {{\ensuremath{\B^0}}\xspace}

\def\Bd      {{\ensuremath{\B^0}}\xspace}

\def\BdorBdbar {\kern \thebaroffset\optbar{\kern -\thebaroffset \Bd}\xspace}
\def\Bu      {{\ensuremath{\B^+}}\xspace}

\def\Bp      {{\ensuremath{\Bu}}\xspace}

\def\Bs      {{\ensuremath{\B^0_\squark}}\xspace}

\def\BsorBsbar {\kern \thebaroffset\optbar{\kern -\thebaroffset \Bs}\xspace}

\def\Bds     {{\ensuremath{\B_{(\squark)}^0}}\xspace}

\def\jpsi     {{\ensuremath{{\PJ\mskip -3mu/\mskip -2mu\Ppsi}}}\xspace}
\def\psitwos  {{\ensuremath{\Ppsi{(2S)}}}\xspace}

\def\Y#1S{\ensuremath{\PUpsilon{(#1S)}}\xspace}

\def\proton      {{\ensuremath{\Pp}}\xspace}

\def\Lz          {{\ensuremath{\PLambda}}\xspace}

\def\LorLbar     {\kern \thebaroffset\optbar{\kern -\thebaroffset \PLambda}\xspace}
\def\Lambdares   {{\ensuremath{\PLambda}}\xspace}

\def\Lb           {{\ensuremath{\Lz^0_\bquark}}\xspace}

\def\BF         {{\ensuremath{\mathcal{B}}}\xspace}

\newcommand{\decay}[2]{\ensuremath{\mathinner{#1\!\to #2}}\xspace}

\def\to                 {\ensuremath{\rightarrow}\xspace}

\newcommand{\tauL}{{\ensuremath{\tau_{\mathrm{ L}}}}\xspace}
\newcommand{\tauH}{{\ensuremath{\tau_{\mathrm{ H}}}}\xspace}

\def\CP                {{\ensuremath{C\!P}}\xspace}

\def\AT#1     {\ensuremath{A_{\mathrm{T}}^{#1}}\xspace}           

\def\C#1      {\ensuremath{\mathcal{C}_{#1}}\xspace}                       
\def\Cp#1     {\ensuremath{\mathcal{C}_{#1}^{'}}\xspace}                    
\def\Ceff#1   {\ensuremath{\mathcal{C}_{#1}^{\mathrm{(eff)}}}\xspace}        
\def\Cpeff#1  {\ensuremath{\mathcal{C}_{#1}^{'\mathrm{(eff)}}}\xspace}       
\def\Ope#1    {\ensuremath{\mathcal{O}_{#1}}\xspace}                       
\def\Opep#1   {\ensuremath{\mathcal{O}_{#1}^{'}}\xspace}                    

\newcommand{\aunit}[1]{\ensuremath{\text{\,#1}}}       

\newcommand{\tev}{\aunit{Te\kern -0.1em V}\xspace}
\newcommand{\gev}{\aunit{Ge\kern -0.1em V}\xspace}
\newcommand{\mev}{\aunit{Me\kern -0.1em V}\xspace}
\newcommand{\kev}{\aunit{ke\kern -0.1em V}\xspace}
\newcommand{\ev}{\aunit{e\kern -0.1em V}\xspace}
 
\newcommand{\mevc}{\ensuremath{\aunit{Me\kern -0.1em V\!/}c}\xspace}
\newcommand{\gevc}{\ensuremath{\aunit{Ge\kern -0.1em V\!/}c}\xspace}
\newcommand{\mevcc}{\ensuremath{\aunit{Me\kern -0.1em V\!/}c^2}\xspace}
\newcommand{\gevcc}{\ensuremath{\aunit{Ge\kern -0.1em V\!/}c^2}\xspace}

\def\fb   {\ensuremath{\aunit{fb}}\xspace}
\def\invfb   {\ensuremath{\fb^{-1}}\xspace}

\def\ns   {\ensuremath{\aunit{ns}}\xspace}
\def\ps   {\ensuremath{\aunit{ps}}\xspace}

\newcommand{\chisq}{\ensuremath{\chi^2}\xspace}
\newcommand{\chisqndf}{\ensuremath{\chi^2/\mathrm{ndf}}\xspace}
\newcommand{\chisqip}{\ensuremath{\chi^2_{\text{IP}}}\xspace}

\def\deriv {\ensuremath{\mathrm{d}}}

\def\gsim{{~\raise.15em\hbox{$>$}\kern-.85em
          \lower.35em\hbox{$\sim$}~}\xspace}
\def\lsim{{~\raise.15em\hbox{$<$}\kern-.85em
          \lower.35em\hbox{$\sim$}~}\xspace}

\def\sqs   {\ensuremath{\protect\sqrt{s}}\xspace}

\def\pt         {\ensuremath{p_{\mathrm{T}}}\xspace}

\def\evtgen     {\mbox{\textsc{EvtGen}}\xspace}

\def\geant      {\mbox{\textsc{Geant4}}\xspace}

\def\photos     {\mbox{\textsc{Photos}}\xspace}

\def\pythia     {\mbox{\textsc{Pythia}}\xspace}

\def\tell1  {TELL1\xspace}
\def\ukl1   {UKL1\xspace}

\newcommand{\eg}{\mbox{\itshape e.g.}\xspace}

\newcommand{\lhcborcid}[1]{\href{https://orcid.org/#1}{\hspace*{0.1em}\raisebox{-0.45ex}{\includegraphics[width=1em]{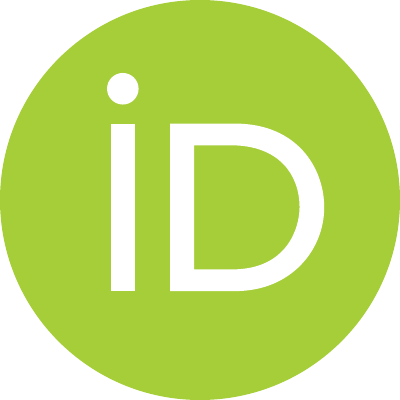}}}}

\hypersetup{
  colorlinks   = true, 
  urlcolor     = blue, 
  linkcolor    = blue, 
  citecolor    = red   
}

\ifEnableSectionTOCLinks
    \usepackage[explicit]{titlesec} 
    
    \let\oldcontentsline\contentsline
    \renewcommand

    \titleformat{\section}{\normalfont\Large\bf}{\hyperlink{tocsection.\thesection}{{\thesection} \parbox[t]{\dimexpr\textwidth-1pc}{#1}}}{1pc}{}

    \titleformat{\subsection}{\normalfont\bf}{\hyperlink{tocsubsection.\thesubsection}{{\thesubsection} \parbox[t]{\dimexpr\textwidth-1pc}{#1}}}{1pc}{}

    \titleformat{name=\section,numberless}[display]{}{}{0pt}{\normalfont\Huge\bfseries #1}
\fi

\usepackage{cite} 
\usepackage{mciteplus}

\usepackage{longtable} 

\begin{document}

\renewcommand{\thefootnote}{\fnsymbol{footnote}}
\setcounter{footnote}{1}


\begin{titlepage}
\pagenumbering{roman}

\vspace*{-1.5cm}
\centerline{\large EUROPEAN ORGANIZATION FOR NUCLEAR RESEARCH (CERN)}
\vspace*{1.5cm}
\noindent
\begin{tabular*}{\linewidth}{lc@{\extracolsep{\fill}}r@{\extracolsep{0pt}}}
\ifthenelse{\boolean{pdflatex}}
{\vspace*{-1.5cm}\mbox{\!\!\!\includegraphics[width=.14\textwidth]{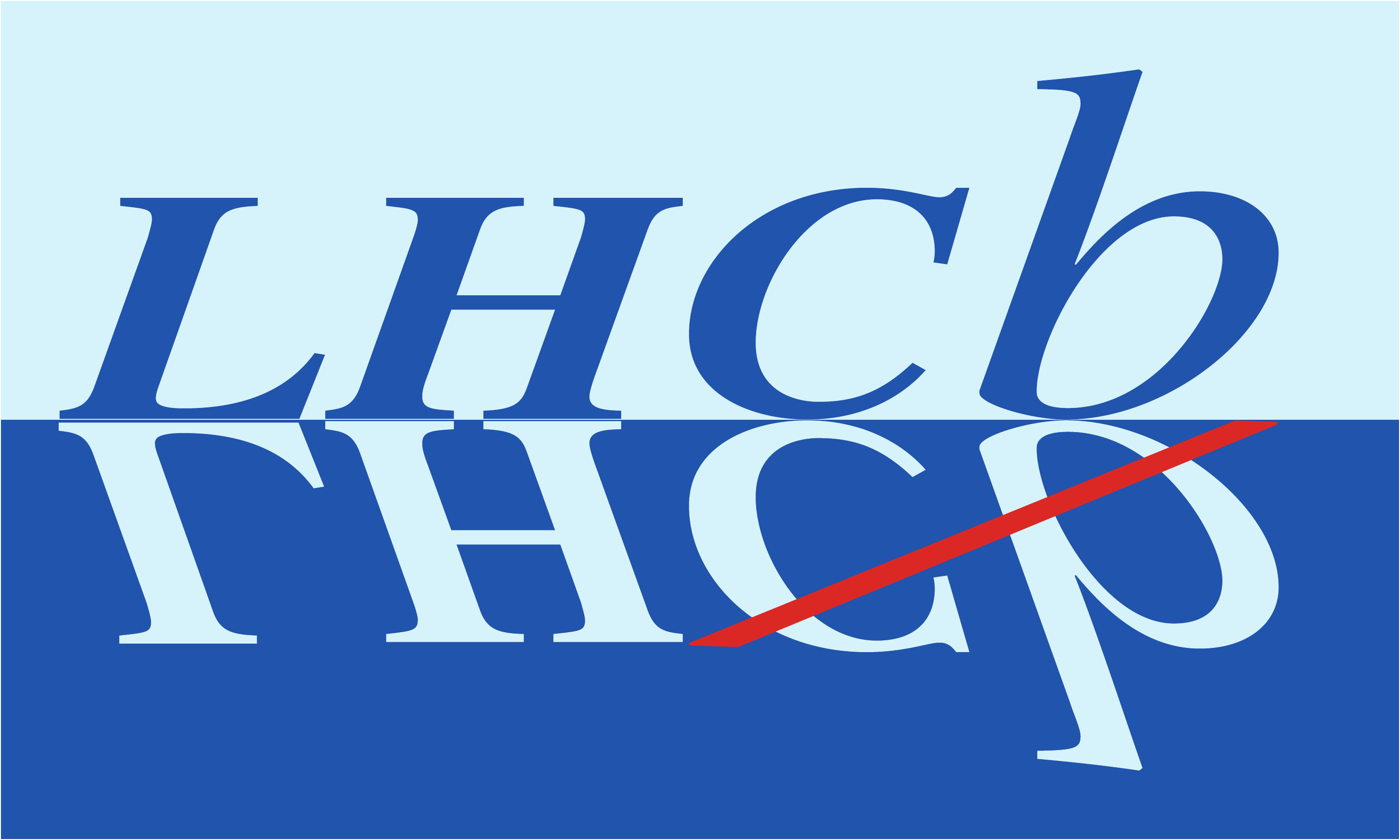}} & &}%
{\vspace*{-1.2cm}\mbox{\!\!\!\includegraphics[width=.12\textwidth]{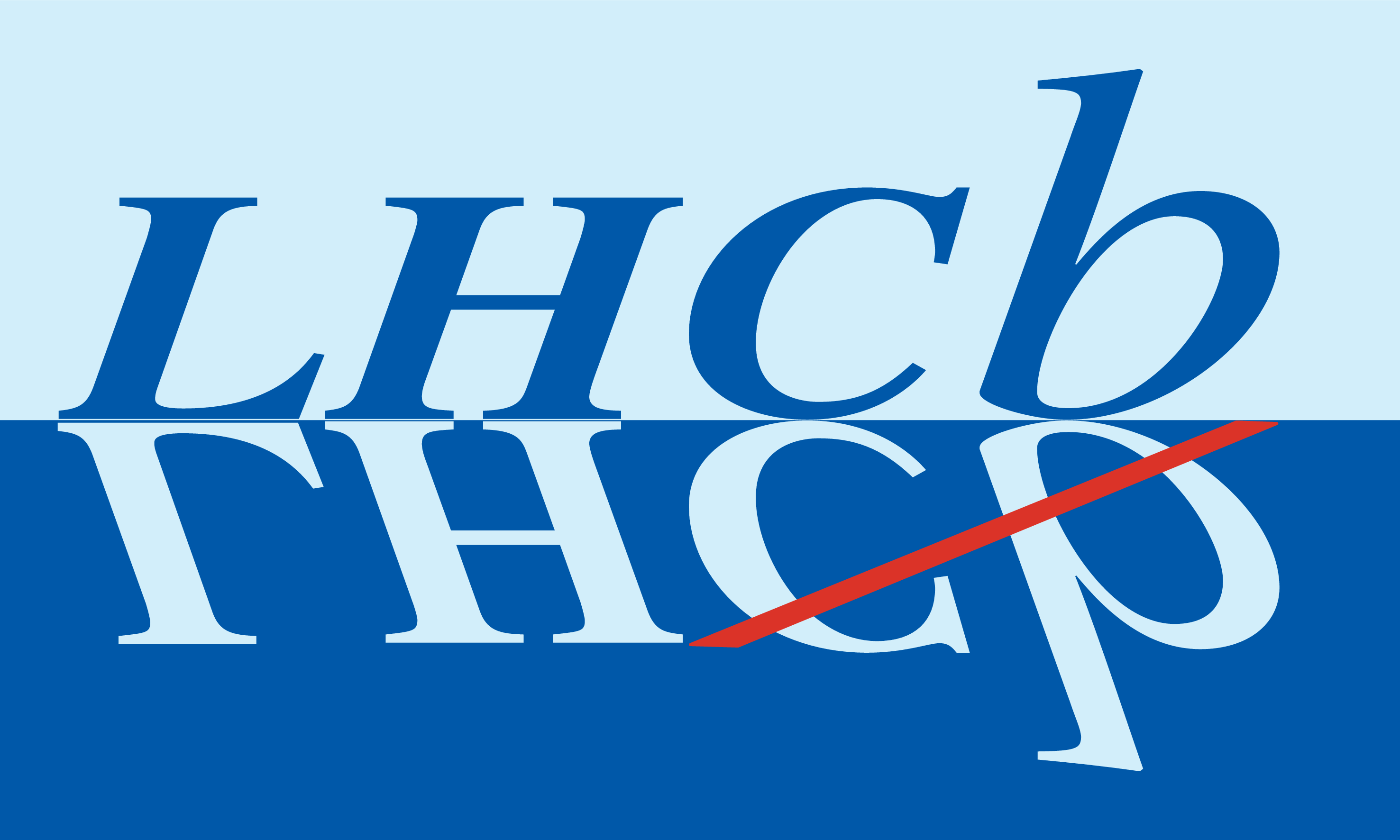}} & &}%
\\
 & & CERN-EP-2026-199 \\  
 & & LHCb-PAPER-2026-014 \\  
 & & 18 August 2026 \\ 
 & & \\
\end{tabular*}

\vspace*{4.0cm}

{\normalfont\bfseries\boldmath\huge
\begin{center}
  \papertitle 
\end{center}
}

\vspace*{2.0cm}

\begin{center}
\paperauthors\footnote{Authors are listed at the end of this paper.}
\end{center}

\vspace{\fill}

\begin{abstract}
  \noindent
  A search for decays of $B$ mesons to final states with four or six muons using $pp$ collision data recorded by the LHCb experiment corresponding to an integrated luminosity of $5.4~\text{fb}^{-1}$ is presented. The decay modes of interest are $B_{(s)}^0 \rightarrow \mu^+\mu^-\mu^+\mu^-$, $B^+ \rightarrow K^+\mu^+\mu^-\mu^+\mu^-$, $B_{(s)}^0 \rightarrow \mu^+\mu^-\mu^+\mu^-\mu^+\mu^-$ and $B^+ \rightarrow K^+\mu^+\mu^-\mu^+\mu^-\mu^+\mu^-$, proceeding via both prompt and long-lived intermediate particles. No evidence for any of the signal modes is found, and upper limits spanning the range of $0.6\times10^{-9}$ to $5.4\times10^{-7}$ at the $95\%$ confidence level are set on their branching fractions, depending on the intermediate-particle masses and lifetimes. In addition, mass-integrated limits across the intermediate-particle lifetime ranges considered in this analysis are determined.
  
\end{abstract}

\vspace*{2.0cm}

\begin{center}
  Submitted to
  JHEP 
\end{center}

\vspace{\fill}

{\footnotesize 
\centerline{\copyright~\papercopyright. \href{\paperlicenceurl}{\paperlicence}.}}
\vspace*{2mm}

\end{titlepage}


\newpage
\setcounter{page}{2}
\mbox{~}
%
%
%
%


\renewcommand{\thefootnote}{\arabic{footnote}}
\setcounter{footnote}{0}


\cleardoublepage


\pagestyle{plain} 
\setcounter{page}{1}
\pagenumbering{arabic}


\section{Introduction}
\label{sec:Introduction}
The Standard Model (SM) of particle physics provides a remarkably successful description of the fundamental particles and their interactions, accurately accounting for a wide range of experimental observations. However, despite its predictive power, it is widely regarded as incomplete. It does not address several open questions, including empirical puzzles such as dark matter and neutrino masses, as well as naturalness problems like the hierarchy problem and the strong-\CP problem. These shortcomings motivate the exploration of theories beyond the Standard Model (BSM).

A possible explanation for the absence of BSM signals is that the corresponding mediators are either too heavy to be produced at current collider energies or too weakly coupled to SM particles to yield detectable signatures. Hierarchical frameworks, like Composite Higgs models, naturally accommodate both features by introducing a new strongly interacting sector with a global symmetry broken at low energies, giving rise to a heavy vector resonance ($V$) and lighter scalar pseudo-Nambu--Goldstone bosons ($a_1, a_2, \ldots$). These can decay either into one another or into SM particles through suppressed couplings to SM fermions~\cite{Blance_2019}. Depending on the specific realisation of the model, these scalar states can be either short- or long-lived~\cite{CidVidal:2022mrx}. Supersymmetric extensions of the SM can give rise to analogous spectra, featuring heavy mediators and lighter states with suppressed couplings, thereby leading to similar experimental signatures \cite{PhysRevD.108.035020, Guchait:2016pes, Parolini:2014rza}.

Decays of $B$ mesons into final states with four and six muons are ideal processes to search for such models when the scalars decay predominantly into leptons.
One possibility is the decay \decay{\Bs}{a a}, where each scalar decays exclusively into a pair of oppositely charged muons, producing a four-muon final state.\footnote{The inclusion of charge-conjugate processes is implied throughout.}
In non-minimal scenarios where more than one scalar is produced by symmetry breaking, six-muon final states can be favoured, arising from decay topologies of the form \decay{\Bs}{a_1 a_2 \to a_1 a_1 a_1 \to\mu^+\mu^-\mu^+\mu^-\mu^+\mu^-}.
Decays featuring muons are typically favoured over those with electrons because of the hierarchy of couplings, whereas a limited phase space restricts the coupling of scalars to tau lepton pairs~\cite{Blance_2019}. 

Such multimuon $B$-meson decays are very rare processes in the SM. The predicted branching fraction for \decay{\Bs}{\mu^+ \mu^- \mu^+ \mu^-} decays is $(0.9$--$1.0) \times 10^{-10}$ \cite{Danilina:2018uzr}, while decays with higher muon multiplicity are suppressed.
However, in the BSM scenarios described above, branching fractions of up to $\mathcal{O}(10^{-8})$ and higher may be obtained for \decay{\Bs}{a_1 a_2} and \decay{\Bp}{K^+ a_1 a_2} decays~\cite{CidVidal:2022mrx}.

Experimentally, the strongest direct constraints on these signatures come from the search for $B$ decays to four muons by the LHCb experiment, finding \mbox{$\BF(\decay{\Bs}{\mu^+ \mu^- \mu^+ \mu^-}) < 8.6 \times 10^{-10}$} and \mbox{$\BF(\decay{\Bz}{\mu^+ \mu^- \mu^+ \mu^-}) < 1.8 \times 10^{-10}$} at the $95\,\%$ confidence level (CL)~\cite{LHCb-PAPER-2021-039}.
Complementary sensitivity arises from searches for muon pairs with the LHCb~\cite{LHCb-PAPER-2020-013} and CMS~\cite{CMSDisplaced} experiments, which can be recast into limits on the branching fraction $\mathcal{B}(\decay{\Bs}{a_1 a_2})<\mathcal{O}(10^{-7})$ for lifetimes of the $a_1$ mediator of up to $10\,\text{ns}$~\cite{CidVidal:2022mrx}. 

\begin{figure}[!bt]
    \centering
    \begin{subfigure}{\linewidth}
        \centering
        \includegraphics[width=0.85\linewidth]{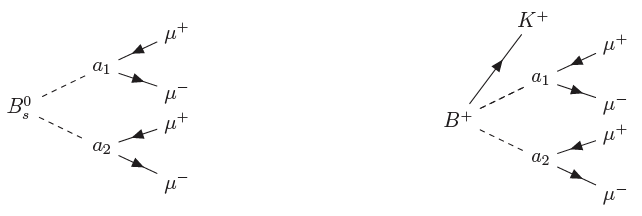}
    \end{subfigure}

    \begin{subfigure}{\linewidth}
        \centering
        \includegraphics[width=0.85\linewidth]{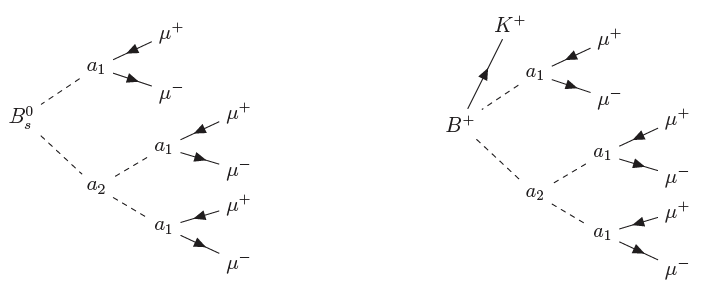}
    \end{subfigure}
    \caption{Decay modes studied in this analysis, including four-muon final states, where $m_{a_2} < 2m_{a_1}$, and six-muon final states, where $m_{a_2} > 2m_{a_1}$.}
    \label{fig:decaymodes}
\end{figure}

This paper reports the search for $B$ decays into four- and six-muon final states at the LHCb experiment, including scenarios where the intermediate scalars can be either short- or long-lived.
Different decay topologies can arise depending on the masses of the scalars, as illustrated in Fig.~\ref{fig:decaymodes}  for $\mumu\mumu$ and $\mumu\mumu\mumu$ states, respectively.
The scalar $a_2$ can either decay into two $a_1$ scalars or directly into two muons.
The scalars are assumed to decay on-shell, as off-shell decays are typically further suppressed~\cite{Blance_2019}.
Concretely, the decays searched for are
\decay{\Bds}{\mu^+ \mu^- \mu^+ \mu^-}, \decay{\Bu}{\Kp \mu^+ \mu^- \mu^+ \mu^-}, \decay{\Bds}{\mu^+ \mu^- \mu^+ \mu^- \mu^+ \mu^-} and \decay{\Bu}{\Kp \mu^+ \mu^- \mu^+ \mu^- \mu^+ \mu^-}.
The reported measurements are performed using $pp$ collision data collected by the LHCb experiment in 2016, 2017 and 2018 at $\sqs=13\tev$, corresponding to an integrated luminosity of 5.4\invfb. Branching fractions are measured relative to the \decay{\Bs}{\jpsi(\to\mu^+ \mu^-)\phi(\to\mu^+ \mu^-)} decay,\footnote{Throughout the paper, $\phi$ is used to mean the $\phi(1020)$ meson.} whose branching fraction has been determined to be $(1.74 \pm 0.14) \times 10^{-8}$ \cite{PDG2024}.

\section{LHCb detector and simulation}
\label{sec:Detector}

The LHCb detector~\cite{LHCb-DP-2008-001,LHCb-DP-2014-002} is a single-arm forward spectrometer covering the pseudorapidity range $2 < \eta < 5$, designed for the study of particles containing \bquark\ or \cquark\ quarks. The detector used to collect the data analysed in this paper includes a high-precision tracking system consisting of a silicon-strip vertex detector (VELO) surrounding the $pp$ interaction region~\cite{LHCb-DP-2014-001}, a large-area silicon-strip detector located upstream of a dipole magnet with a bending power of about $4{\mathrm{\,T\,m}}$, and three stations of silicon-strip detectors and straw drift tubes~\cite{LHCb-DP-2017-001} placed downstream of the magnet. The tracking system
provides a measurement of the momentum, $p$, of charged particles with a relative uncertainty that varies from $0.5\%$ at low momentum to $1.0\%$ at $200\gev$.\footnote{Natural units with $\hbar=c=1$ are used throughout.} The minimum
distance of a track to a primary \proton\proton collision vertex (PV), the impact parameter (IP),
is measured with a resolution of $(15 + 29/\pt) \, \mathrm{\mu m}$, where $\pt$ is the component of the
momentum transverse to the beam measured in \gev. Different types of charged hadrons are distinguished using information from two ring-imaging Cherenkov detectors (RICH)~\cite{LHCb-DP-2012-003}. Photons, electrons and hadrons are identified by a calorimeter system consisting of scintillating-pad and preshower detectors, an electromagnetic calorimeter (ECAL) made from a scintillator/lead structure up to 25 radiation
lengths to fully contain showers from high-energy photons, and a hadronic calorimeter (HCAL), with iron and scintillating tiles running parallel to the beam axis \cite{Amato:494264}. Muons are identified by a system composed of alternating layers of iron and multiwire proportional chambers~\cite{LHCb-DP-2012-002}.

Simulation is required to model the effects of the detector acceptance and the imposed selection requirements. In the simulation, $pp$ collisions are generated using \pythia~\cite{Sjostrand:2007gs} with a specific \lhcb configuration~\cite{LHCb-PROC-2010-056}. Decays of unstable particles are described by \evtgen~\cite{Lange:2001uf}, in which final-state radiation is generated using \photos~\cite{davidson2015photos}. The interaction of the generated particles with the detector, and its response, are implemented using the \geant toolkit~\cite{Allison:2006ve, *Agostinelli:2002hh} as described in Ref.~\cite{LHCb-PROC-2011-006}. 

Simulated signal samples are generated for all relevant decays,
with two mass points concerning the $a_1$ and $a_2$ scalars: configurations \mbox{$(m_{a_1} = 0.25 \, \mathrm{GeV}, m_{a_2} = 0.40\, \mathrm{GeV})$} and \mbox{$(m_{a_1} = 2.00 \, \mathrm{GeV}, m_{a_2} = 2.75\, \mathrm{GeV})$} are generated in $\mumu\mumu$ modes, whereas \mbox{$(m_{a_1} = 0.50 \, \mathrm{GeV}, m_{a_2} = 1.50\, \mathrm{GeV})$} and \mbox{$(m_{a_1} = 1.25 \, \mathrm{GeV}, m_{a_2} = 3.50\, \mathrm{GeV})$} are included for $\mumu\mumu\mumu$ modes. The lifetime of $a_1$ scalars is varied up to $1000\ps$, while $a_2$ resonances are assumed to decay promptly. This strategy follows the phenomenological considerations of Ref.~\cite{CidVidal:2022mrx} and is designed to maximise coverage of the accessible parameter space.
For increased generality, modes with muons not constrained to come from scalars, i.e. \decay{\Bds}{\mu^+\mu^-\mu^+\mu^-}, \decay{\Bu}{\Kp \mu^+\mu^-\mu^+\mu^-}, \decay{\Bds}{\mu^+\mu^-\mu^+\mu^-\mu^+\mu^-}, and \decay{
\Bu}{\Kp \mu^+\mu^-\mu^+\mu^-\mu^+\mu^-} are also studied.
The model used to simulate these decays follows a uniform phase-space~(PHSP) distribution. 
All samples are corrected to account for known data-simulation differences in the $b$-hadron production kinematics, detector occupancy, and trigger efficiencies (see Sec.~\ref{sec:Normalisation}).

\section{Event selection}
\label{sec:EventSelection}

The online event selection is performed by a trigger~\cite{LHCb-DP-2012-004}, which consists of a hardware stage, based on information from the calorimeter and muon systems, followed by a software stage, which applies a full event reconstruction.
In the hardware stage, events containing candidates with at least one muon with transverse momentum above a certain threshold are retained, where the threshold varies across the data-taking period~\cite{LHCb-DP-2019-001}. In the first software trigger stage, candidates must include a high-\pt muon that is inconsistent with having originated at a PV or at least one two-muon combination with a mass larger than $1\gev$. In the second stage, candidates are required to form a two-, three- or four-track vertex inconsistent with the PV, where up to two particles have to be identified as muons.

In the offline reconstruction, particles identified as muons and kaons are combined to produce $B$ candidates formed from between four and seven tracks, with a mass in the range $(4850,6000)$\mev. In the search for decays via short-lived intermediate scalars (referred to as prompt in the following), all tracks are combined in a single step to form a common vertex. For signatures with long-lived scalar intermediates (displaced), scalar candidates are built by combining displaced pairs of muons, which are further combined to form candidate $B$ mesons. 

Muon tracks must satisfy loose requirements on track $\chisqndf$, the probability of reconstructing tracks from unrelated hits in the detector layers, muon identification, and minimum $\pt$. In all cases, muon candidates are required not to point towards any PV.

Multiple categories of track reconstruction are considered, depending on the respective decay scenario.
For the prompt modes, only tracks with hits in all tracking stations are considered; in the following these are referred to  \textit{long}~(L) tracks.

In contrast, displaced candidates are reconstructed either from pairs of \textit{long} tracks or from pairs of \textit{downstream} tracks (D), being defined as tracks reconstructed from hits in the subdetector stations downstream of the VELO subdetector. Muon tracks are combined into dimuon candidates by requiring: (i) a small distance of closest approach between the tracks,  (ii) a low vertex-fit $\chi^{2}$ of the resulting dimuon candidate, which (iii) is required to be well displaced from its associated PV. This is achieved by using the \chisqip\ of the individual tracks,  defined as the difference in the vertex-fit \chisq of a given PV, that is reconstructed with and without the track being considered. These requirements apply to long and downstream muon pairs, although due to the notably worse momentum resolution of downstream tracks, the selection is loosened for the latter. For all reconstructed topologies, at least one pair of long tracks is required, as no trigger algorithm selecting downstream tracks is available for the analysed dataset.
Thus, the different reconstruction categories for the displaced regime are: LL or LD for $\mumu\mumu$ modes or LLL, LLD and LDD for $\mumu\mumu\mumu$ modes.
The muon-pair candidates are further combined to a $B$-meson candidate, which must satisfy requirements on its vertex quality, decay length, and its vertex pointing with respect to their associated PV. 

Finally, in modes with an additional charged kaon, one extra track identified as a kaon is combined with the multimuon candidate. This track must satisfy basic requirements on track quality, $\pt$, and \chisqip with respect to its associated PV.

\begin{figure}[!tbp]
  \begin{center}
    \includegraphics[width=0.5\linewidth]{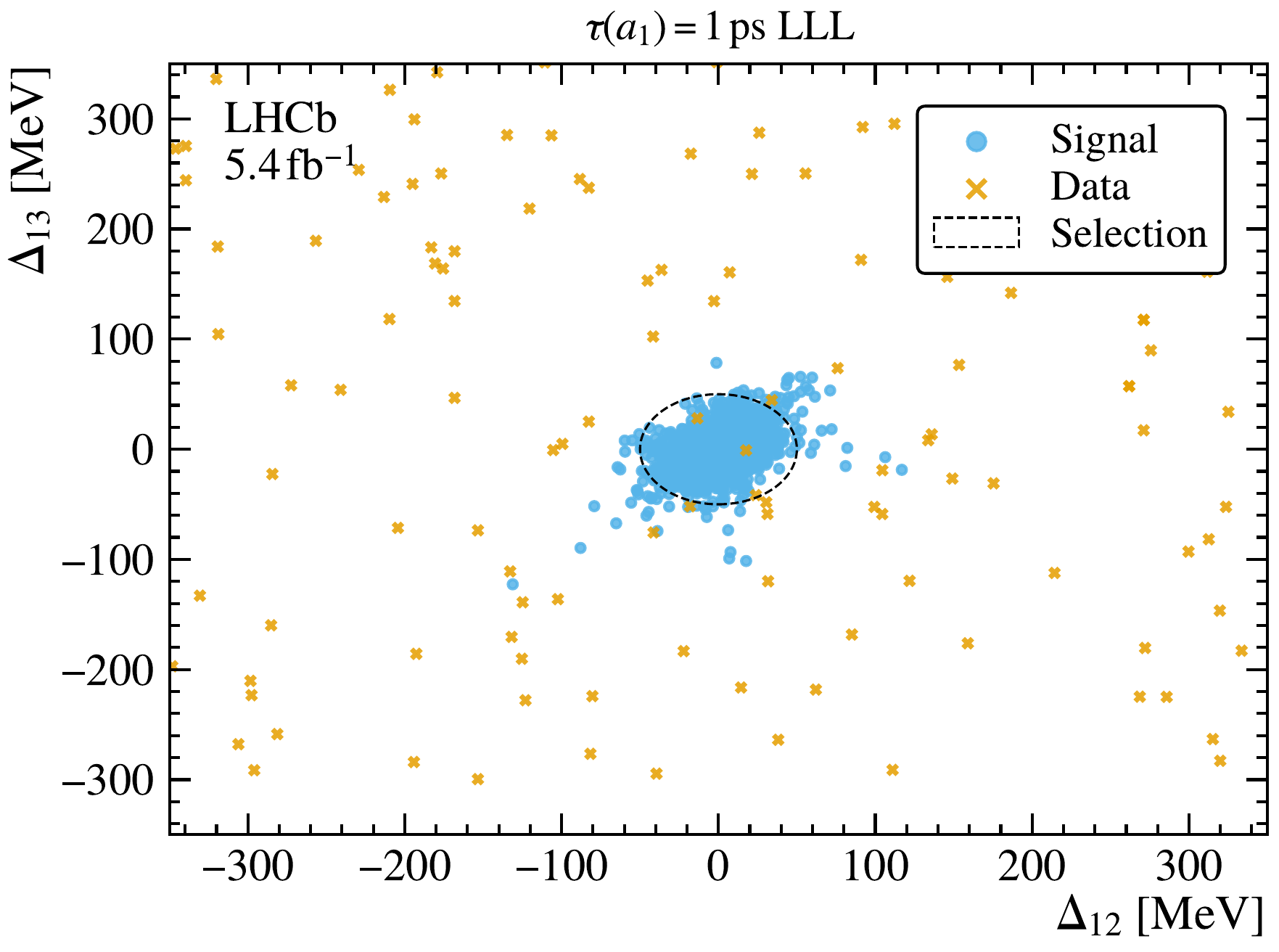}%
    \includegraphics[width=0.5\linewidth]{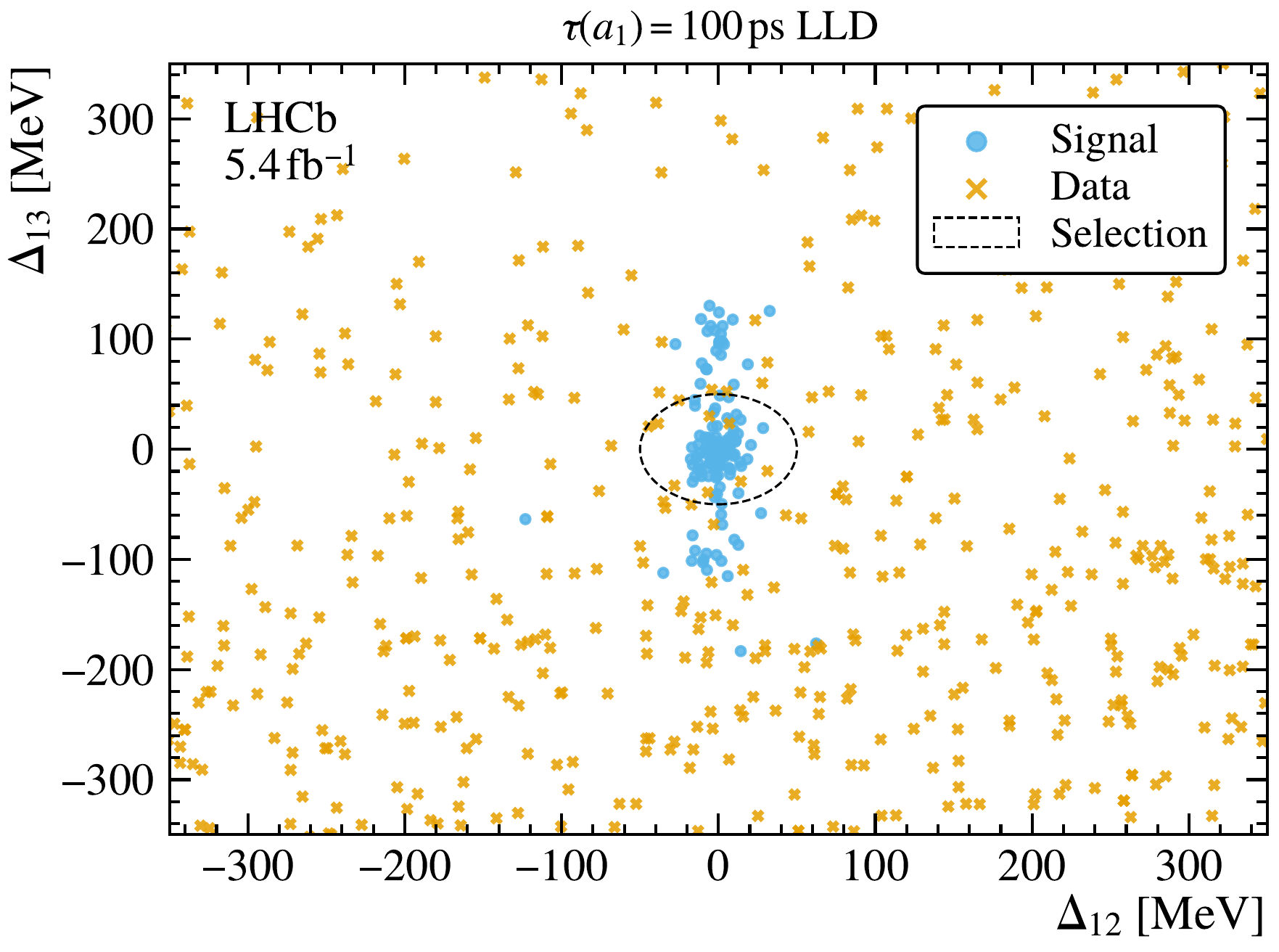}
  \end{center}
  \caption{Representative example of the mass selection for \decay{\Bp}{ \Kp \mu^+ \mu^- \mu^+ \mu^- \mu^+ \mu^-} using simulation generated with $m_{a_1} = 1.25 \gev$, $m_{a_2} = 3.50 \gev$, for the scalar lifetimes (left) $t_{a_1} = 1\ps$  and (right)~$100\ps$, and data from the lower and upper mass sidebands of the $B$-candidate mass distribution. The ellipse encloses the signal selection region. The parameters $\Delta_{1i}$ represent the difference in masses between the $a_1$ scalar coming from the \B vertex and $a_i$, each of the scalars coming from the $a_2$.}
    \label{fig:B6mu_Displ_MassCut}
\end{figure}

Hadronic background is suppressed using particle identification (PID) information provided by the muon system, calorimeters and RICH detectors, which is used to select muons. Particle-identification requirements are also placed on kaon candidates in modes with an additional charged kaon.

Track pairs, where the hits of one particle are used to reconstruct two different tracks, are rejected by imposing a minimum angle requirement between each pair of tracks. In displaced regimes, 
duplicate candidates may also arise due to the same set of muon candidates being arranged in different sets of pairs. They are removed based on the quality of the vertex fit of the long-lived scalar in $\mumu\mumu$ modes or the sum of the fit qualities in $\mumu\mumu\mumu$ modes; retaining only the candidate with the highest quantity per event.

To suppress contamination of the prompt signal modes from well-known SM resonances, such as the $\phi$, $\jpsi$  or $\psitwos$, requirements on the masses of pairs of oppositely charged muon candidates are imposed. Mass regions around these resonances are explicitly vetoed, where the regions are defined as \mbox{$|m(\mu^+\mu^-) - M_{\phi}| < 70 \mev$}, \mbox{$|m(\mu^+\mu^-) - M_\jpsi| < 100 \mev$} and \mbox{$|m(\mu^+\mu^-) - M_\psitwos| < 100\mev$}, respectively, where $M_{\phi}$, $M_\jpsi$ and $M_\psitwos$ are the known $\phi$, $\jpsi$ and $\psitwos$ masses~\cite{PDG2024}. Additionally, to suppress potential contributions from \decay{\Bp}{\Kp \jpsi(\to \mumu\mumu)} decay, candidates with a four-muon mass near the $\jpsi$ resonance are vetoed in the \decay{\Bp}{\Kp \mu^+ \mu^- \mu^+ \mu^-} prompt channel. 

An equivalent approach is employed for displaced samples. In four-muon modes, mass vetoes are included in those dimuon pairs used to build the prompt scalar candidates. For the six-muon modes, mass requirements are applied to exploit the fact that all three reconstructed dimuon pairs are to be mutually consistent, satisfying
\begin{equation}
  S = \sqrt{(\Delta_{12})^{2} + (\Delta_{13})^{2}} < 50 \mev,
\end{equation}
where $\Delta_{1i}$ stands for the difference in masses between the $a_1$ scalar coming from the \B vertex and $a_i$, each of the scalars coming from the $a_2$ . This requirement effectively suppresses background from random track combinations while retaining high signal efficiency.
A representative example of this selection for \decay{\Bs}{\mu^+ \mu^- \mu^+ \mu^- \mu^+ \mu^-} decays with a given mass configuration can be found in Fig.~\ref{fig:B6mu_Displ_MassCut} for scalar lifetimes of $t_{a_1} = 1\ps$ and $t_{a_1} = 100\ps$.

To ensure a clean separation between prompt (muon tracks being originated at the B meson vertex) and long-lived (dimuon tracks compatible with an intermediate displaced vertex) regimes, a decay time of $t > 0.2 \ps$ is required for the displaced muon-pair candidates.
The threshold is determined such that the contribution from decays \decay{\Bs}{\jpsi(\to \mu^+\mu^-)\phi(\to \mu^+\mu^-)} surviving the signal reconstruction criteria is less than 1 in the long-lived regime. In accordance with the prompt regime, this selection functions as a natural veto to prompt resonances like the $\phi$, $\jpsi$ and $\psitwos$ mesons. This selection is illustrated in Fig.~\ref{fig:fom_dectime}, from \ref{sec:fom_dectime}.

The control decay modes, \decay{\Bs}{\jpsi(\to \mu^+\mu^-)\phi(\to \mu^+\mu^-)} and \decay{\Bs}{\jpsi(\to \mu^+\mu^-)\phi(\to K^+K^-)}, are selected in a complementary way to the prompt signal samples. Candidate \decay{\Bs}{\jpsi(\to \mu^+\mu^-)\phi(\to \mu^+\mu^-)} decays are selected by requiring that the mass of one of the muon-pair candidates falls within the $\jpsi$ mass region whilst the other muon pair candidate falls in the $\phi$ mass region. Candidate \decay{\Bs}{\jpsi(\to \mu^+\mu^-)\phi(\to K^+K^-)} decays are selected similarly, with the mass of the kaon pair being consistent with the known $\phi$ mass.

In addition to the aforementioned selections, a multivariate classifier (MVA) is applied to further suppress background from random track combinations in the four-muon modes. A boosted decision tree (BDT) classifier implemented in the \textsc{scikit-learn} toolkit~\cite{py:scikit-hep:2020}, applying the \texttt{AdaBoost} technique~\cite{Breiman,AdaBoost}, is giving the best separation between signal and background.
Unbiased training is achieved via applying the $k$-folding method~\cite{kFold}. Separate BDTs are trained for each of the prompt, displaced-LL and displaced-LD reconstruction categories, and for the different decay channels. The training uses signal simulation and background candidates taken from both low- and high-mass sideband regions in data as signal and background proxies, respectively. Here, the sidebands are defined as the region enclosing a reconstructed B candidate mass between $4850 \mev$ and $6000 \mev$, with the window $(M_B - 75, M_B + 75) \mev$ ($M_B$ is the known mass of the corresponding \B meson) being subtracted.\footnote{The defined signal region contains about $95\%$ of the simulated signal.} To avoid experimenter’s bias, the signal region was not examined while defining and optimising the selection.
The optimal set of input variables for each BDT is determined by recursively training the BDT, removing the training feature with the least importance until a considerable decrease of the area under the receiver operating characteristic (ROC) curve is observed.
The requirement on the BDT response is optimised in terms of background rejection and efficiency preservation with the Punzi figure-of-merit $\varepsilon_{\mathrm{BDT}}/{({3}/{2} + \sqrt{N_{\mathrm{bkg}}})}$~\cite{Punzi:2003bu},
where $\varepsilon_{\mathrm{BDT}}$ is the efficiency of the BDT requirement and $N_{\mathrm{bkg}}$ is the estimated number of background events within the signal region interpolated from the sidebands.

\section{Backgrounds}
\label{sec:Backgrounds}

The use of different reconstruction techniques leads to the presence of different peaking background sources that could mimic the signal processes. In order to identify the dominant background sources, a broad survey of possible background channels is made, for which: i) corresponding branching fractions are taken from their known values~\cite{PDG2024}, ii) the effects of prompt mass or decay-time selections are included in the corresponding lifetime regime, and iii) hadron-to-muon misidentification rates~\cite{LHCb-DP-2013-001} are included.\footnote{The \decay{\Bs}{\jpsi(\to \mu^+\mu^-)\phi(\to \mu^+\mu^-)} control mode reconstructed as \decay{\Bs}{a_1 a_2 \to \mu^+ \mu^- \mu^+ \mu^-} is used to test their influence on signal decays.} Those with a non-negligible yield are studied according with detailed simulation. After evaluation, no background component has revealed any clear effect on signal estimates.

For prompt-reconstructed samples, the most significant physical backgrounds mainly come from two sources: i) nonresonant $B$ decays, and ii) unflavoured and charmonium decays from \B mesons, including \decay{\phi}{\mu^+ \mu^-}, \decay{\jpsi}{\mu^+ \mu^-} and \decay{\psitwos}{\mu^+ \mu^-}, or resonances like the $\omega$, $\rho^0$ and $\eta$ in the low dimuon mass spectrum. Additionally, semileptonic or hadronic $b$-hadron decays, \eg \decay{\Bs}{\pi^+ \pi^- \mu^+ \mu^-}, \decay{\Bz}{\pi^+ \pi^- \pi^+ \pi^-}, \decay{\Lb}{p^+ K^- \mu^+ \mu^-}, with two or more hadrons misidentified as muons are another possible source of background. In general, prompt-mass vetoes and PID requirements are sufficient to effectively veto contamination from prompt hadron decays, with processes involving \decay{\jpsi}{\mu^+\mu^-(\mu^+\mu^-)} decays like \decay{\Bs}{\jpsi (\to \mu^+ \mu^-) K^+ K^-} or \decay{B^+}{\Kp \jpsi (\to \mu^+ \mu^- \mu^+ \mu^-)}.

Backgrounds affecting non-prompt signals mainly proceed from two sources: i) prompt decays that, due to finite vertex resolution, are erroneously classified as displaced, and ii) decays involving pure long-lived resonances like the $\KS$ and $\Lambdares$ hadrons. Decays of the first type are likely to contaminate displaced categories with long tracks, whereas decays of the second type pose a threat to both displaced categories with long and downstream tracks. Likewise, but in a residual way, $\Lb$ candidates with a similar topology to the signal modes (\eg \decay{\Lb}{\Lambdares( \to p\pi^-)\mu^+ \mu^-}) can be pushed to the region of interest because of their proximity to the $B$ meson mass. Examples of decays arising in the displaced regime are \decay{\Bz}{\jpsi(\to \mu^+\mu^-)\KS(\to\pi^+\pi^-)} or \decay{\Bz}{\jpsi(\to \mu^- \mu^+) K^{*}(892)^0(\to K^+ \pi^-)}.

While numerous processes can mimic four-muon signatures, very few can produce six reconstructed muons. For example, one possible source within the SM is given by the decay \decay{\Bs}{\jpsi(\to \mu^+ \mu^-)\phi(\to \mu^+ \mu^-)\phi(\to \mu^+ \mu^-)}, whose branching fraction is at the level of $10^{-13}$, orders of magnitude below the signal sensitivity of this analysis. Other potential six-muon backgrounds are similarly suppressed and, after applying the requirements described in Sec. \ref{sec:EventSelection}, are found to have negligible contributions in all prompt- and displaced-reconstruction categories, with expected yields of less than $10^{-4}$.

Once the modes requiring dedicated studies are identified, the prompt, displaced-LL and displaced-LD offline selection criteria are applied over simulated samples and the number of expected candidates is computed as

\begin{equation}
    N_{\text{bkg}} = \BF_\text{bkg} \times \frac{N_{\text{norm}}}{\BF_{\text{norm}}} \times \frac{\varepsilon_{\text{bkg}}}{\varepsilon_{\text{norm}}} \times \frac{f_{\text{bkg}}}{f_{\text{norm}}},
\end{equation}
where $\BF_i$, $N_i$, $\varepsilon_i$ and $f_{i}$ are the known branching ratios~\cite{PDG2024}, yields, total efficiency of the selection criteria, and $b$-quark fragmentation fraction to each $b$-hadron species of the background and normalisation channels. The integrated ratio of $f_s/f_d$ has been measured at $\sqrt{s} = 13\tev$ for $0.5\gev < \pt < 4.0\gev$ and $2.0 < \eta < 6.4$ and was found to be \mbox{0.2539 $\pm$ 0.0079}~\cite{LHCb-PAPER-2020-046}. In addition,  $f_u \approx f_d$, due to isospin symmetry.

For the \decay{\Bs}{\mumu\mumu} searches, contributions from processes with two pions in the final state can populate regions of the mass spectrum in the signal region, and modes such as \decay{\Bs}{\jpsi K^+K^-} appear in the low-mass sidebands. A potential background contribution across all reconstruction categories is the decay \decay{\Bz}{\jpsi(\to \mu^+\mu^-)\KS(\to \pi^+\pi^-)}, which after the full selection requirements, yields a contamination compatible with zero. In the displaced categories, the most important contribution arises from \decay{\Bs}{\jpsi \pi^+\pi^-}. However, these decays are also suppressed to a negligible level through the decay-time and PID criteria.

For the \decay{B^+}{\Kp \mu^+\mu^- \mu^+\mu^-} search, the background composition is similar. Processes with two pions in the final state dominate the sidebands, with partially reconstructed decays populating the low-mass region in the prompt, LL, and LD categories. The decay mode \decay{B^+}{\Kp \jpsi(\to\mu^+\mu^-)\eta(\to\mu^+\mu^-\gamma)} contributes at a low level but does not form a peaking structure. The most significant peaking background arises from \decay{B^+}{\Kp \jpsi(\to \mumu\mumu)}, which is efficiently removed by vetoing candidates with a four-muon mass near the $\jpsi$ resonance. This requirement eliminates the only non-negligible peaking background in any reconstruction category.

In addition, partially reconstructed backgrounds are strongly suppressed by kinematic and PID requirements. Semileptonic decays such as \decay{\Bz}{ D^- \mu^+\nu_\mu}, with \decay{D^-}{K^- \pi^+ \pi^-}, can mimic signal-mode vertices in the LD and LDD categories if hadrons are misidentified as muons. However, the combined misidentification rates required for these topologies reduce their expected contribution to negligible levels. Partially reconstructed signal decays with higher muon multiplicities, \eg \decay{B}{\mu^+\mu^-\mu^+\mu^-\mu^+\mu^-}, can produce apparent four-muon candidates if one or more muons fail the reconstruction criteria. However, these decays populate the low-mass region and do not lie within the signal windows in either prompt or displaced searches.

To summarise, a detailed study of potential SM backgrounds determines that none of the processes considered might contaminate the signal region, with combinatorial background as the only remaining component.

\section{Normalisation}
\label{sec:Normalisation}

The signal branching ratio is measured with respect to the branching ratio of the normalisation resonant decay \decay{\Bs}{\jpsi(\to \mu^+\mu^-)\phi(\to \mu^+\mu^-)}, and is parameterised in terms of signal yield as $N_{\text{sig}} = \mathcal{B}_{\text{sig}}/\alpha_{\text{sig}}$, with the single-event sensitivity $\alpha_{\text{sig}}$ given by
\begin{equation}
        \alpha_{\text{sig}} = \frac{\BF_{\text{norm}}}{N_{\text{norm}}} \times \frac{\varepsilon_{\text{norm}}}{\varepsilon_{\text{sig}}} \times \frac{f_{\text{norm}}}{f_{\text{sig}}} \,,
        \label{eq:alpha_sig}
\end{equation}
where the labels ``sig'' and ``norm'' refer to the signal and normalisation modes.

Efficiencies are calculated using simulated decays, to which weights are applied in order to minimise discrepancies between data and simulation. These weights are calculated by comparing \decay{\Bs}{\jpsi(\to\mu^+\mu^-)\phi(\to K^+K^-)} decays in data and simulation, for which the branching fraction is three orders of magnitude larger than for \decay{\Bs}{\jpsi(\to \mu^+\mu^-)\phi(\to \mu^+\mu^-)}, allowing for a more precise determination of differences between data and simulation. The trigger and offline selections for these decays are similar to those used for \decay{\Bs}{\jpsi(\to \mu^+\mu^-)\phi(\to \mu^+\mu^-)} decays, but no PID requirement is applied on the kaons. The distributions of the variables of interest for \decay{\Bs}{\jpsi(\to\mu^+\mu^-)\phi(\to K^+K^-)} decays are separated from background in data using the \textit{sPlot} method \cite{Pivk:2004ty}, with the $B$-candidate mass used as the discriminating variable.

The first set of weights, $w_{\text{gen}}$, referred to as the generator weights, corrects the $\pt$ and pseudorapidity distributions of the $B$ mesons, along with the multiplicity of the underlying events. The weights $w_{\text{L0}}$, referred to as the trigger weights, correct the efficiency of the hardware (L0) trigger. The weights $w_{\text{rec}}$, referred to as the reconstruction weights, correct the vertex $\chi^2$, $\chisqip$ and flight-distance significance of the $B$ mesons. The efficiency for a given decay mode can be expressed as
\begin{equation}
    \varepsilon_{\mathrm{sig}} = \varepsilon_{\text{geo}} \times \frac{\sum\limits_{\text{presel}} w_{\text{gen}}}{\sum\limits_{\text{gen}}w_{\text{gen}}}\times \frac{\sum\limits_{\text{trig}}w_{\text{gen}}w_{\text{L0}}}{\sum\limits_{\text{presel}} w_{\text{gen}}}\times \frac{\sum\limits_{\text{sel}}w_{\text{gen}}w_{\text{L0}}w_{\text{rec}}}{\sum\limits_{\text{trig}} w_{\text{gen}} w_{\text{L0}}w_{\text{rec}}} \,,
    \label{eq:weights}
\end{equation}
where $\varepsilon_{\text{geo}}$ is the detector geometrical acceptance efficiency, the label ``gen'' refers to the generated sample before any reconstruction or selection, while ``presel'', ``trig'' and ``sel'' refer to the offline reconstruction sample passing the preselection requirements, the preselection and trigger requirements, and the full selection requirements, respectively. 
In the above equation, the reconstruction weights are normalised such that $\sum\limits_{\text{trig}}w_{\text{gen}}w_{\text{L0}} = \sum\limits_{\text{trig}} w_{\text{gen}} w_{\text{L0}}w_{\text{rec}}$.
Efficiencies are calculated individually for each year of data taking and are then combined in a weighted average according to the integrated luminosity corresponding to each sample. 

The dependence of the selection efficiency on the scalar-particle decay time is studied, which in turn drives the shape of the excluded branching-fraction curves. This study is not used to extract efficiencies directly, but rather as a consistency check of the lifetime dependence and of the reconstruction categories included in the analysis. To model this dependence, per-event weights are applied at both generator and reconstruction levels according to
\begin{equation}
    \omega_i = \frac{\tau_{\text{gen}}}{\tau_{\text{tar}}} \times \frac{e^{-t_i/\tau_{\text{tar}}}}{e^{-t_i/\tau_{\text{gen}}}
    } \,,
\end{equation}
and then included in Eq.~\ref{eq:weights}. Here, $t_i$ is the decay time of the scalar, $\tau_{\text{gen}}$ is the lifetime used to generate the simulated sample, and $\tau_{\text{tar}}$ is the target lifetime. For target decay time lying between two generated lifetime points, $g_1 < t < g_2$, the corresponding efficiency is obtained through a linear interpolation following a finite-elements approach,
\begin{equation}
    \varepsilon_{\mathrm{sig}} = w_1\,\varepsilon^{g_1} + w_2\,\varepsilon^{g_2},
    \label{eq:finitelements}
\end{equation}
with weights defined as
\begin{equation}
    \begin{split}
        w_1 & = 1 - f, \\
        w_2 & = f = \frac{t - g_1}{g_2 - g_1}.
    \end{split}
\end{equation}

\begin{figure}[!tb]
    \centering
    \includegraphics[width=0.5\linewidth]{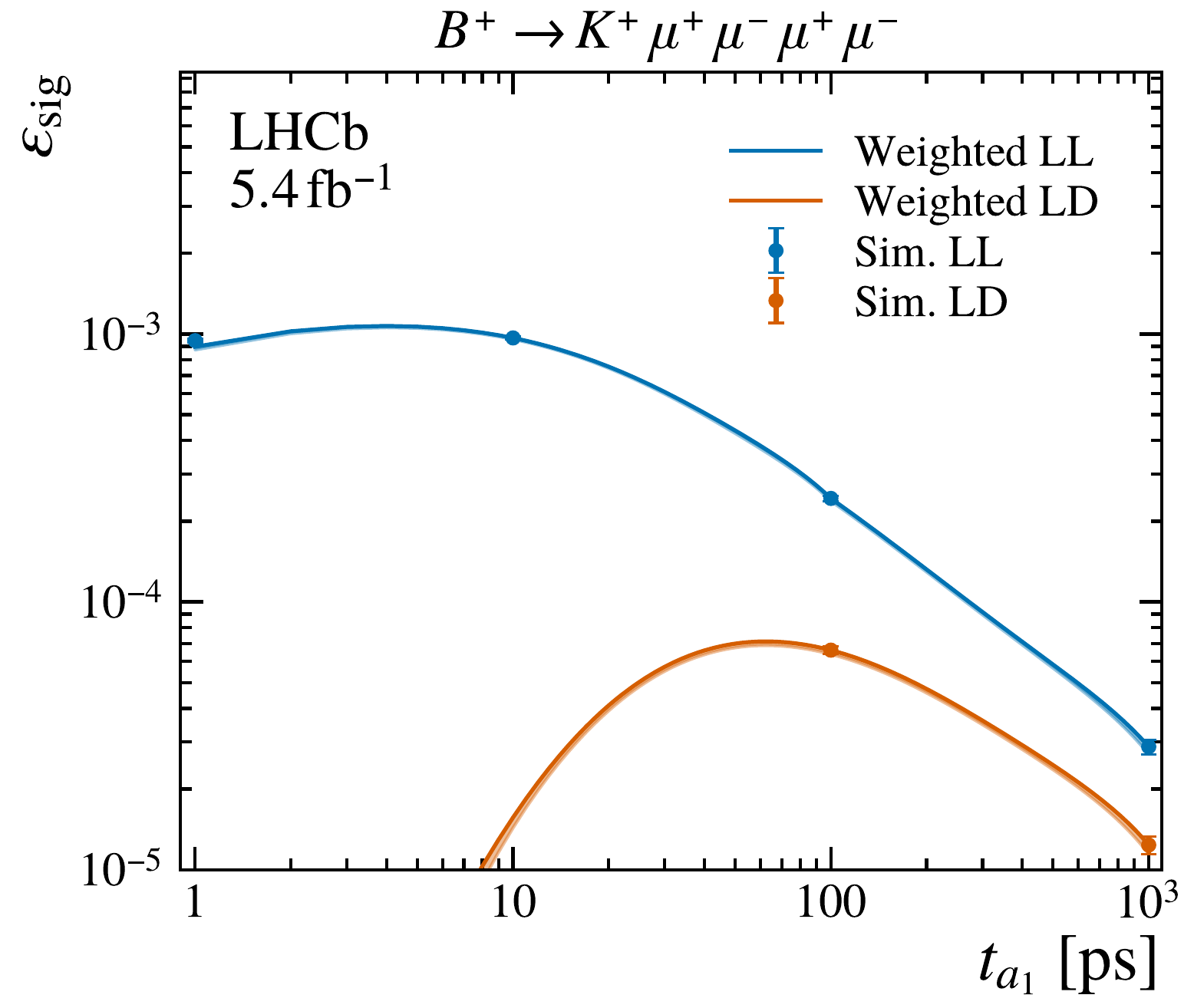}%
    \includegraphics[width=0.5\linewidth]{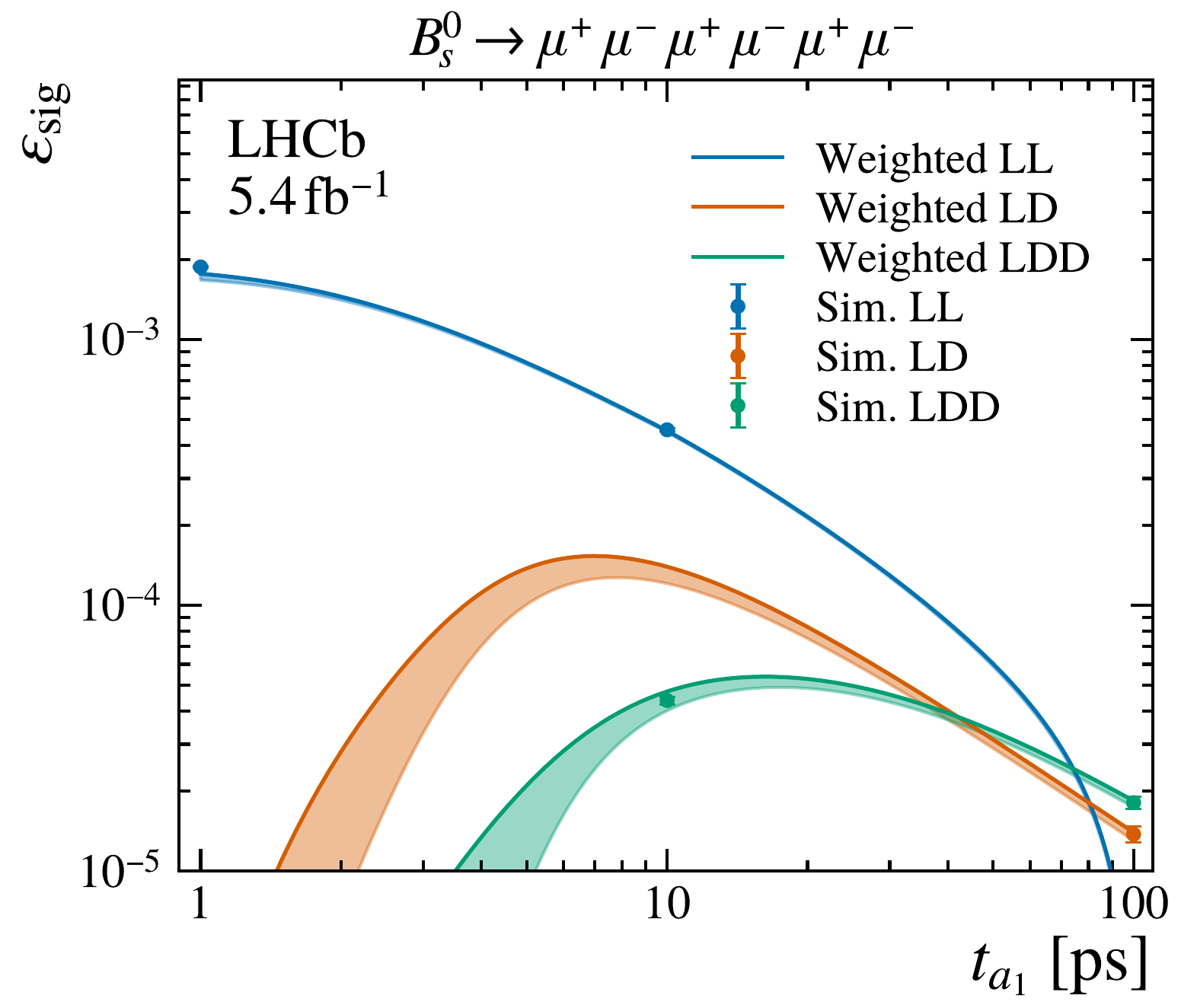}
    \caption{Extrapolation of total efficiencies across lifetime for  (left) \decay{B^+}{\Kp \mu^+ \mu^- \mu^+ \mu^-} with \mbox{$(m_{a_1}, m_{a_2}) = (2.00, 2.75)\gev$}, and (right)  \decay{\Bs}{\mu^+ \mu^- \mu^+ \mu^- \mu^+ \mu^-} with  \mbox{$(m_{a_1}, m_{a_2}) = (0.50, 1.50) \gev$}. Colour bands represent statistical uncertainties on the efficiency that are propagated through the finite-elements approach in Eq. \ref{eq:finitelements}.}
    \label{fig:lf_rew}
\end{figure}

Figure~\ref{fig:lf_rew} shows representative examples of the resulting total-efficiency behaviour as a function of the scalar decay time for long and downstream track reconstruction criteria in the \decay{B^+}{\Kp \mu^+ \mu^- \mu^+ \mu^- } and \decay{\Bs}{\mu^+ \mu^- \mu^+ \mu^- \mu^+ \mu^-} signal modes. The smooth evolution of the efficiencies with lifetime provides a qualitative understanding of the lifetime dependence observed in the excluded branching-fraction curves. In addition, it is confirmed that at large lifetimes, the inclusion of downstream track reconstruction is well behaved within the analysis.

The single-event sensitivity $\alpha_{\text{sig}}$, which is defined in Eq.~\ref{eq:alpha_sig}, sets the scale at which $\BF_\text{sig}$ can be measured if signal candidates are observed. The quantity $N_{\text{norm}}$ is determined from a fit to the mass distribution of the normalisation mode,  $\BF_{\text{norm}}$ is determined from the product of existing branching fraction measurements~\cite{PDG2024}, namely \decay{\Bs}{\jpsi\phi}, \decay{\jpsi}{\mu^+\mu^-} and \decay{\phi}{\mu^+\mu^-}, yielding $\BF_{\text{norm}}\;=\;(1.74 \pm 0.14) \times 10^{-8}$.

\begin{figure}[t!]
    \centering

    \includegraphics[width=0.85\textwidth]{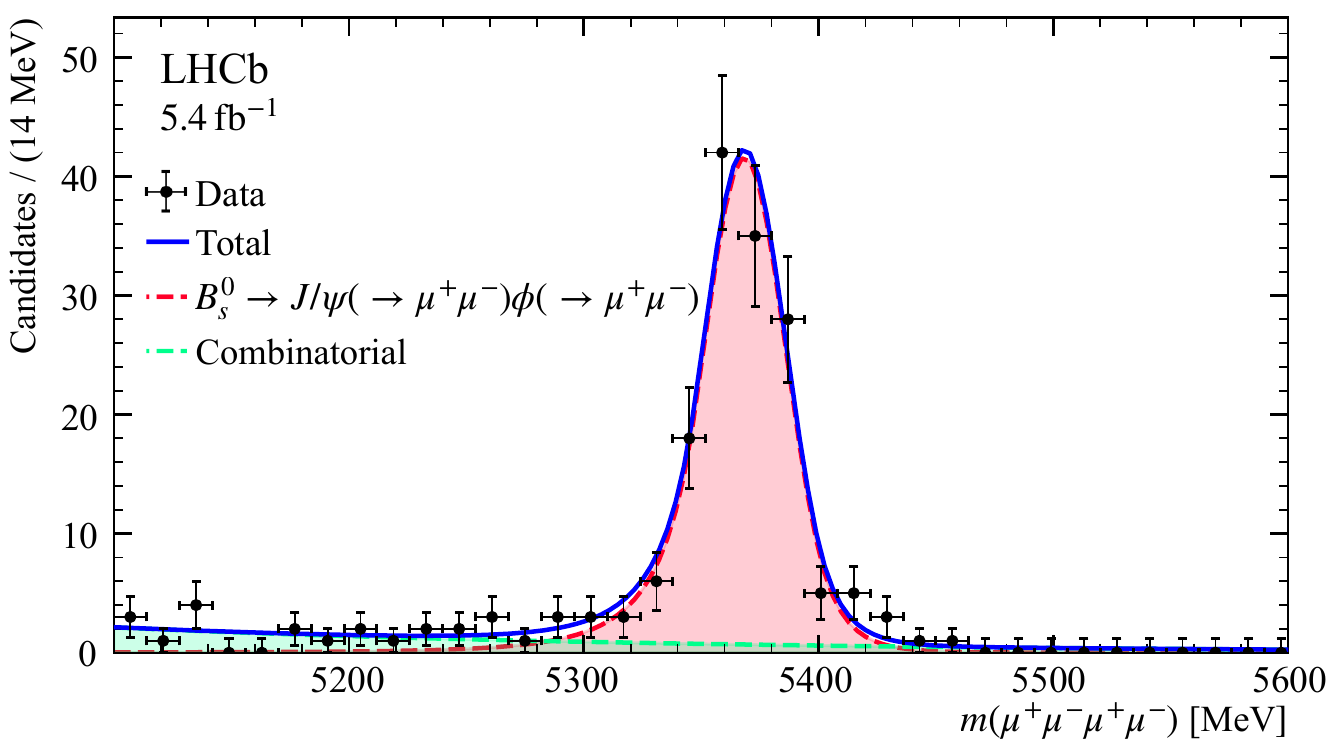}
    \caption{Distribution of the $\mu^+\mu^-\mu^+\mu^-$ mass of candidates passing the \decay{\Bs}{\jpsi(\to \mu^+\mu^-)\phi(\to \mu^+\mu^-)} selection procedure with the result of the fit also shown.}
    \label{fig:ControlChannelEvents}
\end{figure}

The yield of the normalisation mode $N_{\text{norm}}$ is determined from an extended unbinned maximum-likelihood fit to the mass distribution of \decay{\Bs}{\jpsi(\to \mu^+\mu^-)\phi(\to \mu^+\mu^-)} candidates with the full normalisation-mode selection criteria applied, as shown in Fig.~\ref{fig:ControlChannelEvents}. The fit model is composed of a double-sided Crystal Ball~(DSCB) function \cite{Skwarnicki:1986xj} for the signal, and an exponential function for the combinatorial background. The parameters of the DSCB function are determined from an unbinned maximum-likelihood fit to simulated \decay{\Bs}{\jpsi(\to \mu^+\mu^-)\phi(\to \mu^+\mu^-)} decays. The values obtained for the parameters in the fit to simulation are used to constrain with Gaussian priors the parameters of the DSCB function, except for the peak position $\mu$ and width parameter $\sigma$, which are allowed to vary freely, accounting for differences in resolution and mass scale between data and simulation. The slope of the background model is also allowed to vary freely. The fitted yield of normalisation candidates is \mbox{$N_{\text{norm}}$ = 145 $\pm$} 13, where the uncertainty is statistical only.

\section{Fit procedure}
\label{sec:FitProcedure}

Extended maximum-likelihood fits to the mass of the $B$-meson candidates in the range $(4850,6000)\mev$ are used to search for the signal decays and to extract the signal branching ratio. In the case that no significant signal is observed, 95\% CL upper limits on the branching fractions are set using the $\mathrm{CL}_\mathrm{s}$ method~\cite{CLs}. For prompt searches and displaced modes with $\tau_{a_1} = 1\ps$, individual fits are performed to each mass component, since no effect from displaced reconstruction (neither long nor downstream tracks) is present. On the other hand, simultaneous fits to LL and LD samples, or LLL, LLD and LDD samples for very displaced modes ($t_{a_1}> 1\, \mathrm{ps}$) are performed, sharing the branching-ratio parameter and weighted by the corresponding single-event sensitivities.

The signal component is modelled using a DSCB function. Its shape parameters are obtained by fitting to simulation candidates that remain after the complete selection criteria. In particular, mismodelling is often observed, impacting the stability of some fit parameters. Due to strong correlations between some parameters, the power-law exponents of the DSCB  are fixed during the fit, while the transition points are allowed to float. This procedure has been shown to improve fit stability. The signal shape parameters are subsequently fixed in the mass fit, and the single-event sensitivities for each category are constrained with Gaussian priors, where the width of the Gaussian comprises both statistical and systematic uncertainties. The background is dominated by a combinatorial contribution, whose shape is modelled by an exponential function. Thus, the fitting strategy relies on an unbinned maximum-likelihood fit whose signal and background components are normalised to $N_{\rm sig}$ and $N_{\rm bkg}$, respectively. 

As statistical uncertainties are large due to the low event yields, validation studies are crucial to reaffirm the fitting approach. Therefore, pseudoexperiments are generated under the signal-plus-background hypothesis. For each pseudoexperiment, the total number of events is fixed to 
$N = N_{\text{bkg}} + N_{\text{sig}}$, and \B-candidate mass values are randomly sampled from the combined signal and background probability density functions (PDFs). To test the robustness of the fit model, $N_{\text{sig}}$ is chosen at various points corresponding to branching ratios up to $10^{-8}$, ensuring that the model is evaluated under meaningful conditions.
Each pseudoexperiment is then fitted using the baseline model. From the fit results, standardised residuals of the extracted signal branching fractions
$\BF$ are computed for all signal modes, where $N$ denotes the number of muons. In addition, validation tests with $\BF = 0$ are performed and confirming the reliability of the model in the absence of signal. In all cases, $1000$ samples are generated, and the occurrence of failed fits does not exceed $1\%$. No particular deviation is observed in any of the signal modes, with the standardised-residual distribution being compatible with a normal Gaussian distribution.

Finally, in order to extrapolate the sensitivity reach across the lifetime spectrum of the $a_1$ scalar, the core resolution of the signal PDF is studied for different reconstruction categories to validate the assumption that the signal mass resolution is similar across different scalar lifetimes. For all categories, the fitted resolutions are found to be compatible within uncertainties for different lifetimes.

\section{Systematic uncertainties}
\label{sec:SystematicUncertainties}

A number of systematic uncertainties have the potential to affect the measurements reported in this paper, with dominant contributions arising from the potential mismodelling of the PID efficiencies and the background distribution. The full list can be found in Tables~\ref{tab:combined_systematics_Bs} and~\ref{tab:combined_systematics_Bu} of Appendix~\ref{sec:systematics_app}.

Due to the underlying decay topology, the signal decay modes can exhibit a nontrivial dependence on the momentum transfer, $q^2$. However, this is not modelled in the simulation samples used in this analysis, which are generated under a phase-space model assumption. In particular, it has been shown that for \decay{B^+}{K^+ a_1 a_2} decays, the differential width $\deriv\Gamma/\deriv q^2$ follows a nonuniform, physically motivated distribution~\cite{Blance_2019}. Currently, no theoretical prediction for the decay dynamics of \decay{\Bs}{a_1 a_2} decays is available. To assess the impact of this mismatch on selection efficiency, the simulated signal sample is weighted to follow the aforementioned $q^2$ distribution for the \decay{B^+}{K^+ a_1 a_2} decays. The selection efficiency is then recomputed using the weighted samples, and the difference with respect to the baseline phase-space efficiency is assigned as a systematic uncertainty. This results in systematic uncertainties ranging from $0.1\%$ to $10\%$.

The effective lifetime of any $\Bs$ signal mode is unknown a priori and can vary between the lifetimes of the light and heavy eigenstates, denoted as $ \tauL $ and $ \tauH $, respectively~\cite{DeBruyn:2012wj}. These lifetimes are measured to be $\tauL$ = $1.429 \pm 0.006\ps$ and $\tauH$ = $1.622 \pm 0.008\ps$~\cite{PDG2024}. Simulation samples used in this analysis for $\Bs$ signal modes are generated assuming the mean $\Bs$ lifetime. Since the selection efficiencies depend on the $\Bs$ flight distance, the effective lifetime introduces a source of systematic uncertainty on the efficiencies. To quantify this, the simulated candidates are weighted such that their true decay times correspond to an effective lifetime of either $\tauL$ or $\tauH$. The selection efficiencies are recomputed using these weighted samples, and the largest difference with respect to the baseline efficiency is assigned as a systematic uncertainty. This results in systematic uncertainties of up to 7.1\%.

The parametrisation of the combinatorial background constitutes the dominant source of model dependence in the limit-setting procedure. To assess its impact, the baseline exponential description is replaced by a linear function and the $95\%$ CL upper limits are re-evaluated using the $\mathrm{CL}_\mathrm{s}$ method~\cite{CLs}. Unlike the other systematic uncertainties, which act multiplicatively on the signal efficiency and are propagated to the single-event sensitivity through Gaussian constraints, this term reflects how much the extracted branching fraction changes under a different background hypothesis. The resulting variation of the upper limits ranges from $2\%$ up to $17\%$, the largest values being found in the lowest-yield channels, where the limited number of candidates in the sidebands leaves the background shape poorly constrained. 

Further sources of systematic uncertainties arise from differences between data and simulation. It is known that the performance of the PID algorithms differs between data and simulation.
To correct for this, per-track PID weights are computed using calibration samples in data and simulation that map the PID performance as a function of track kinematics and event multiplicity.
These weights are applied to simulated candidates to correct the PID efficiencies.
However, the calibration samples have a limited size and do not cover the full kinematic range, particularly at low muon momentum.
As a result, the PID performance in simulation may not be fully corrected, leading to potential biases in the selection efficiencies. To quantify the potential bias this introduces, per-muon PID weights are calculated as a function of the muon transverse momentum, $p_{\rm T}^{\mu}$, with the high-yield \decay{\Bs}{\jpsi(\to \mu^+\mu^-)\phi(\to K^+K^-)} control mode.
Weights are then applied to each muon in the simulation samples based on $p_{\rm T}^{\mu}$ and a per-event weight is calculated as the product of these weights.
The selection efficiencies are recomputed with these weights applied, and the difference with respect to the baseline efficiency is assigned as a systematic uncertainty.
These weights are not included in the baseline efficiencies because it is only an approximate estimate of the possible size of the PID mismodelling, parametrised in a single variable, and therefore does not constitute a robust correction.
These systematic uncertainties range from $0.3\%$ to $21\%$, the largest values occurring for the lightest scalar configurations of the four-muon modes, where the softer muon spectrum is most affected by the limited calibration coverage at low momentum.

\begin{figure}[htbp]
    \centering
    \includegraphics[width=0.5\linewidth]{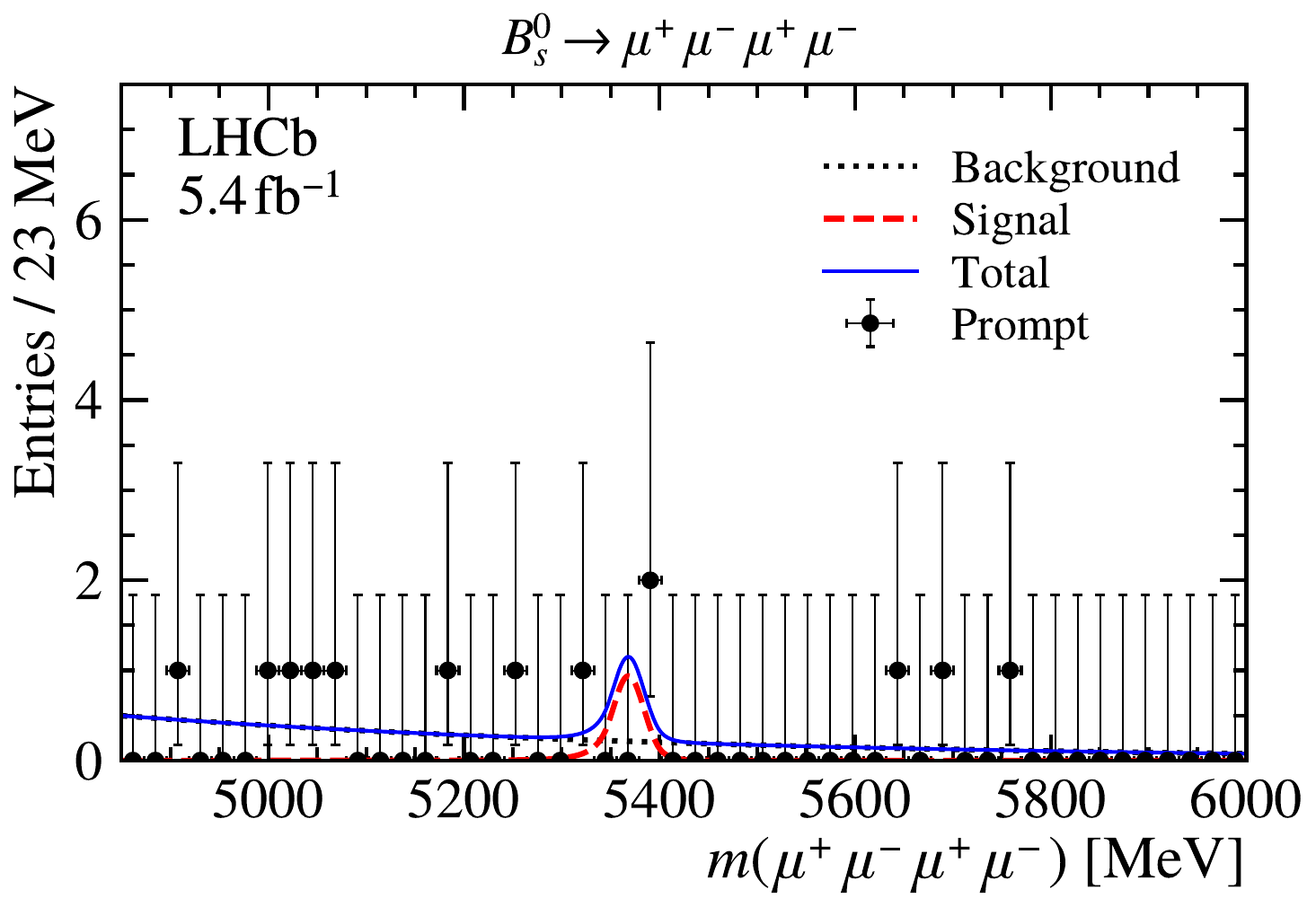}%
    \includegraphics[width=0.5\linewidth]{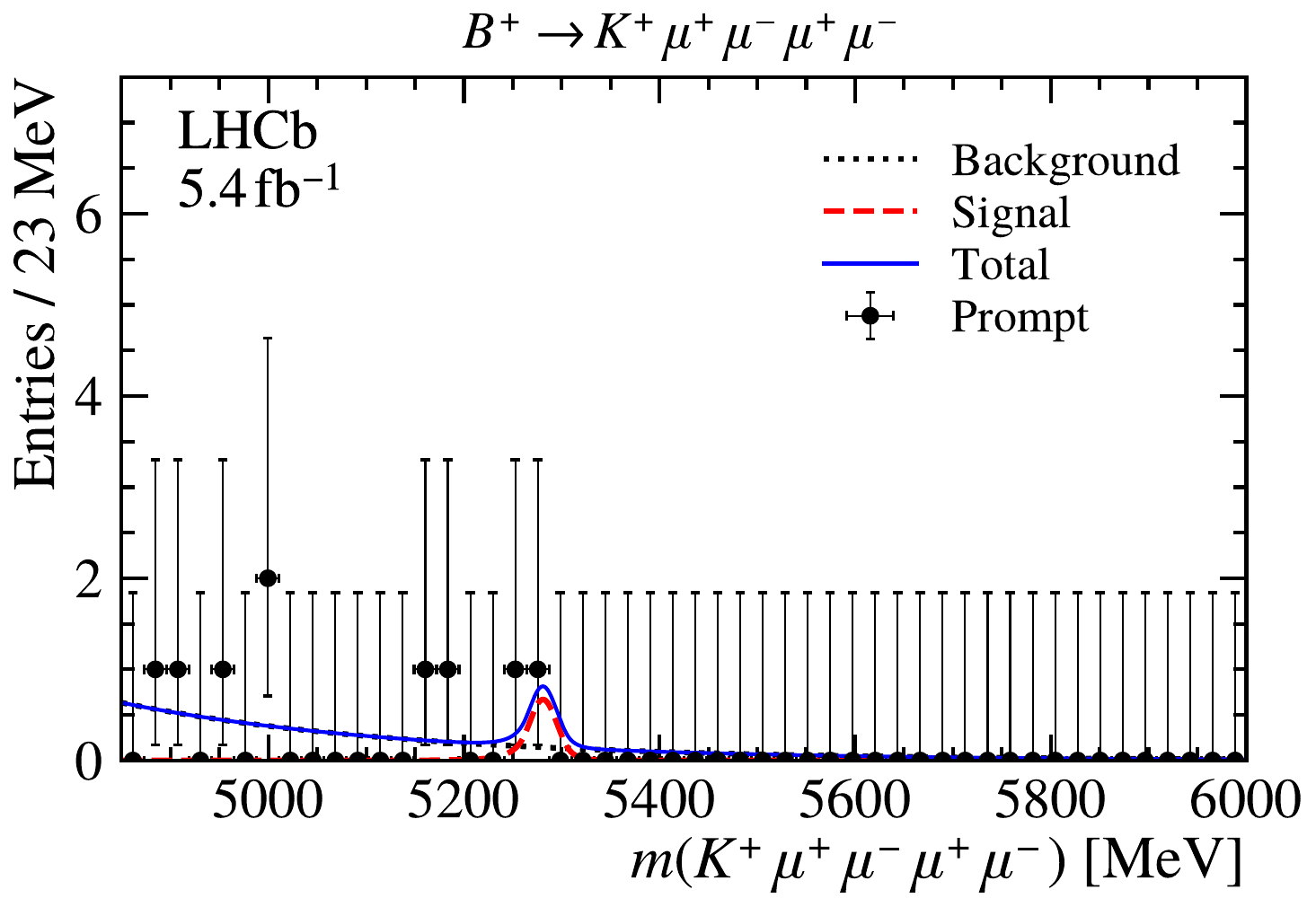}

    \includegraphics[width=0.5\linewidth]{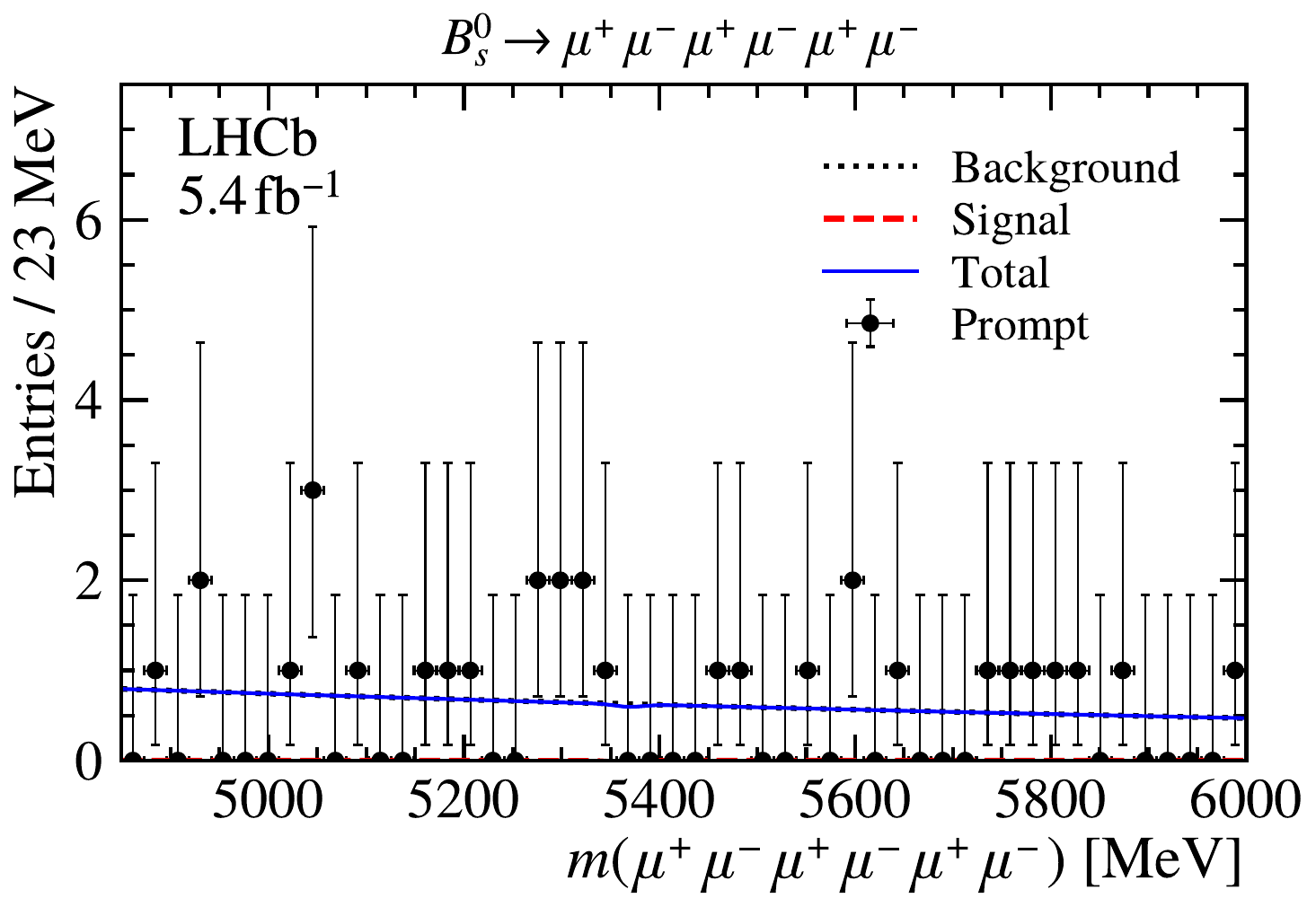}%
    \includegraphics[width=0.5\linewidth]{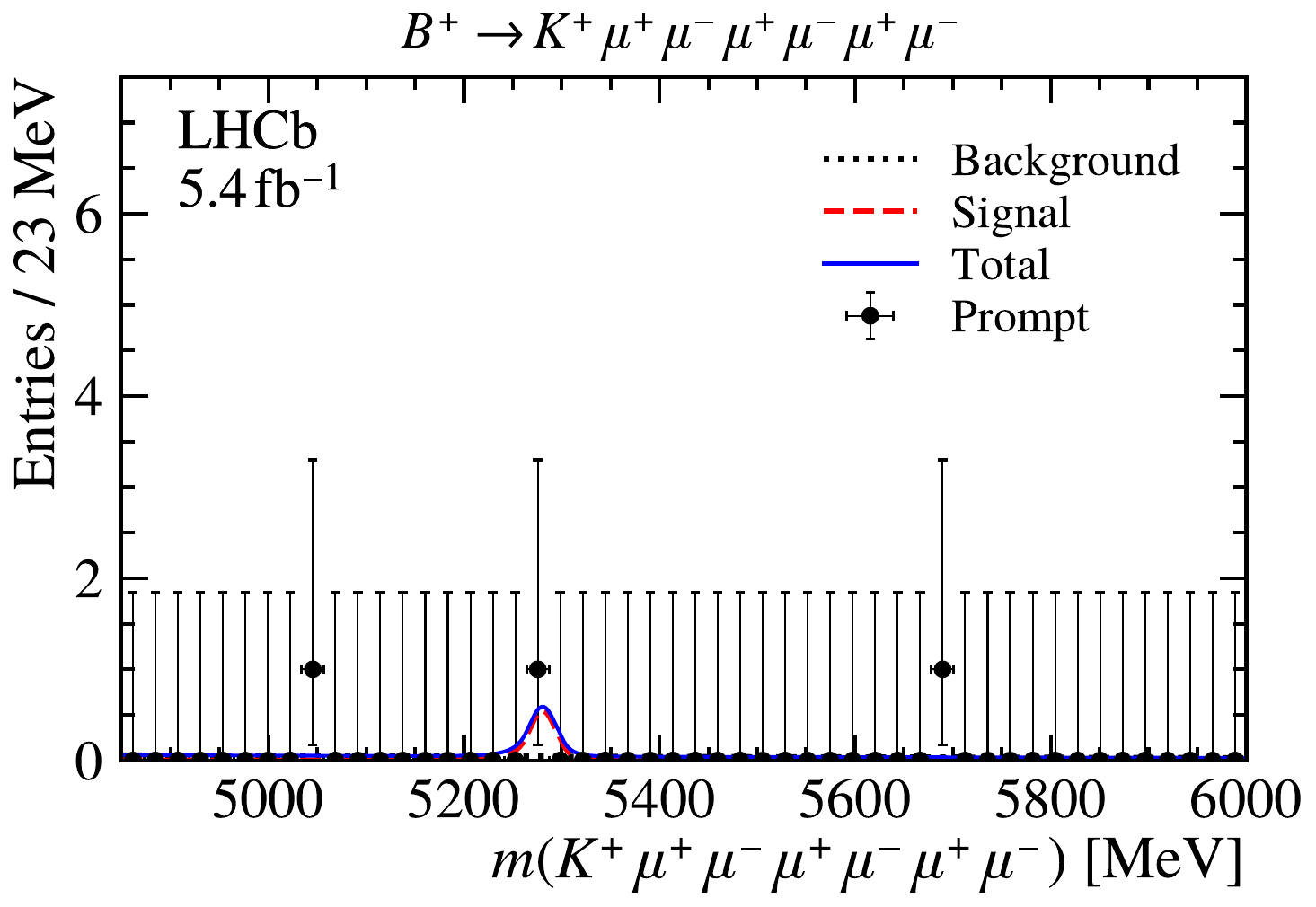}

    \caption{Fits to the \B-candidate mass for prompt signal channels: (top left) \decay{\Bs}{\mu^+ \mu^- \mu^+ \mu^-}, (top right) \decay{\Bp}{\Kp \mu^+ \mu^- \mu^+ \mu^-}, (bottom left) \decay{\Bs}{\mu^+ \mu^- \mu^+ \mu^- \mu^+ \mu^-} and (bottom right) \decay{\Bp}{ \Kp \mu^+ \mu^- \mu^+ \mu^- \mu^+ \mu^-}.}
    \label{fig:ResPrompt}
\end{figure}

Similarly, tracking efficiencies, which differ between data and simulation, are not corrected. To quantify the potential bias this introduces, calibration samples which provide corrections in bins of track momentum $p$ and pseudorapidity $\eta$ are used to assign weights to each track. A per-event weight is calculated as the product of these weights. The selection efficiencies are then calculated with these weights applied, and the difference with respect to the baseline efficiency is assigned as a systematic uncertainty. These systematic uncertainties range from $0.5\,\%$ to $1.8\,\%$. To quantify the uncertainty arising from the corrections applied to the simulated samples, the control sample used to compute these corrections is bootstrapped~\cite{efron:1979} multiple times to generate alternative sets of correction weights. The selection efficiencies were recomputed using each set of alternative corrections, and the standard deviation of the resulting efficiency distribution is assigned as a systematic uncertainty. The systematic uncertainties range from about 1\% to 10\%.

\section{Results}
\label{sec:Results}
\begin{figure}[htbp]
    \centering
    \includegraphics[width=\linewidth]{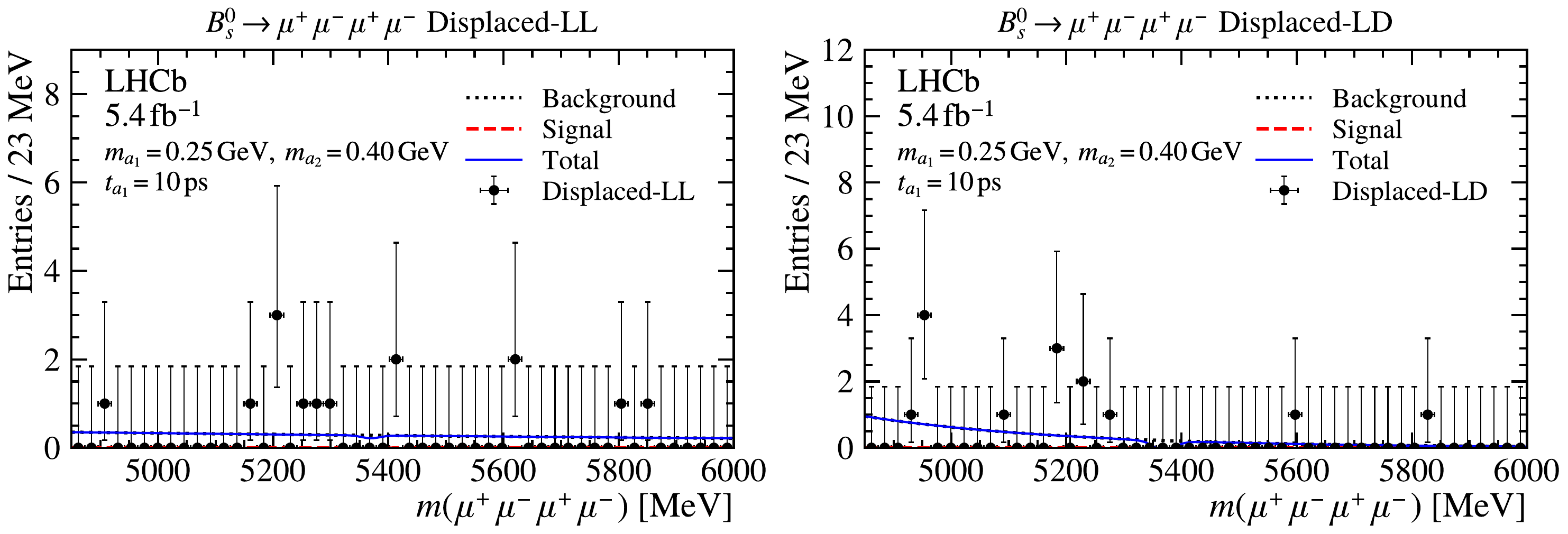}
    \includegraphics[width=\linewidth]{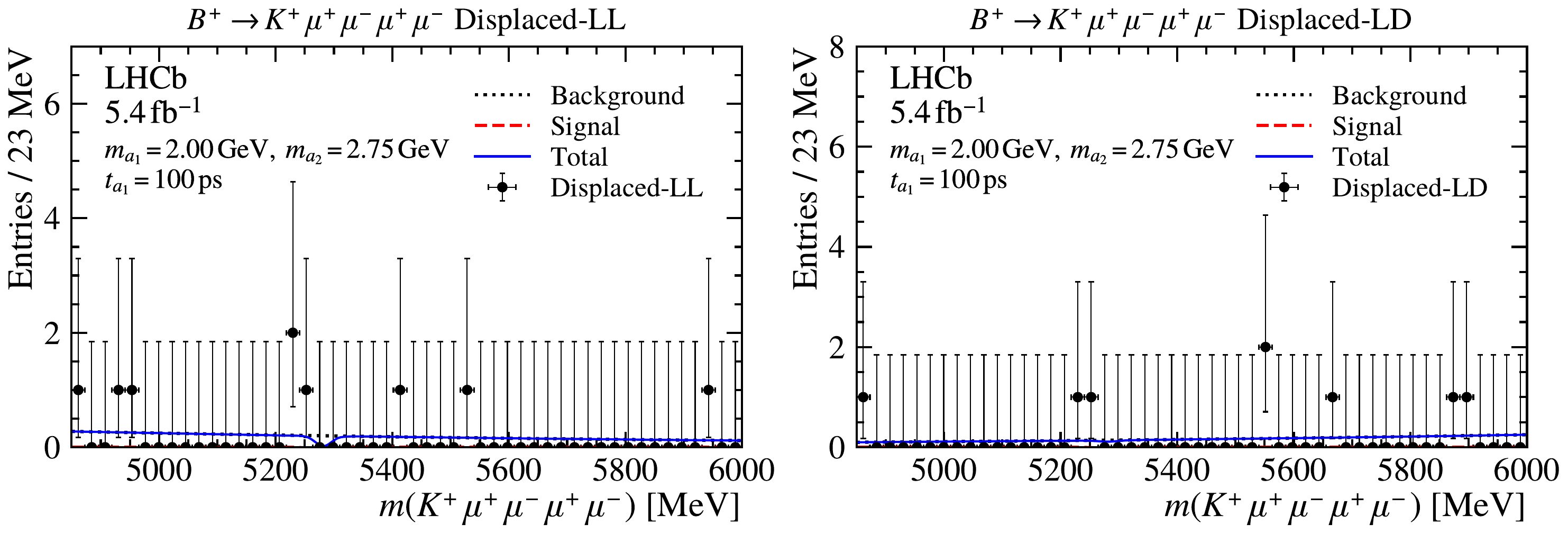}
    \caption{Simultaneous fits to the $B$-candidate meson mass in the (left) LL and (right) LD categories for (top) \decay{\Bs}{\mu^+\mu^-\mu^+\mu^-} and (bottom) \decay{B^+}{\Kp \mu^+ \mu^- \mu^+ \mu^-} decays for the indicated displaced channels.}
    \label{fig:Res4muDispl}
\end{figure}

\begin{figure}[b!]
    \centering
    \includegraphics[width=\linewidth]{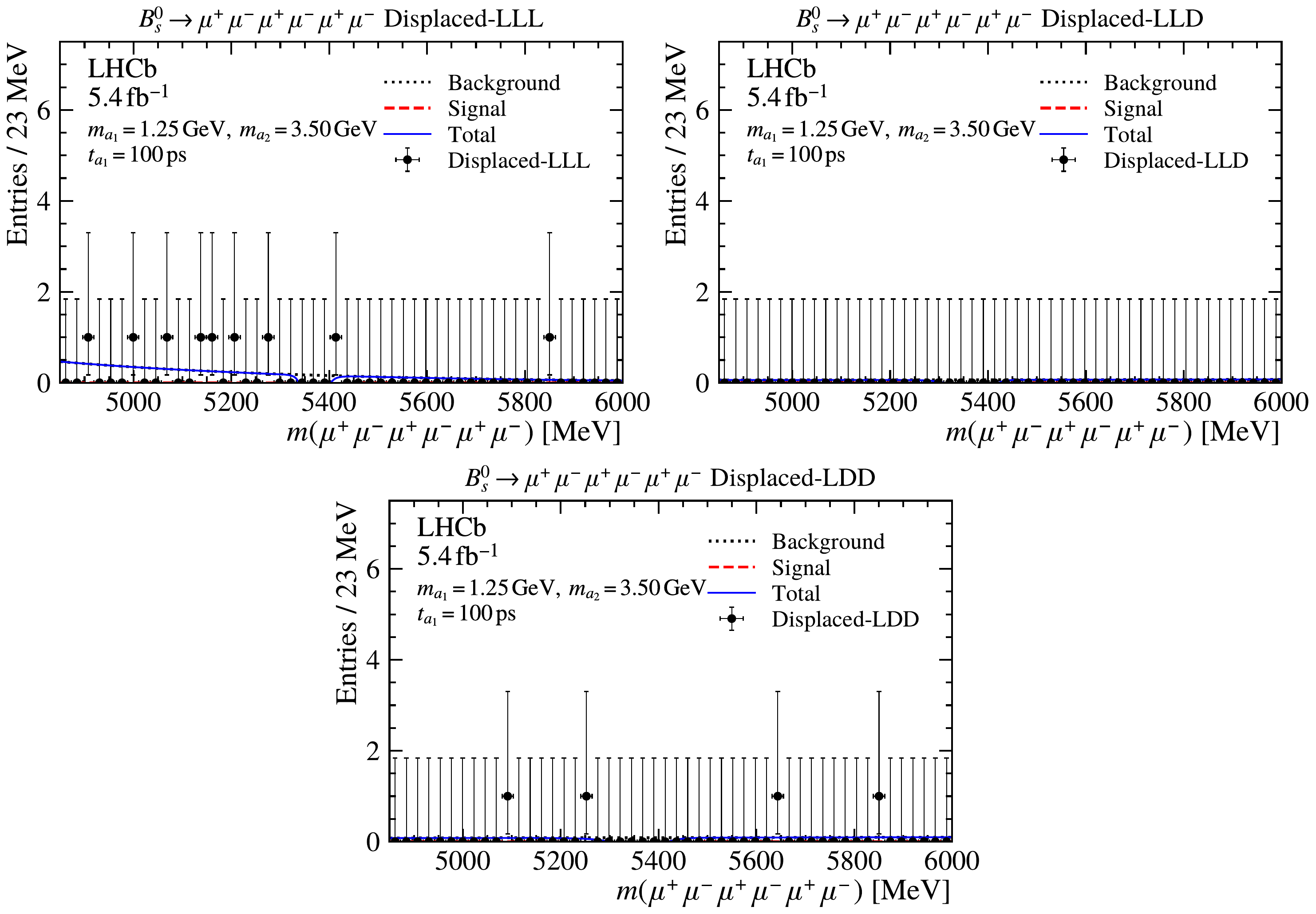}
    \caption{Simultaneous fits to the $B$-candidate meson mass in the (top left) LLL, (top right)~LLD and (bottom) LDD categories for \decay{\Bs}{\mu^+ \mu^- \mu^+ \mu^- \mu^+ \mu^-} decays for the indicated displaced samples.}
    \label{fig:Res6muDispl}
\end{figure}

\begin{figure}[t!]
    \centering
    \includegraphics[width=\linewidth]{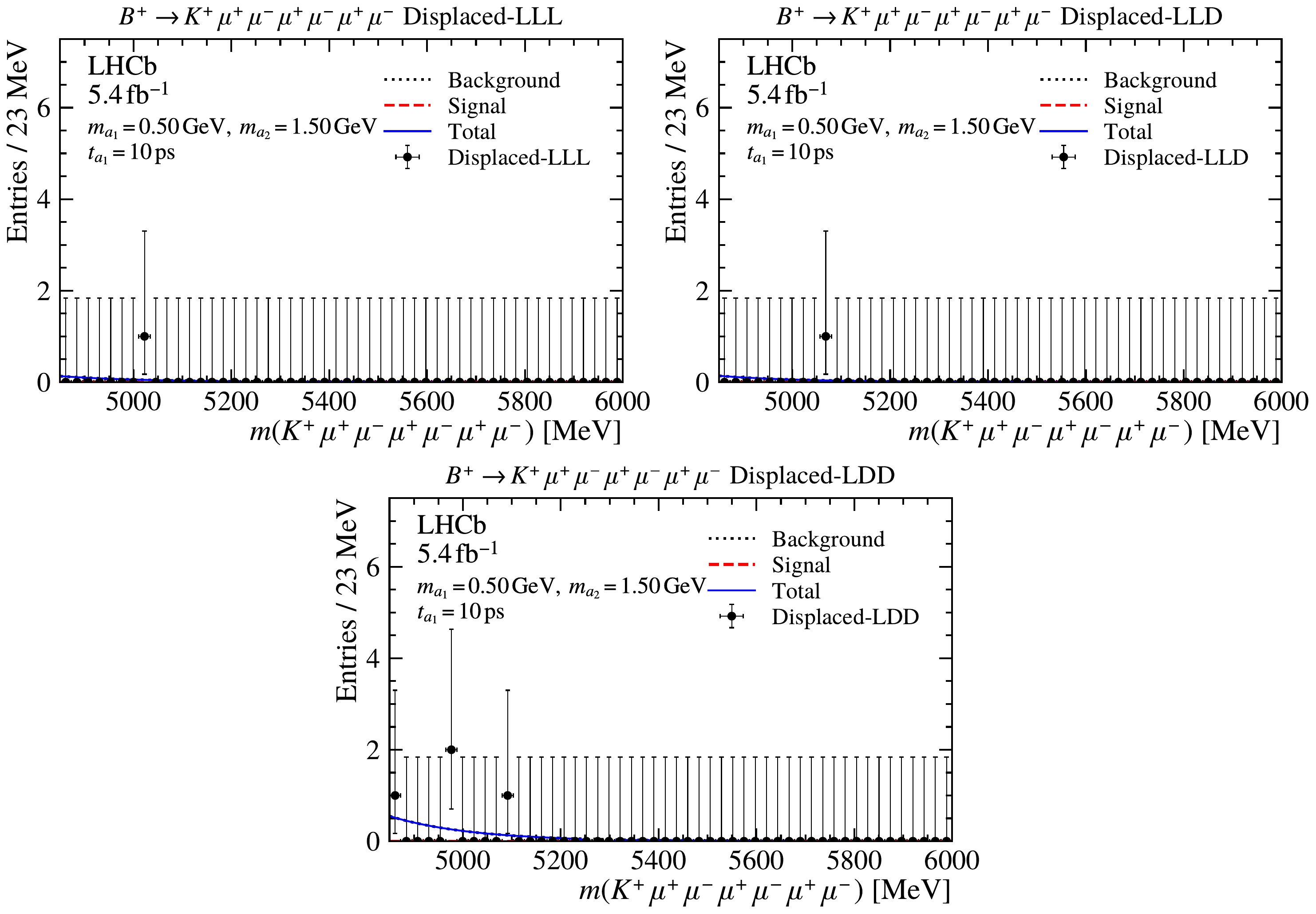}
    \caption{Simultaneous fits to the $B$-candidate meson mass in the (top left) LLL, (top right)~LLD and (bottom) LDD categories for \decay{\Bp}{K^+ \mu^+ \mu^- \mu^+ \mu^- \mu^+ \mu^-} decays for the indicated displaced samples.}
    \label{fig:Res6muKDispl}
\end{figure}

Fits to the fully selected data for a set of mass and lifetime configurations are displayed in Figs.~\ref{fig:ResPrompt}--\ref{fig:Res6muKDispl}. No signal excess is found in any of the different categories considered in this analysis.
Extending the physics reach, the data are also fit for the \decay{\Bz}{\mumu\mumu} and \decay{\Bz}{\mumu\mumu\mumu} decays. Given the very similar kinematics between $\Bz$ and $\Bs$ modes, the following approach is used: i) signal PDF shape parameters and background templates are taken from the corresponding $\Bs$ cases, and the peak position is shifted towards the known \Bz mass~\cite{PDG2024}, ii) efficiencies are considered as in the analogous $\Bs$ modes, but including the corresponding hadronisation factor for $\Bz$ mesons.

The \textsc{GammaCombo} package \cite{GammaCombo,LHCb-PAPER-2016-032} is used to compute the expected and observed limits on the branching fractions with the CL$_\text{s}$ method~\cite{CLs}, employing a one-sided test-statistic~\cite{Cowan:2010js}. The limit-setting procedure is run with 2500 pseudoexperiments each to achieve smooth $p$-value distributions. To present upper limits across the lifetime reach, the \texttt{PchipInterpolator}~\cite{Fritsch_Butland_1984} algorithm is used to compute exclusion limits in the range limited by the samples coverage. Such a method allows the interpolation of discrete data while keeping monotonicity and positivity, avoiding nonphysical oscillations. This decision is justified by three reasons: i) single-event sensitivities smoothly follow an exponential-decay law, ii) the final mass spectrum after selection does not depend on the decay time of the intermediate scalar particles and iii) the width of the DSCB is kept constant within an uncertainty window.

\begin{table}[tb!]
\centering
\caption{Observed (expected) upper limits at 95$\%$ CL on the
$\BF(\decay{B_{(s)}^0}{\mu^+ \mu^- \mu^+ \mu^- })$,
$\BF(\decay{B^+}{\Kp \mu^+ \mu^- \mu^+ \mu^-})$,
$\BF(\decay{B_{(s)}^0}{\mu^+ \mu^- \mu^+ \mu^- \mu^+ \mu^-})$ and
$\BF(\decay{B^+}{\Kp \mu^+ \mu^- \mu^+ \mu^- \mu^+ \mu^-})$ decays.}
{\setlength{\tabcolsep}{11pt}
\resizebox{1\textwidth}{!}{
\begin{tabular}{ccccc}
\toprule
$m_{a_1}$, $m_{a_2}$ & $\tau_{a_1}$ &
\multicolumn{3}{c}{Observed (expected) upper limit at 95\% CL} \\

[\gev] & [\ps] & \decay{\Bs}{\mumu\mumu} &
$\Bz \rightarrow \mumu\mumu$ &
\decay{B^+}{\Kp \mumu\mumu} \\
\hline
PHSP & 0 &
$1.95\,(0.98)\times10^{-9}$ &
$0.49\,(0.63)\times10^{-9}$ &
$1.33\,(0.68)\times10^{-9}$ \\
\hline
0.25, 0.40 & 1 &
$1.36\,(1.42)\times10^{-8}$ &
$6.73\,(3.00)\times10^{-9}$ &
$3.27\,(2.44)\times10^{-8}$ \\
 & 10 &
$3.58\,(4.83)\times10^{-9}$ &
$2.28\,(1.33)\times10^{-9}$ &
$4.29\,(3.46)\times10^{-8}$ \\
 & 100 &
$7.00\,(8.84)\times10^{-9}$ &
$3.18\,(2.38)\times10^{-9}$ &
$12.8\,(9.38)\times10^{-8}$ \\
 & 1000 &
$3.43\,(4.40)\times10^{-8}$ &
$1.59\,(1.19)\times10^{-8}$ &
$5.63\,(5.40)\times10^{-7}$ \\
\hline
2.00, 2.75 & 1 &
$2.04\,(2.23)\times10^{-9}$ &
$0.71\,(0.53)\times10^{-9}$ &
$2.70\,(2.50)\times10^{-9}$ \\
 & 10 &
$1.35\,(1.49)\times10^{-9}$ &
$0.73\,(0.42)\times10^{-9}$ &
$1.59\,(1.50)\times10^{-9}$ \\
 & 100 &
$2.97\,(4.22)\times10^{-9}$ &
$2.30\,(1.13)\times10^{-9}$ &
$6.29\,(5.21)\times10^{-9}$ \\
 & 1000 &
$2.23\,(2.96)\times10^{-8}$ &
$16.6\,(7.75)\times10^{-9}$ &
$5.52\,(4.43)\times10^{-8}$ \\
\hline
 & &
\decay{\Bs}{\mumu\mumu\mumu} &
$\Bz \rightarrow \mumu\mumu\mumu$ &
\decay{B^+}{\Kp \mumu\mumu\mumu} \\
\hline
PHSP & 0 &
$6.65\,(7.85)\times10^{-9}$ &
$3.56\,(1.86)\times10^{-9}$ &
$16.9\,(8.44)\times10^{-9}$ \\
\hline
0.50, 1.50 & 1 &
$3.55\,(4.43)\times10^{-9}$ &
$1.36\,(1.00)\times10^{-9}$ &
$2.83\,(3.20)\times10^{-9}$ \\
 & 10 &
$4.26\,(5.18)\times10^{-9}$ &
$1.71\,(1.59)\times10^{-9}$ &
$2.70\,(3.16)\times10^{-9}$ \\
 & 100 &
$0.97\,(1.30)\times10^{-7}$ &
$3.58\,(3.17)\times10^{-8}$ &
$4.98\,(5.17)\times10^{-8}$ \\
\hline
1.25, 3.50 & 1 &
$2.61\,(3.13)\times10^{-9}$ &
$0.96\,(0.78)\times10^{-9}$ &
$4.08\,(4.75)\times10^{-9}$ \\
 & 10 &
$5.58\,(9.11)\times10^{-9}$ &
$1.45\,(1.44)\times10^{-9}$ &
$5.01\,(5.52)\times10^{-9}$ \\
 & 100 &
$2.33\,(2.88)\times10^{-7}$ &
$8.48\,(8.04)\times10^{-8}$ &
$2.21\,(2.42)\times10^{-7}$ \\
\bottomrule
\end{tabular}
}}
\label{tab:Obs_limits}
\end{table}

Expected and observed upper limits are shown in Table~\ref{tab:Obs_limits}, and interpolated across the lifetime range considered in this analysis in Fig.~\ref{fig:Lf_lims}. The corresponding profile-likelihood curves are provided in appendix~\ref{app:NLLs}. While a comprehensive grid of mass configurations goes beyond the scope of this analysis due to the large number of possible combinations, two mass configurations for each decay and lifetime are provided to illustrate the spread of sensitivities across the $(m_{a_1},m_{a_2})$ spectrum.
The limit obtained for the nonresonant \decay{\Bs}{\mu^+ \mu^- \mu^+ \mu^-} decays are in full agreement, yet less sensitive than the previously published search~\cite{LHCb-PAPER-2021-039}.
The weaker sensitivity is driven by the fact that this paper analyses only a subset of the data used in Ref.~\cite{LHCb-PAPER-2021-039}.

The trends in the obtained limits can be largely understood with the following considerations. First, the larger available phase space in the $\mumu\mumu$ final states allows for wider variations in the transverse momentum of the decay products, resulting in more prominent behaviour depending on the decay time of the long-lived scalar. At higher lifetimes, the \decay{\Bs}{\mu^+ \mu^- \mu^+ \mu^-} channel tends to provide improved sensitivity compared to the \decay{B^+}{\Kp \mu^+ \mu^- \mu^+ \mu^-} mode, which can be attributed to a more effective background suppression. Within each decay channel, the differences observed between the lower- and higher-mass hypotheses can be understood in terms of the corresponding sensitivity profiles. Second, the overall reduced sensitivity reach in $\mumu\mumu\mumu$ channels can be explained by the more restricted phase space available for transverse momentum, due to the higher multiplicity of the final state. In these channels, the generally lower background levels imply the behaviour of the exclusion-limit curves being fully explained by kinematic features. At high lifetimes, the reach of decays involving lighter resonances surpasses their counterparts with heavier resonances because of the impact of highly boosted muons originating from momentum balance. At shorter lifetimes, a systematic difference is observed between the mass hypotheses considered in the \decay{\Bs}{\mu^+ \mu^- \mu^+ \mu^- \mu^+ \mu^-} and \decay{B^+}{\Kp \mu^+ \mu^- \mu^+ \mu^- \mu^+ \mu^-} channels. This behaviour can be interpreted as a consequence of momentum-balance effects, which increasingly hinder the higher-mass configurations from satisfying the trigger requirements as the number of final-state particles increases.

\begin{figure}[tb!]
    \centering
    \includegraphics[width=0.5\linewidth]{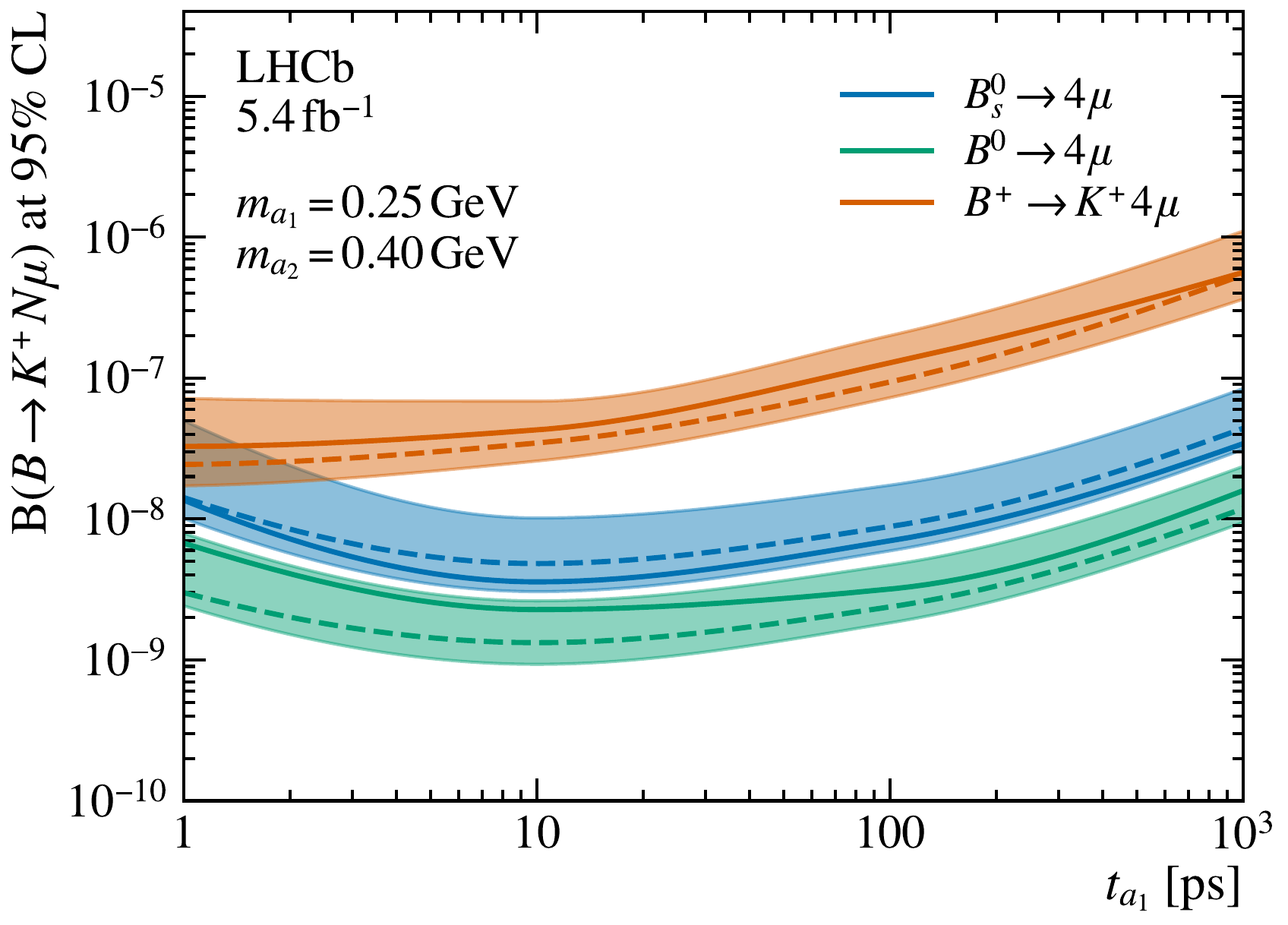}%
    \includegraphics[width=0.5\linewidth]{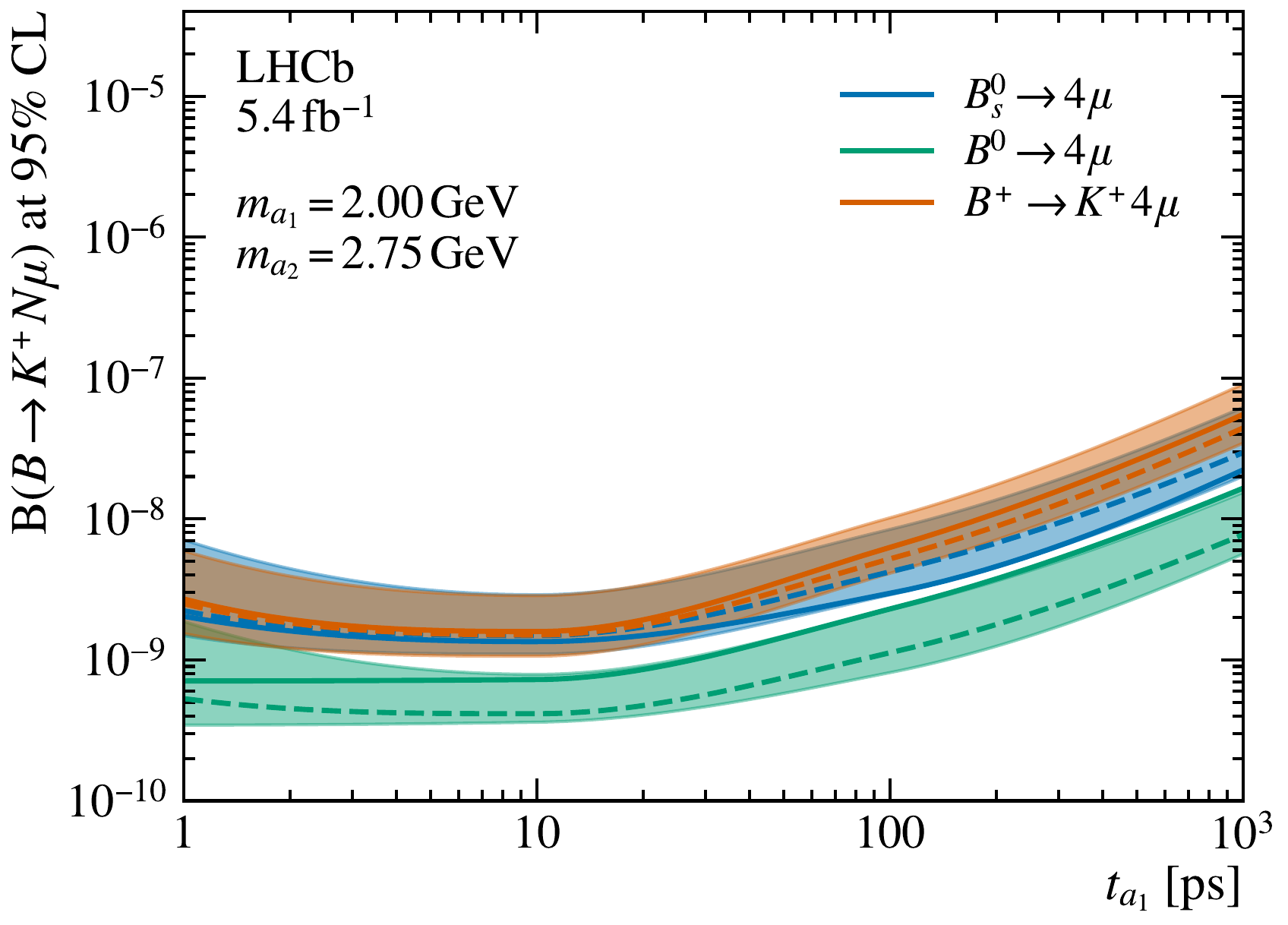}
    \includegraphics[width=0.5\linewidth]{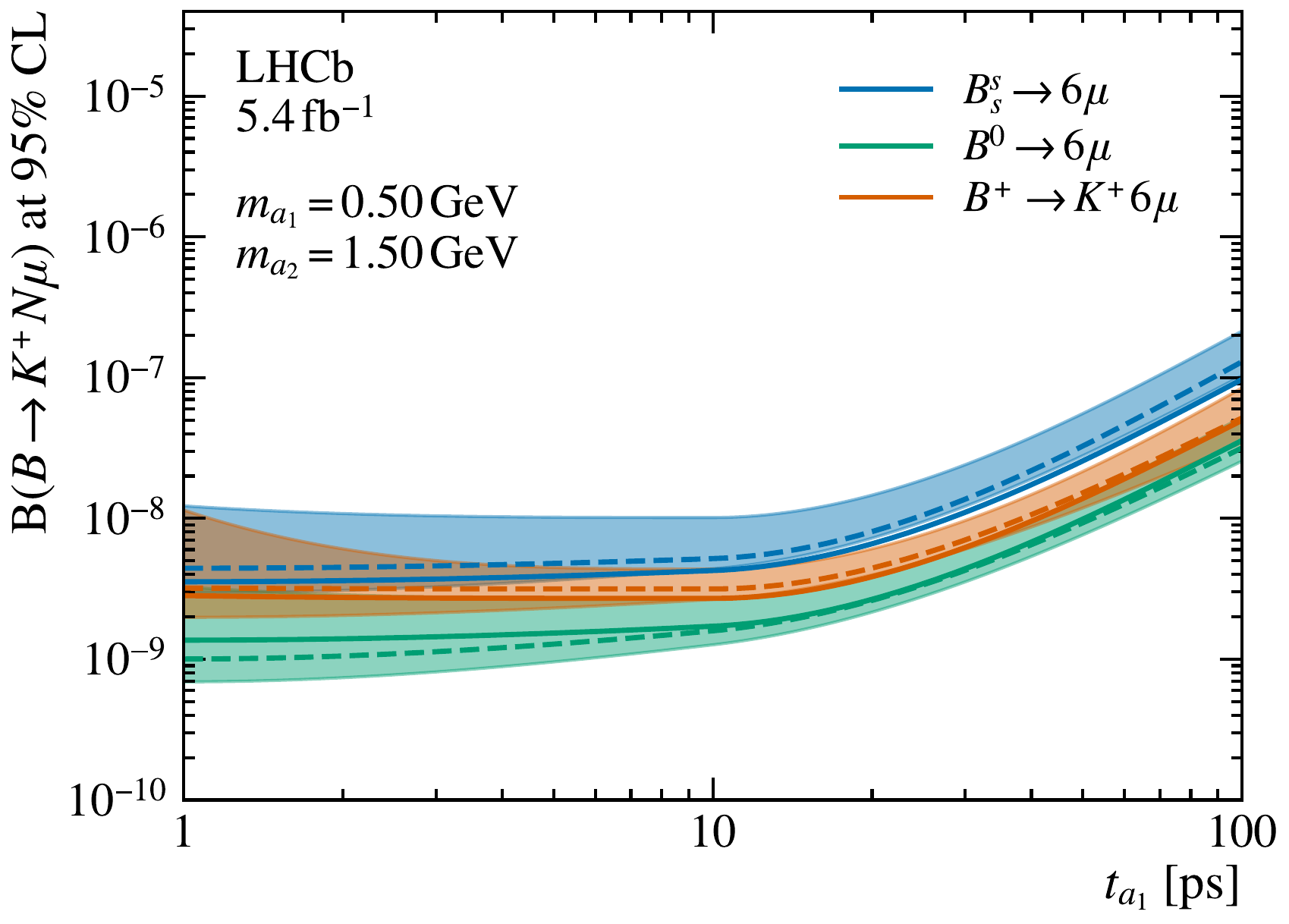}%
    \includegraphics[width=0.5\linewidth]{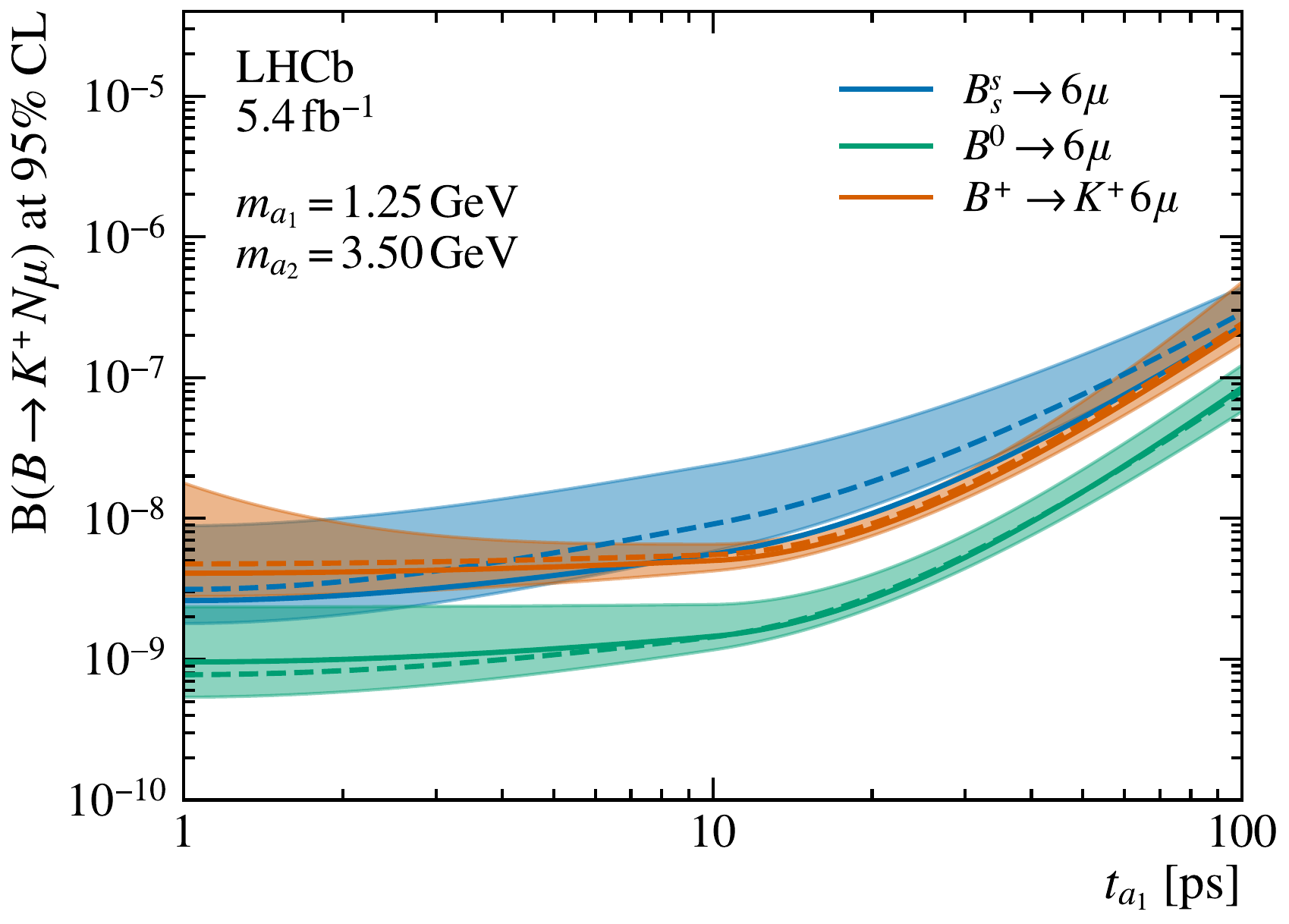}
    \caption{Upper limits on the branching ratio across the lifetime of the $a_1$ scalar for the indicated mass configurations. Expected and observed upper limits are displayed with a dashed and solid line, respectively. The shaded regions correspond to the $2\sigma$ error bands with respect to the null hypothesis.}
    \label{fig:Lf_lims}
\end{figure}

\section{Conclusions}
\label{sec:conclusion}
This paper presents a comprehensive search for decays of $B$ mesons into four- and six-muon final states, proceeding either directly or through intermediate scalar states with lifetimes up to $1\ns$. The sample used in this analysis corresponds to $pp$ collision data recorded with an integrated luminosity of $5.4\invfb$, collected by the LHCb experiment at $\sqs = 13\tev$. The decays \decay{\Bds}{\mu^+\mu^-\mu^+\mu^-}, \decay{\Bu}{\Kp\mu^+\mu^-\mu^+\mu^-}, \decay{\Bds}{\mu^+\mu^-\mu^+\mu^-\mu^+\mu^-} and \decay{\Bu}{\Kp\mu^+\mu^-\mu^+\mu^-\mu^+\mu^-} are considered, in both prompt and displaced topologies.

No significant signal is observed, and upper limits on the branching fractions are set at the $95\%$ CL, spanning the range from $0.6\times10^{-9}$ to $5.4\times10^{-7}$ depending on the decay topology, and on the mass and lifetime of the intermediate mediators. With the exception of the \decay{\Bds}{\mu^+\mu^-\mu^+\mu^-} mode with prompt mediators~\cite{LHCb-PAPER-2021-039}, these are the most stringent limits on these decays, improving the constraints obtained from the recasting of dimuon resonance searches by up to two orders of magnitude~\cite{CidVidal:2022mrx}. These results are complementary to direct searches for long-lived dimuon resonances at the LHC~\cite{CMS:EXO-24-016, CMSDisplaced}, and constrain the couplings of nonminimal models with hierarchical sectors, such as composite Higgs models and supersymmetric extensions of the SM, over a wide range of mediator masses and lifetimes.

\section*{Acknowledgements}
%
%
\noindent We express our gratitude to our colleagues in the CERN
accelerator departments for the excellent performance of the LHC. We
thank the technical and administrative staff at the LHCb
institutes.
We acknowledge support from CERN and from the national agencies:
ARC (Australia);
CAPES, CNPq, FAPERJ and FINEP (Brazil); 
MOST and NSFC (China); 
CNRS/IN2P3 and CEA (France);  
BMFTR, DFG and MPG (Germany);
NKFIH (Hungary);              
INFN (Italy); 
NWO (Netherlands); 
MNiSW and NCN (Poland); 
MEC/IFA (Romania); 
MICIU and AEI (Spain);
SNSF and SER (Switzerland); 
NASU (Ukraine); 
STFC (United Kingdom); 
DOE NP and NSF (USA).
We acknowledge the computing resources that are provided by ARDC (Australia), 
CBPF (Brazil),
CERN, 
IHEP and LZU (China),
IN2P3 (France), 
KIT and DESY (Germany), 
INFN (Italy), 
SURF (Netherlands),
Polish WLCG (Poland),
IFIN-HH (Romania), 
PIC (Spain), CSCS (Switzerland), 
GridPP (United Kingdom),
and NSF (USA).  
We are indebted to the communities behind the multiple open-source
software packages on which we depend.
Individual groups or members have received support from
RTP (Australia), 
FWO Odysseus grant G0ASD25N (Belgium), 
Key Research Program of Frontier Sciences of CAS, CAS PIFI, CAS CCEPP (China); 
Minciencias (Colombia);
EPLANET, Marie Sk\l{}odowska-Curie Actions, ERC and NextGenerationEU (European Union);
A*MIDEX, ANR, IPhU and Labex P2IO, and R\'{e}gion Auvergne-Rh\^{o}ne-Alpes (France);
Alexander-von-Humboldt Foundation (Germany);
ICSC (Italy); 
Severo Ochoa and Mar\'ia de Maeztu Units of Excellence, GVA, XuntaGal, GENCAT, InTalent-Inditex and Prog.~Atracci\'on Talento CM (Spain);
the Leverhulme Trust, the Royal Society and UKRI (United Kingdom).

\newpage
\clearpage

\appendix
\FloatBarrier
\section{Decay Time Selection}
\label{sec:fom_dectime}

\begin{figure}
    \centering
    \includegraphics[width=0.5\linewidth]{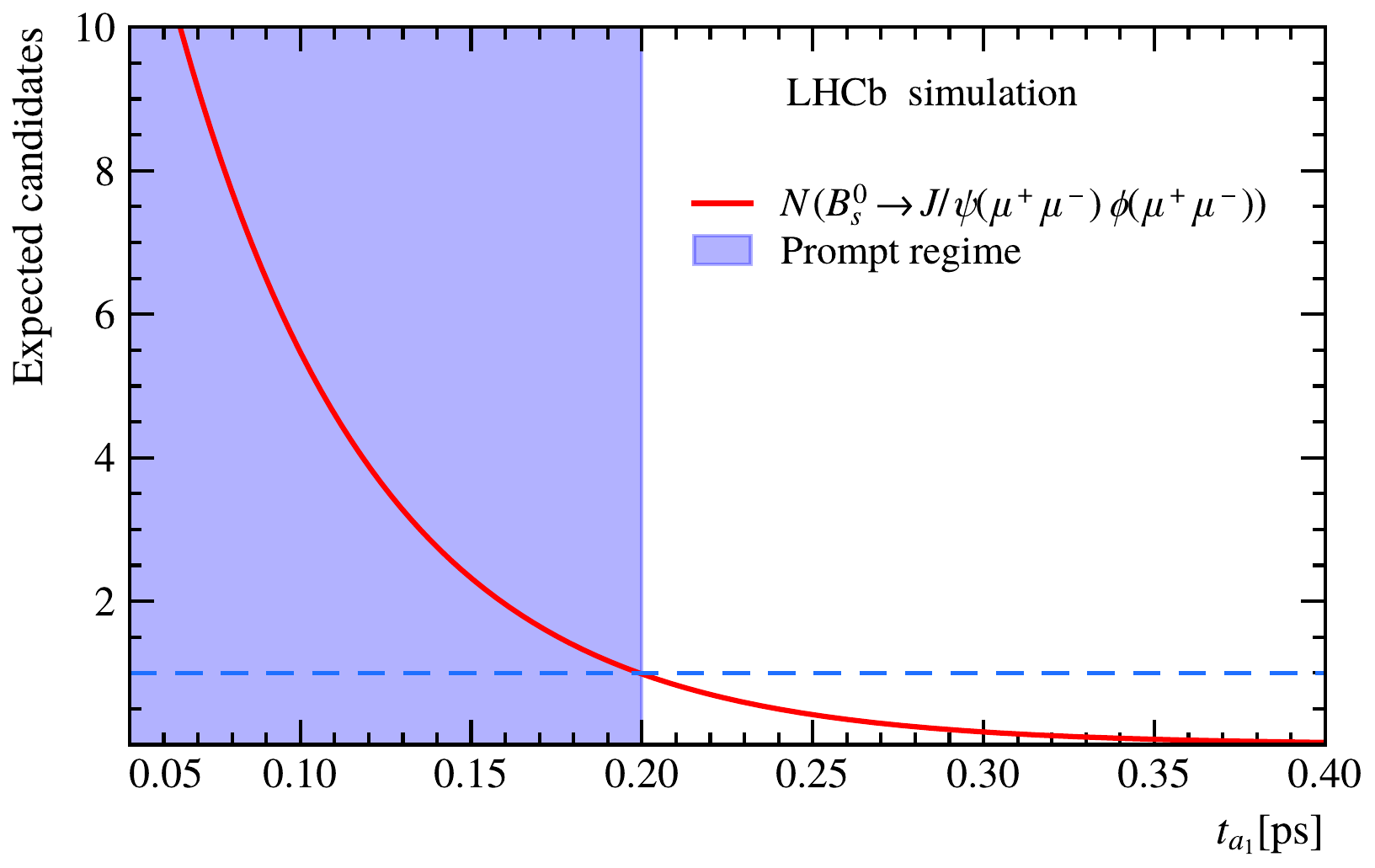}
    \caption{Total number of expected \decay{\Bs}{\jpsi(\to \mu^+\mu^-)\phi(\to \mu^+\mu^-)} candidates in the analysed dataset as a function of the requirement on the decay time of the intermediate scalars. The blue-dashed line marks the threshold of one expected candidate, whereas the shaded region delimits the prompt regime.}
    \label{fig:fom_dectime}
\end{figure}

The decay-time selection is optimized according to a figure of merith, see Fig. \ref{fig:fom_dectime}, which represents the number of expected \decay{\Bs}{\jpsi(\to \mu^+\mu^-)\phi(\to \mu^+\mu^-)} surviving signal selection up to the decay-time cut. The chosen point ensures that less than 1 candidate is expected to survive such a selection criteria, delimiting the long-lived regime.

\FloatBarrier
\section{Systematic uncertainties}
\label{sec:systematics_app}

Systematic uncertainties for 
\decay{\Bs}{a_1 a_2} and
\decay{B^+}{K^+ a_1 a_2} decays are shown in Tables~\ref{tab:combined_systematics_Bs} and~\ref{tab:combined_systematics_Bu}, where the combined systematic uncertainty results from adding the systematic uncertainties of each category in quadrature.

\begin{table}[h]
\centering
\caption{Systematic uncertainties on $\BF(\decay{\Bs}{\mu^+ \mu^- \mu^+ \mu^-})$ and $\BF(\decay{\Bs}{\mu^+ \mu^- \mu^+ \mu^- \mu^+ \mu^-})$. The systematic uncertainty related to the background model used is separated from the rest of the systematic uncertainties (combined). The combined systematic uncertainty is given as a relative uncertainty on the efficiency ratio, whereas the background-model systematic uncertainty is reported as an absolute uncertainty on the branching-fraction upper-limit estimate.}
{\setlength{\tabcolsep}{14pt}
\begin{tabular}{cccc}
\toprule
$m_{a_1}$, $m_{a_2}$ [\gev]  & $\tau_{a_1}$ [\ps] & Combined systematic [\%] & Background model \\
\midrule
\multicolumn{4}{c}{\decay{\Bs}{\mu^+ \mu^- \mu^+ \mu^-}} \\
\midrule
PHSP           & 0    & 2.1  & $\phantom{<}6.6 \times 10^{-14}$ \\
0.25, 0.40     & 1    & 9.9  & $\phantom{<}1.3 \times 10^{-9\phantom{0}}$ \\
0.25, 0.40     & 10   & 3.1  & $<$$1.0 \times 10^{-11}$ \\
0.25, 0.40     & 100  & 3.3  & $\phantom{<}2.6 \times 10^{-15}$ \\
0.25, 0.40     & 1000 & 3.4  & $\phantom{<}1.4 \times 10^{-9\phantom{0}}$ \\
2.00, 2.75     & 1    & 3.1  & $\phantom{<}5.3 \times 10^{-11}$ \\
2.00, 2.75     & 10   & 2.8  & $<$$1.0 \times 10^{-11}$ \\
2.00, 2.75     & 100  & 3.4  & $<$$1.0 \times 10^{-11}$ \\
2.00, 2.75     & 1000 & 4.4  & $\phantom{<}1.1 \times 10^{-10}$ \\
\midrule
\multicolumn{4}{c}{$\decay{\Bs}{\mu^+ \mu^- \mu^+ \mu^- \mu^+ \mu^-}$} \\
\midrule
PHSP           & 0    & 2.4  & $\phantom{<}1.1 \times 10^{-9\phantom{0}}$ \\
0.50, 1.50     & 1    & 2.4  & $\phantom{<}5.5 \times 10^{-11}$ \\
0.50, 1.50     & 10   & 3.7  & $<$$1.0 \times 10^{-11}$ \\
0.50, 1.50     & 100  & 3.4  & $\phantom{<}3.8 \times 10^{-9\phantom{0}}$ \\
1.25, 3.50     & 1    & 2.3  & $\phantom{<}1.6 \times 10^{-12}$ \\
1.25, 3.50     & 10   & 2.5  & $<$$1.0 \times 10^{-11}$ \\
1.25, 3.50     & 100  & 6.5  & $\phantom{<}3.3 \times 10^{-8\phantom{0}}$ \\
\bottomrule
\end{tabular}
}
\label{tab:combined_systematics_Bs}
\end{table}
\begin{table}[h!]
\renewcommand{\arraystretch}{1.0}
\centering
\caption{Systematic uncertainties on $\BF(\decay{B^+}{\Kp \mu^+ \mu^- \mu^+ \mu^-})$ and $\BF(\decay{B^+}{\Kp \mu^+ \mu^- \mu^+ \mu^- \mu^+ \mu^-})$. The systematic uncertainty related to the background model used is separated from the rest of the systematic uncertainties (combined). The combined systematic uncertainty is given as a relative uncertainty on the efficiency ratio, whereas the background-model systematic uncertainty is reported as an absolute uncertainty on the branching-fraction upper-limit estimate.}
\small{\setlength{\tabcolsep}{14pt}
\begin{tabular}{cccc}
\toprule
$m_{a_1}$, $m_{a_2}$ [\gev] & $\tau_{a_1}$ [\ps] & Combined systematic [\%] & Background model \\
\midrule
\multicolumn{4}{c}{$\decay{B^+}{\Kp \mu^+ \mu^- \mu^+ \mu^-}$} \\
\midrule
PHSP           & 0    & 8.8  & $\phantom{<}2.1 \times 10^{-17}$ \\
0.25, 0.40     & 1    & 13.0 & $\phantom{<}7.8 \times 10^{-10}$ \\
0.25, 0.40     & 10   & 16.0 & $\phantom{<}1.1 \times 10^{-10}$ \\
0.25, 0.40     & 100  & 16.0 & $\phantom{<}2.2 \times 10^{-9\phantom{0}}$ \\
0.25, 0.40     & 1000 & 19.0 & $\phantom{<}8.4 \times 10^{-10}$ \\
2.00, 2.75     & 1    & 2.4  & $\phantom{<}1.4 \times 10^{-12}$ \\
2.00, 2.75     & 10   & 2.2  & $<$$1.0 \times 10^{-11}$ \\
2.00, 2.75     & 100  & 2.4  & $<$$1.0 \times 10^{-11}$ \\
2.00, 2.75     & 1000 & 4.7  & $\phantom{<}9.6 \times 10^{-10}$ \\
\midrule
\multicolumn{4}{c}{$\decay{B^+}{\Kp \mu^+ \mu^- \mu^+ \mu^- \mu^+ \mu^-}$} \\
\midrule
PHSP           & 0    & 2.2  & $\phantom{<}6.2 \times 10^{-10}$ \\
0.50, 1.50     & 1    & 2.4  & $\phantom{<}4.2 \times 10^{-11}$ \\
0.50, 1.50     & 10   & 2.7  & $<$$1.0 \times 10^{-11}$ \\
0.50, 1.50     & 100  & 4.5  & $\phantom{<}3.7 \times 10^{-10}$ \\
1.25, 3.50     & 1    & 2.1  & $\phantom{<}2.7 \times 10^{-11}$ \\
1.25, 3.50     & 10   & 2.0  & $<$$1.0 \times 10^{-11}$ \\
1.25, 3.50     & 100  & 7.7  & $\phantom{<}3.7 \times 10^{-8\phantom{0}}$ \\
\bottomrule
\end{tabular}
}
\label{tab:combined_systematics_Bu}
\end{table}

\clearpage
\section{Profile likelihoods}
\label{app:NLLs}

\label{sec:profile_likelihoods}
\begin{figure}[h!]
    \centering
    \includegraphics[width=0.33\linewidth]{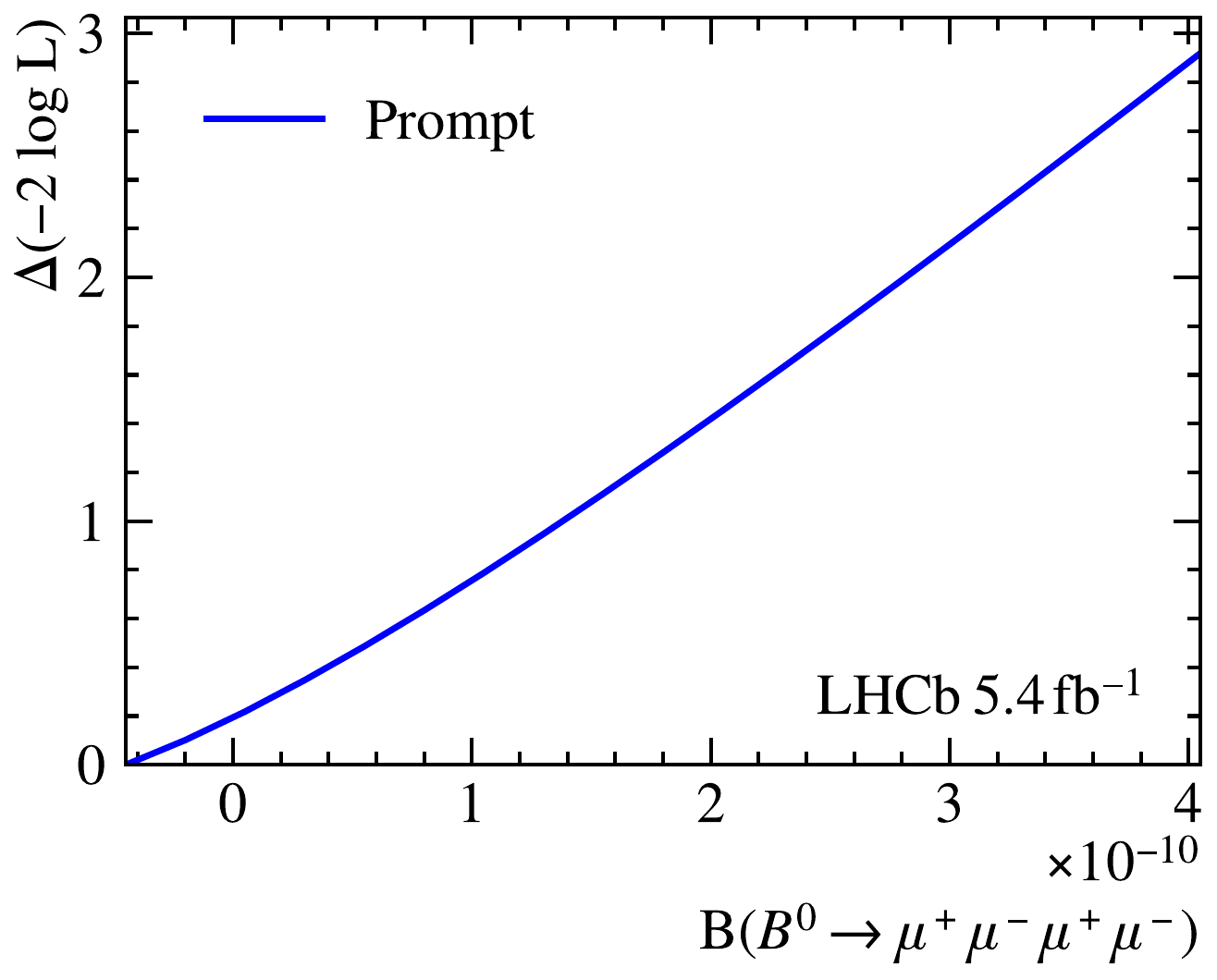}%
    \includegraphics[width=0.33\linewidth]{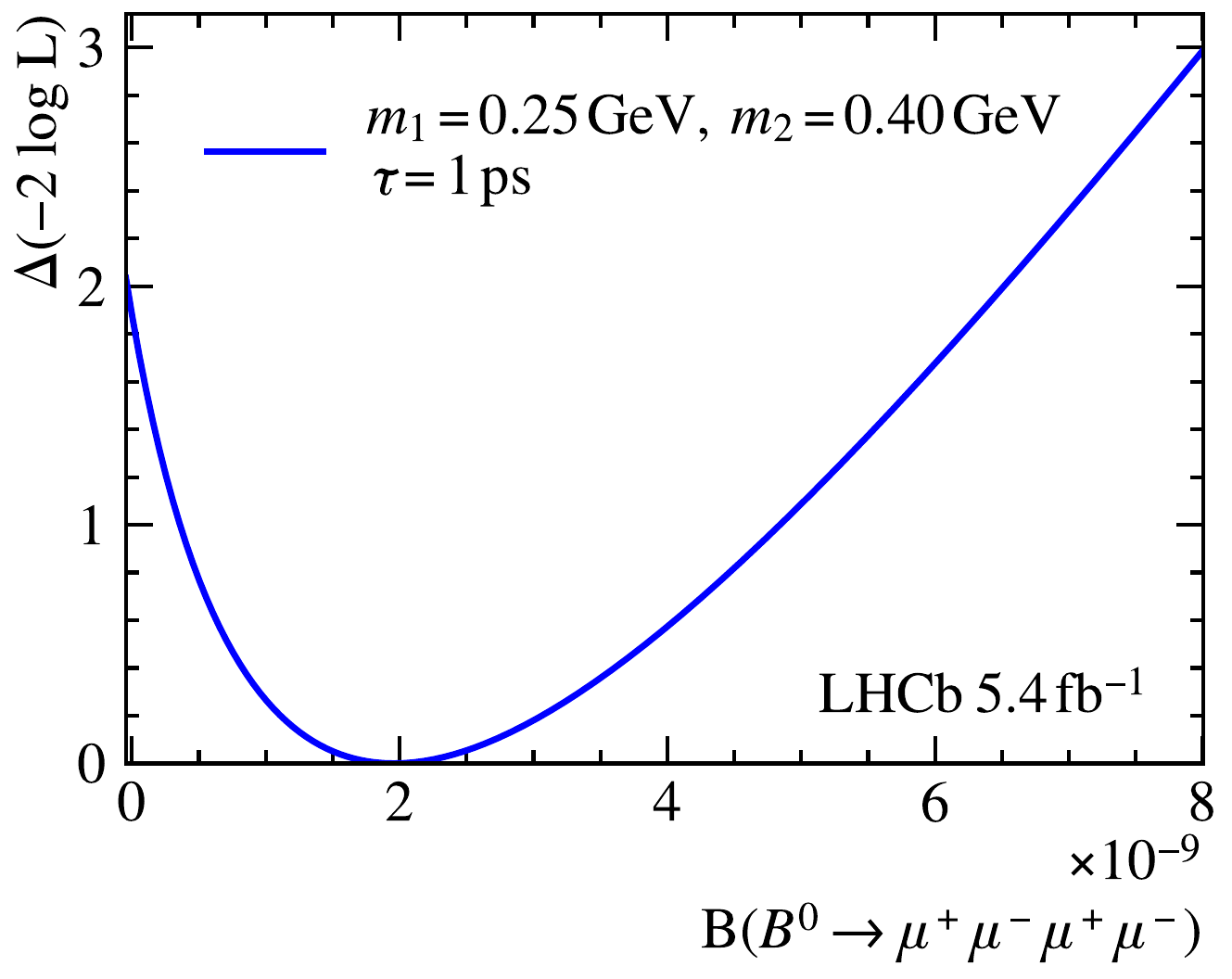}%
    \includegraphics[width=0.33\linewidth]{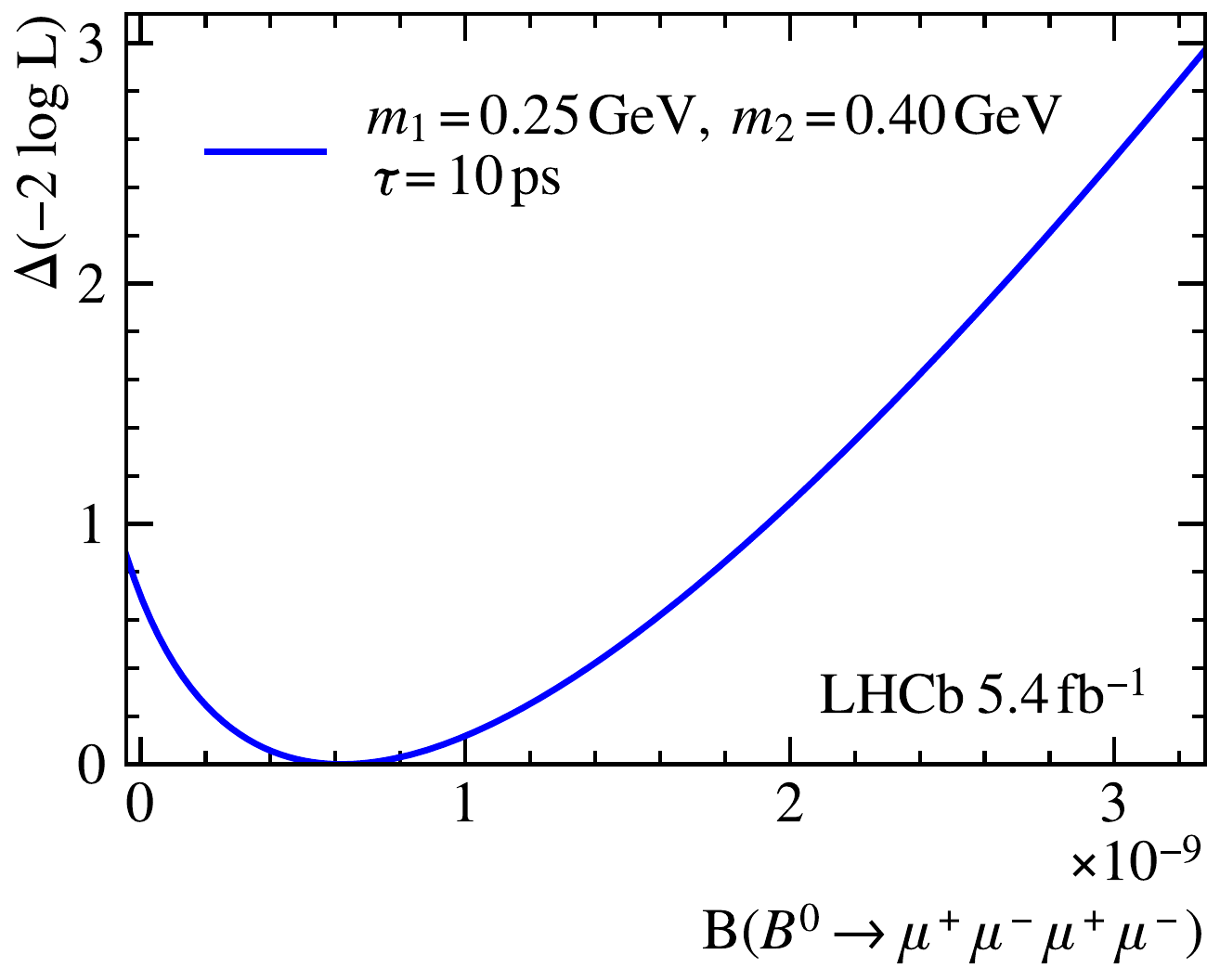}
    \includegraphics[width=0.33\linewidth]{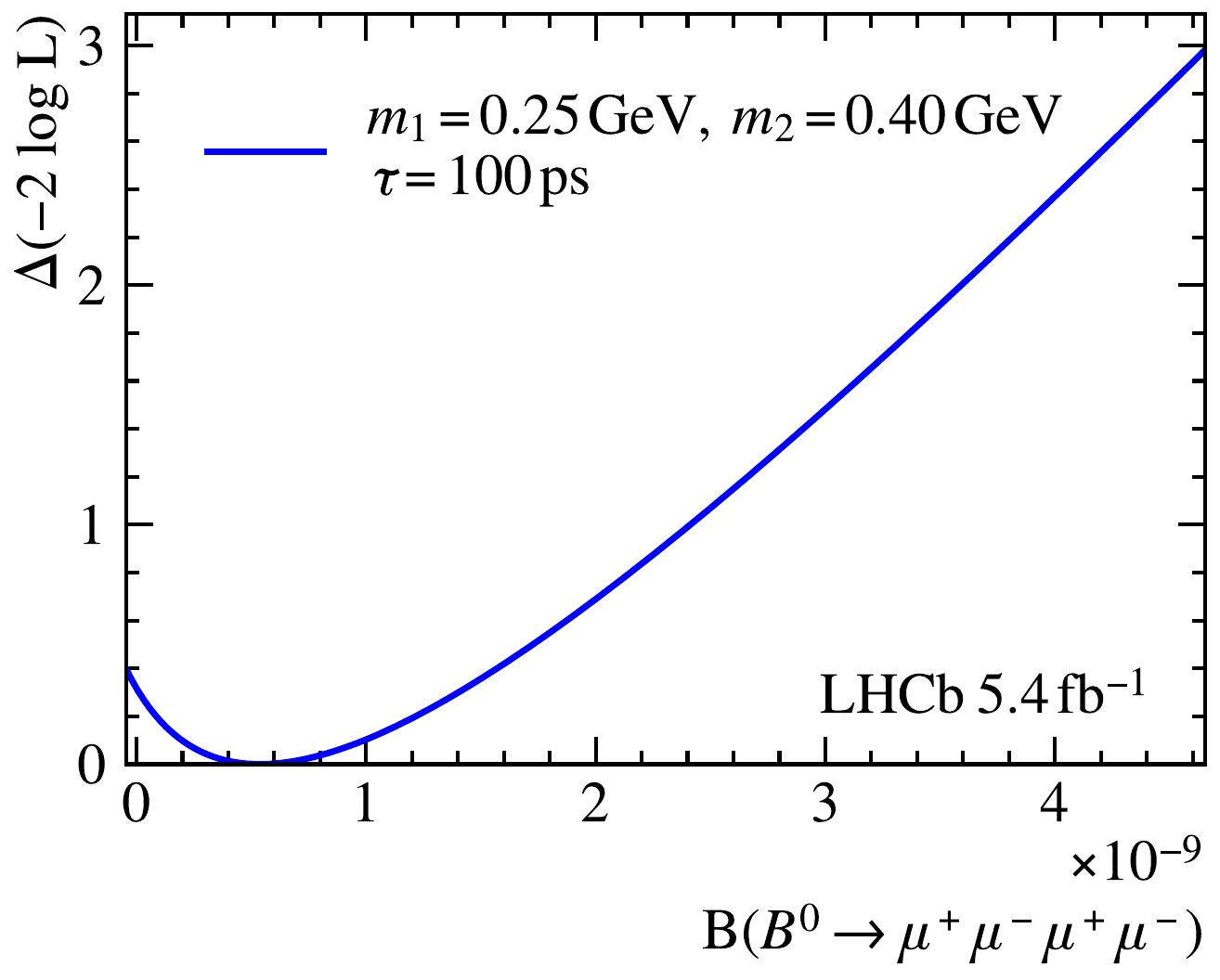}%
    \includegraphics[width=0.33\linewidth]{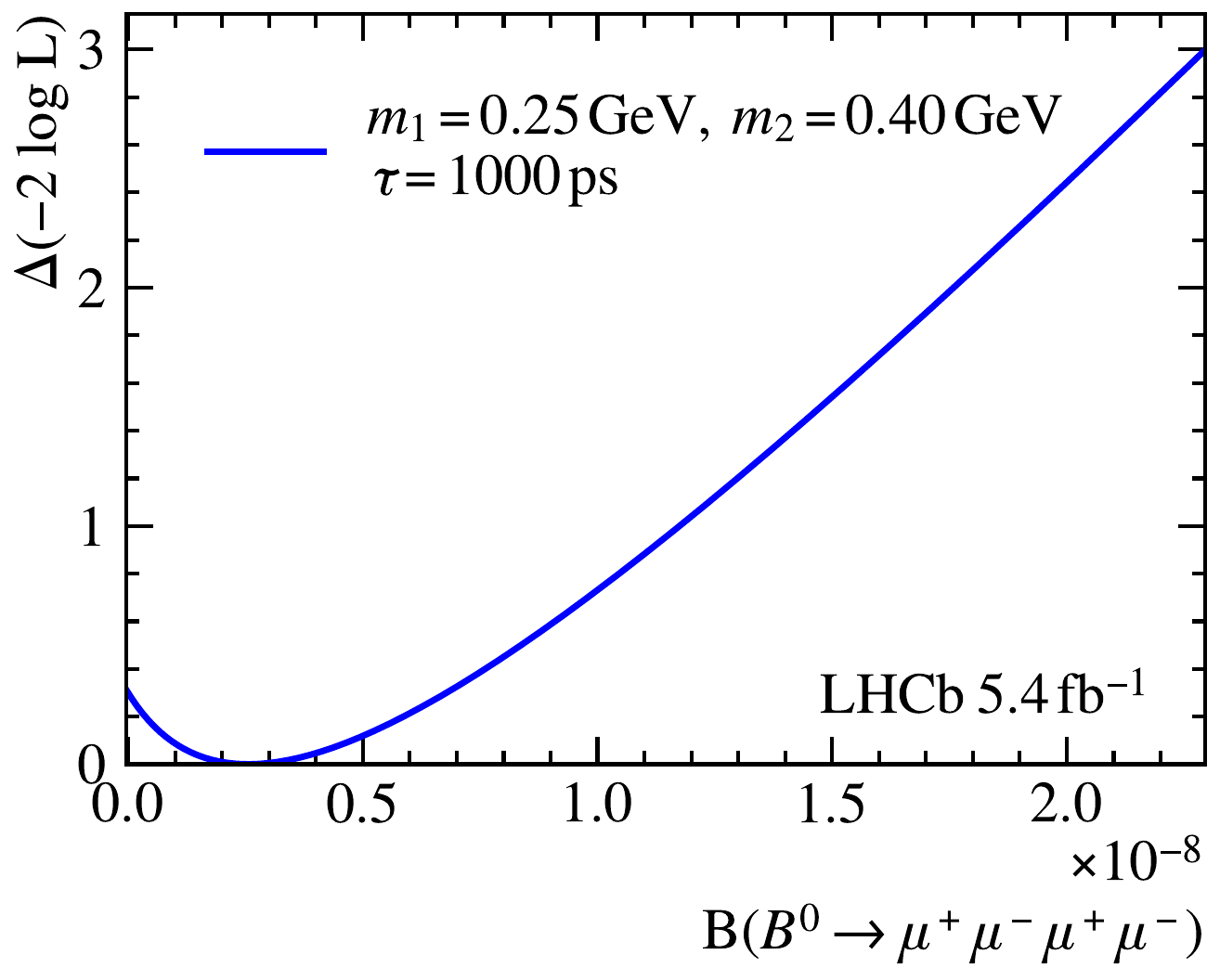}%
    \includegraphics[width=0.33\linewidth]{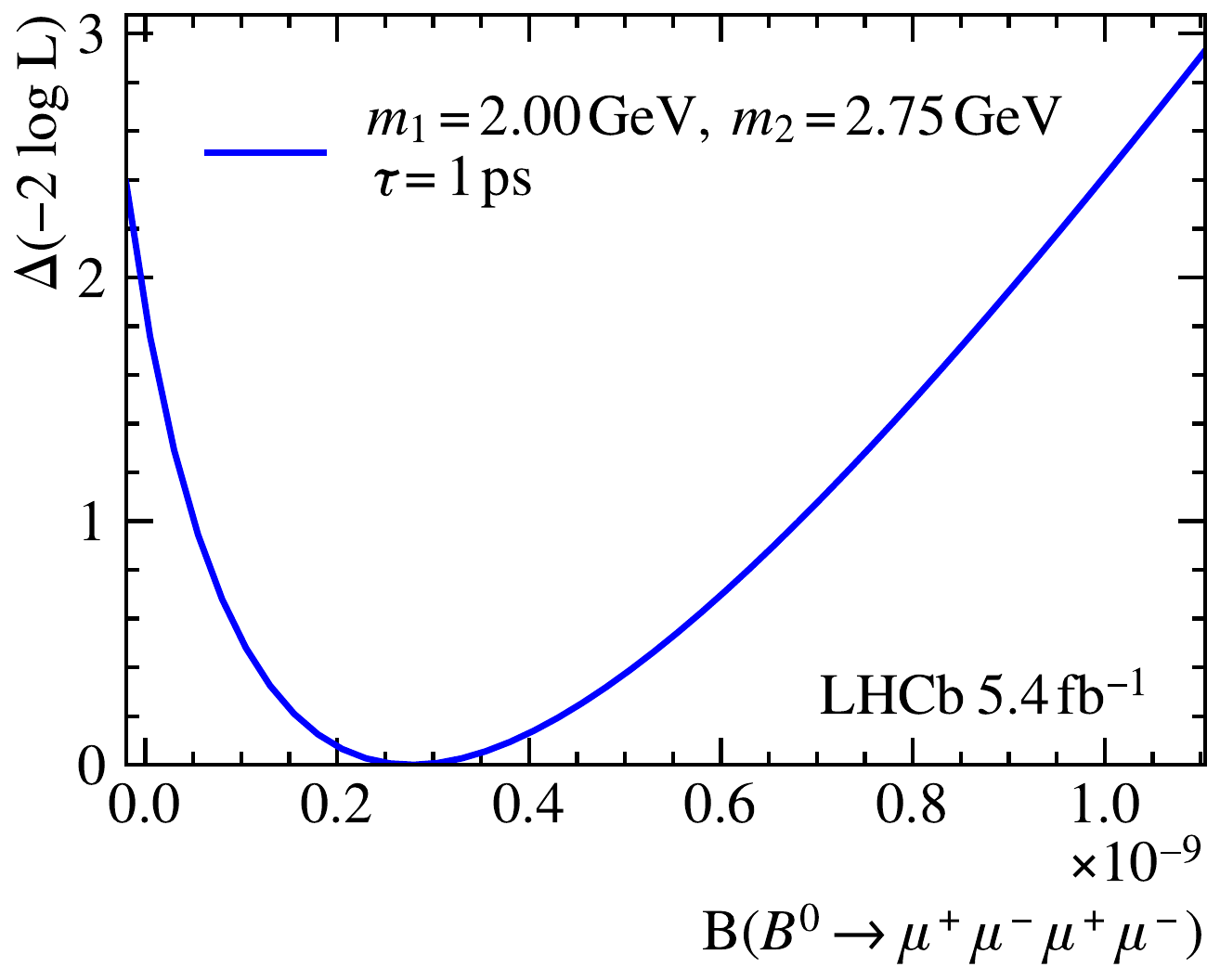}
    \includegraphics[width=0.33\linewidth]{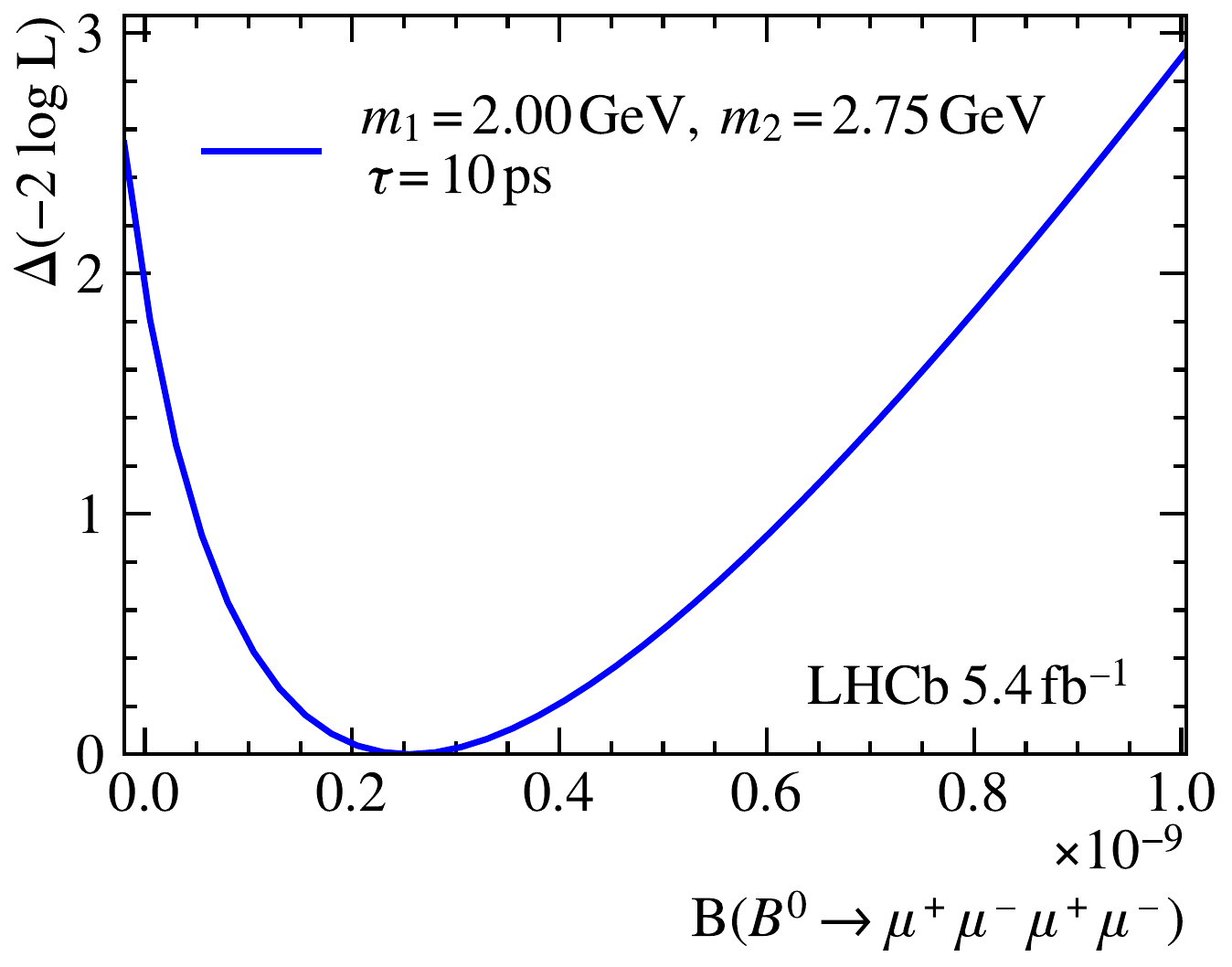}%
    \includegraphics[width=0.33\linewidth]{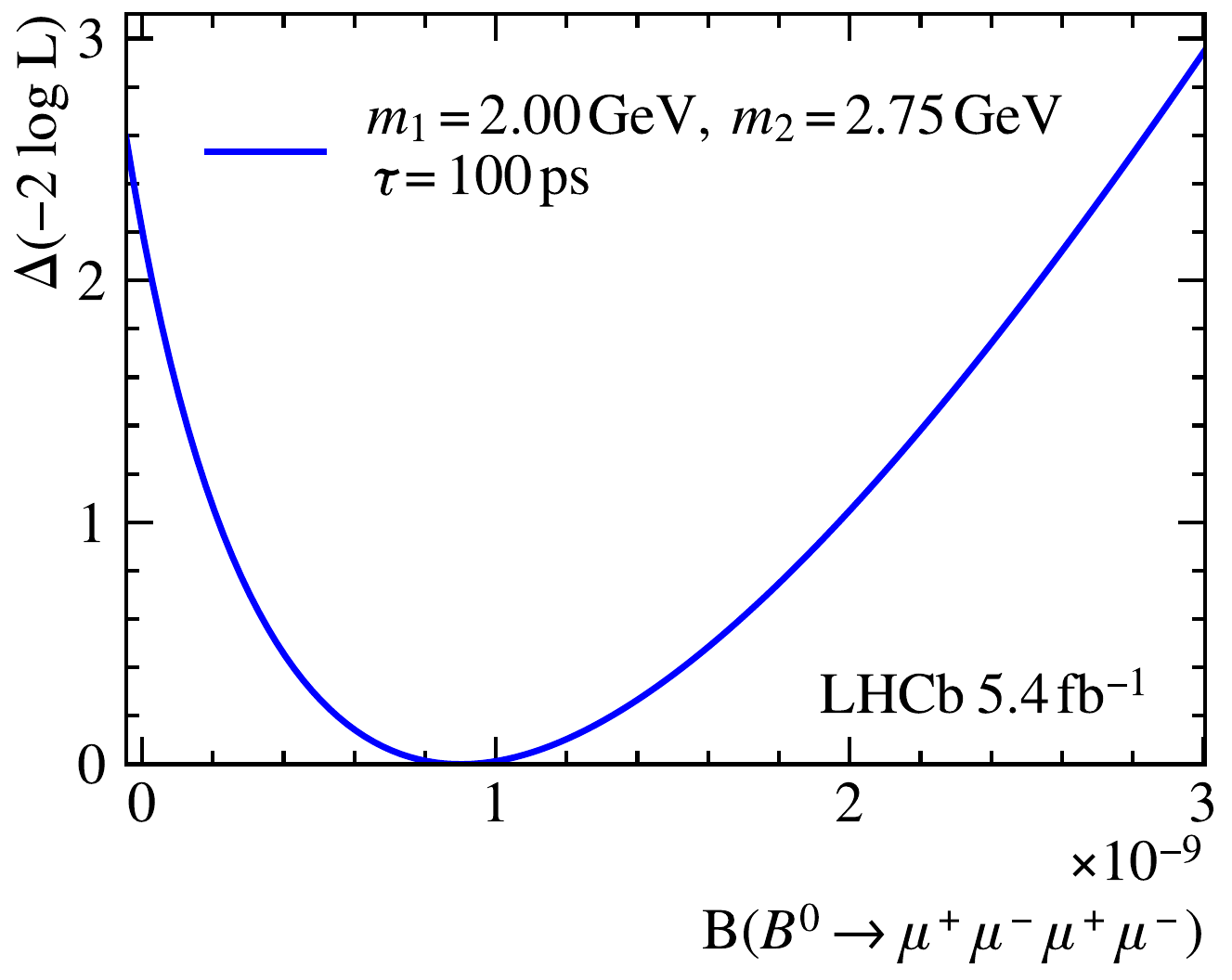}%
    \includegraphics[width=0.33\linewidth]{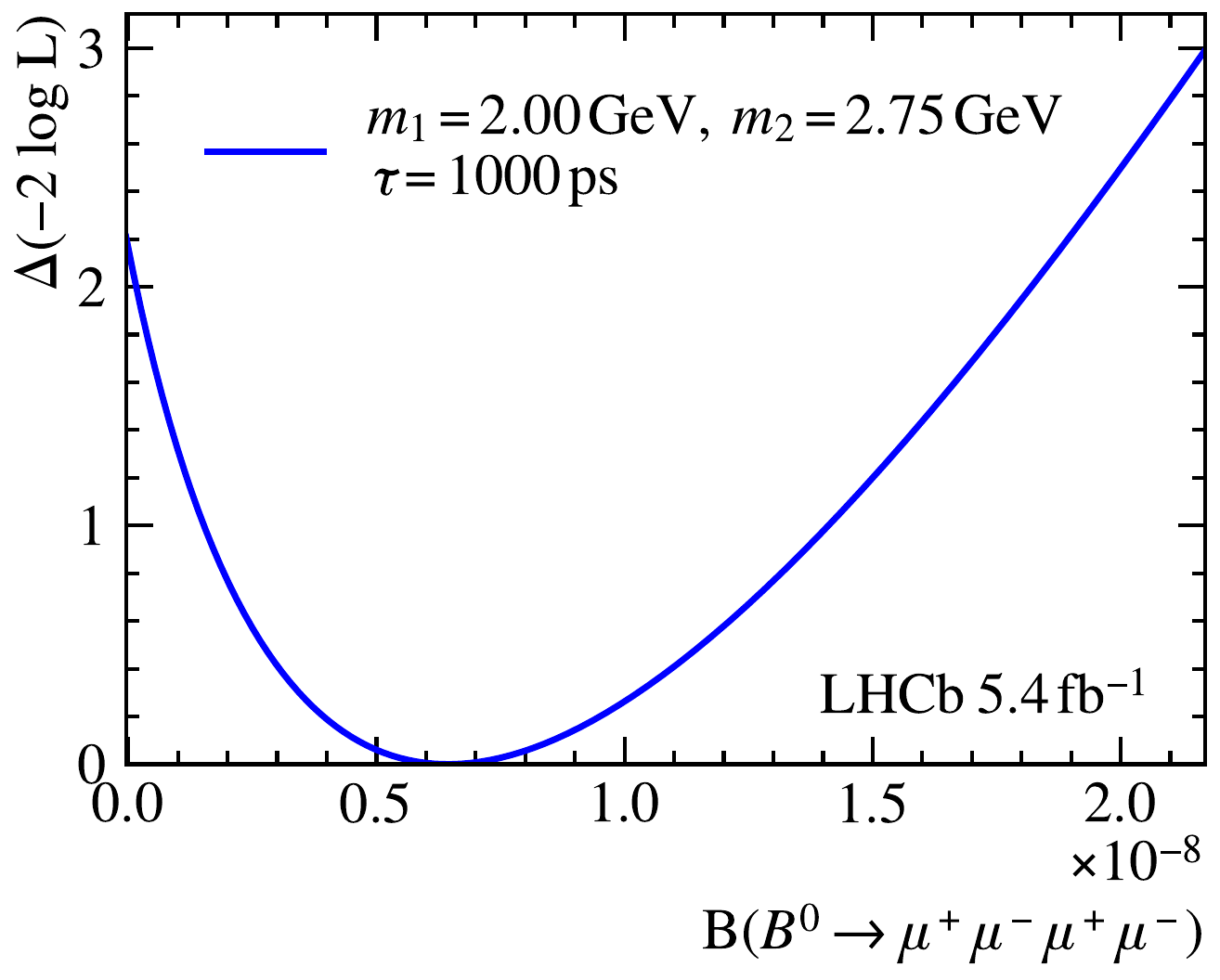}
    \caption{Profile likelihoods for the fits for \decay{\Bz}{\mu^+ \mu^- \mu^+ \mu^-}. From left to right, top to bottom, the fits correspond to the prompt scenario, the lifetimes $[1,10,100,1000]\ps$ for the mass configuration $(m_{a_1}=0.25\gev,m_{a_2}=0.40\gev)$ and the lifetimes $[1,10,100,1000]\ps$ for the mass configuration $(m_{a_1}=2.0\gev,m_{a_2}=2.75\gev)$.}
    \label{fig:NLL_Bd24mu}
\end{figure}

\begin{figure}
    \centering
    \includegraphics[width=0.33\linewidth]{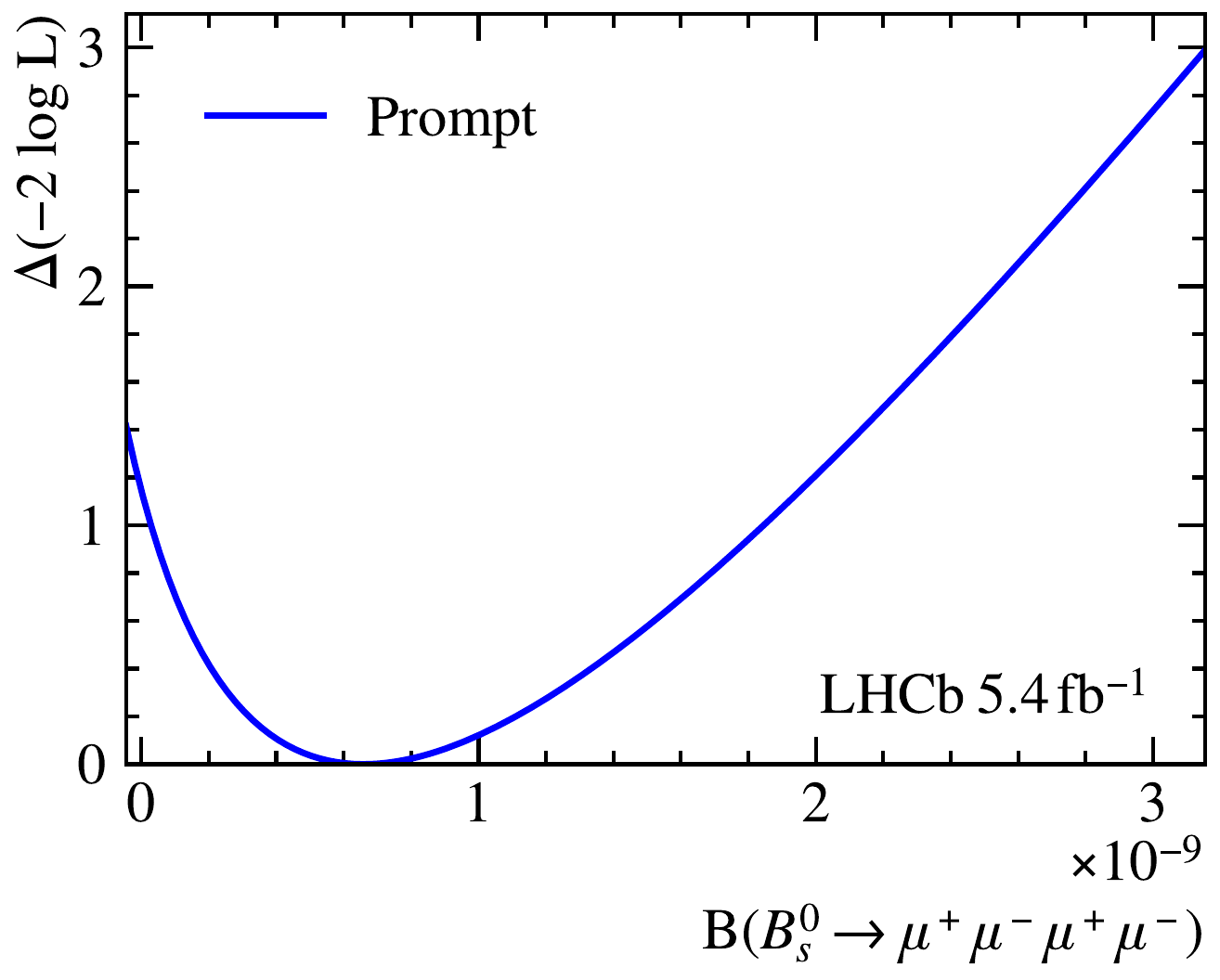}%
    \includegraphics[width=0.33\linewidth]{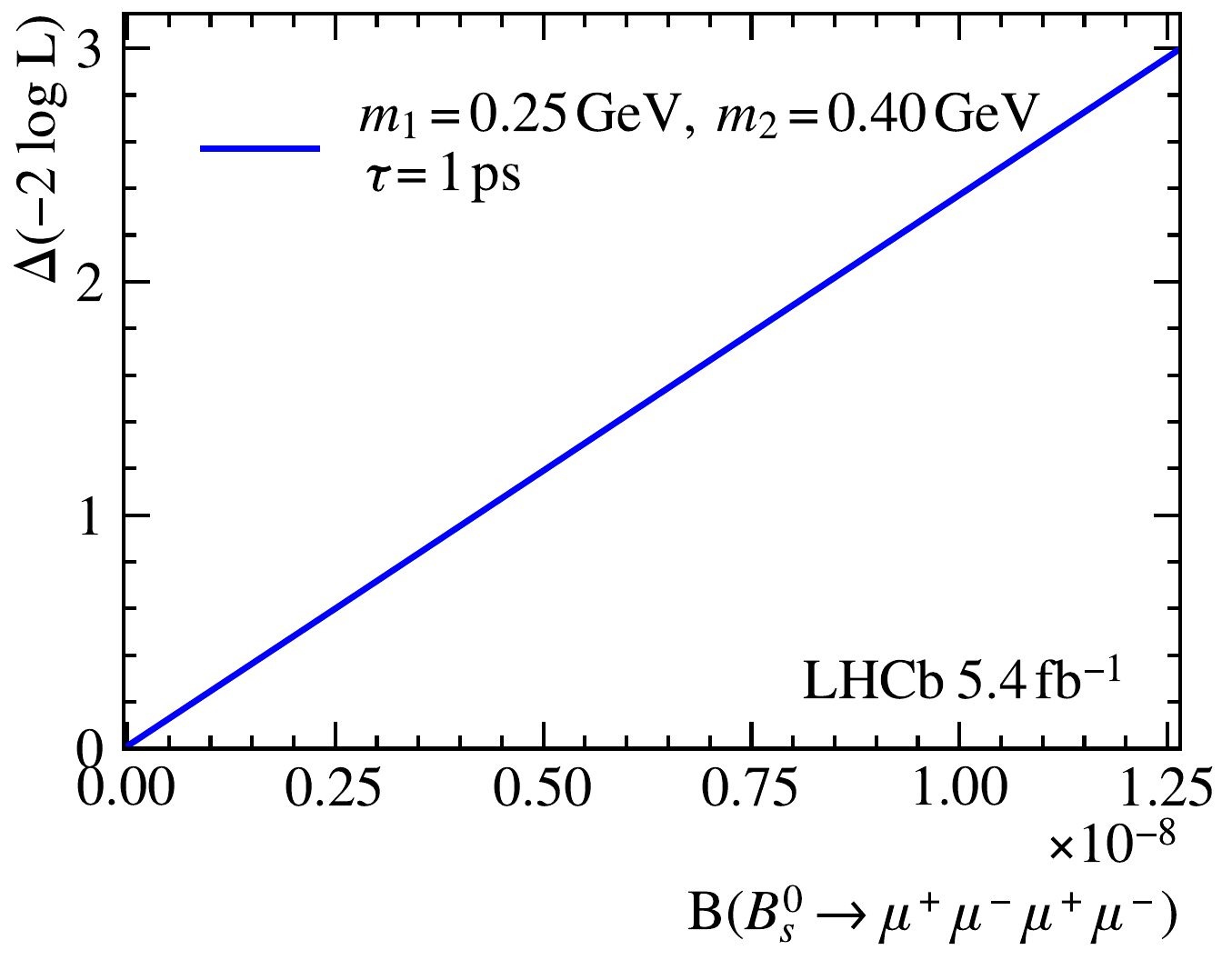}%
    \includegraphics[width=0.33\linewidth]{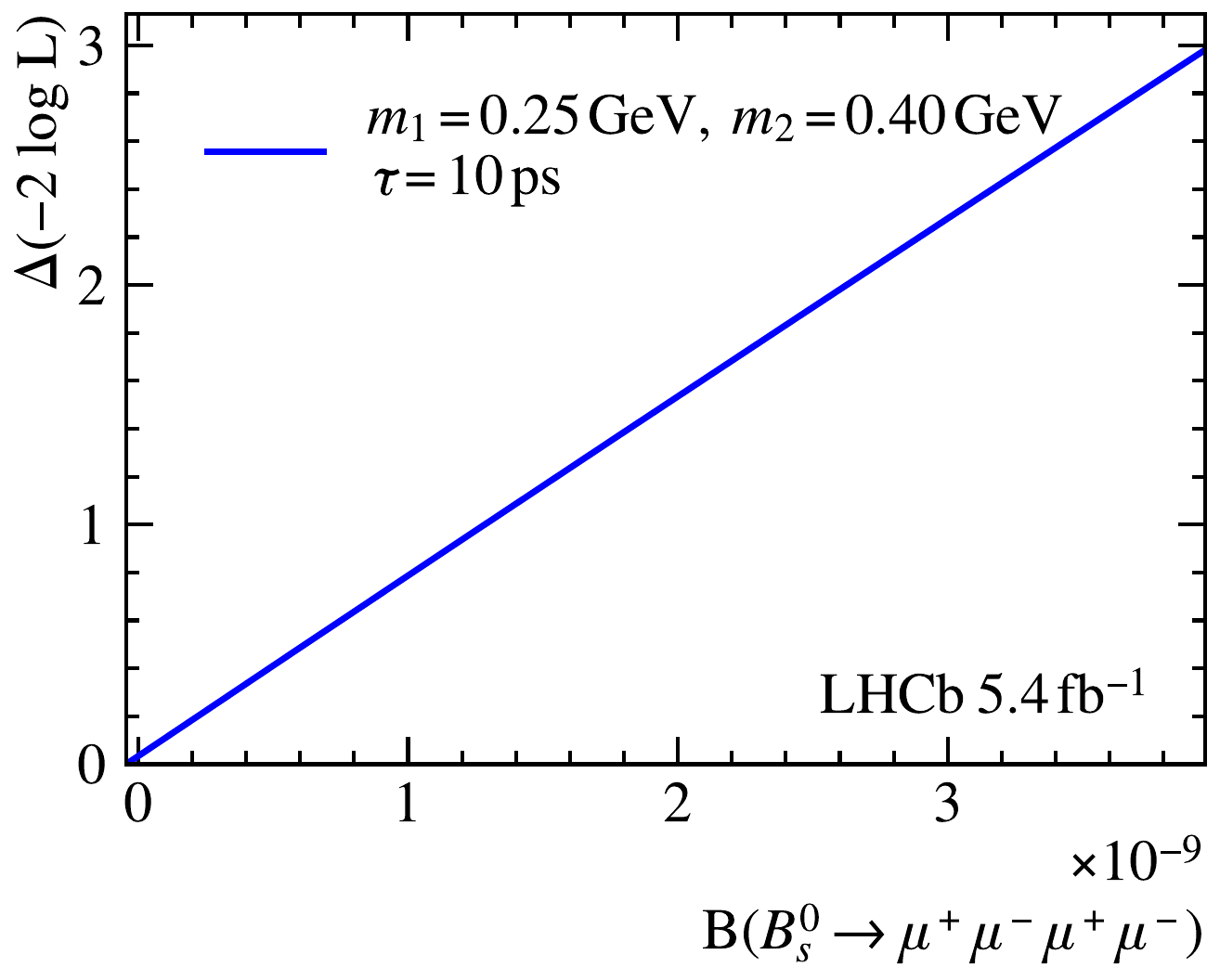}
    \includegraphics[width=0.33\linewidth]{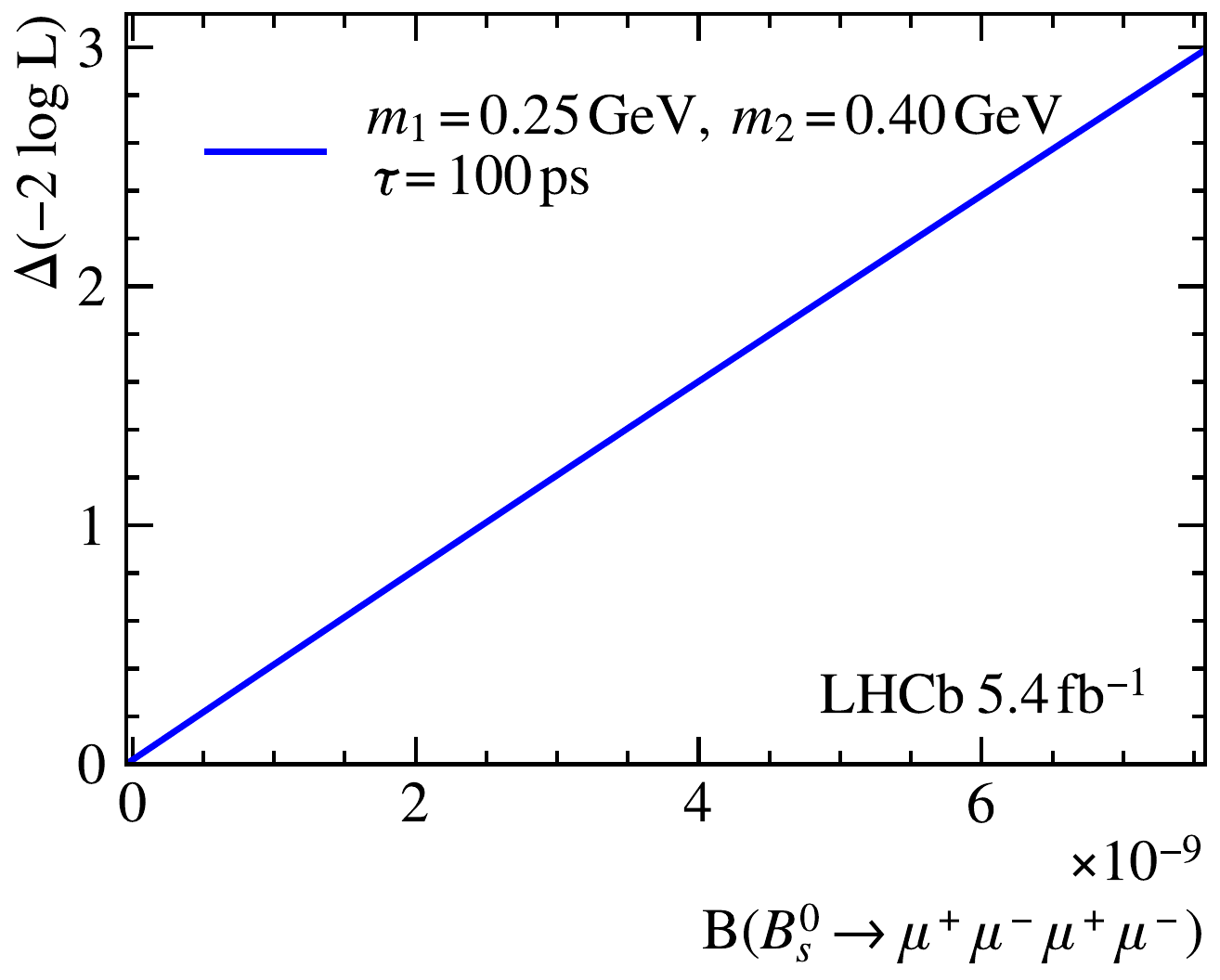}%
    \includegraphics[width=0.33\linewidth]{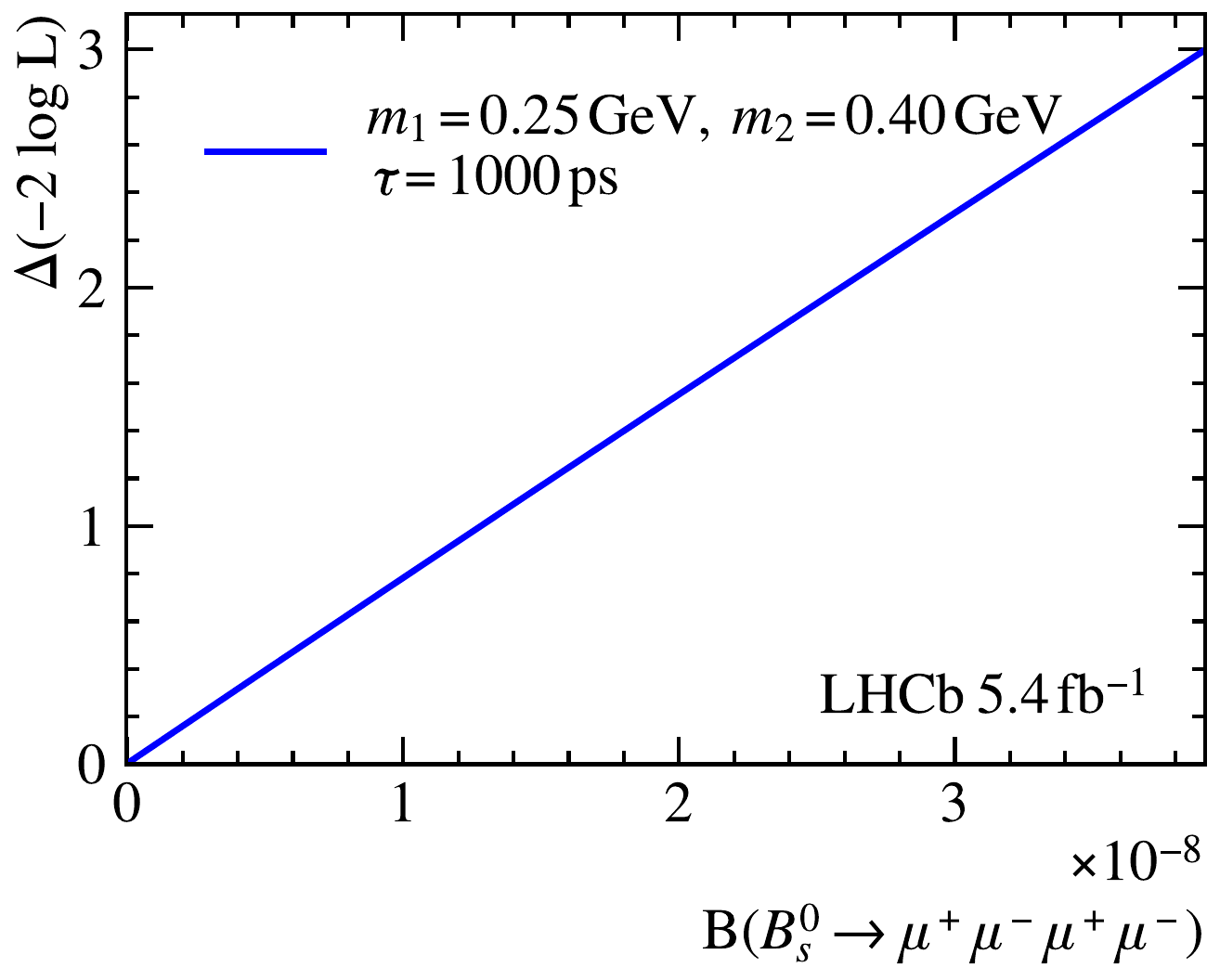}%
    \includegraphics[width=0.33\linewidth]{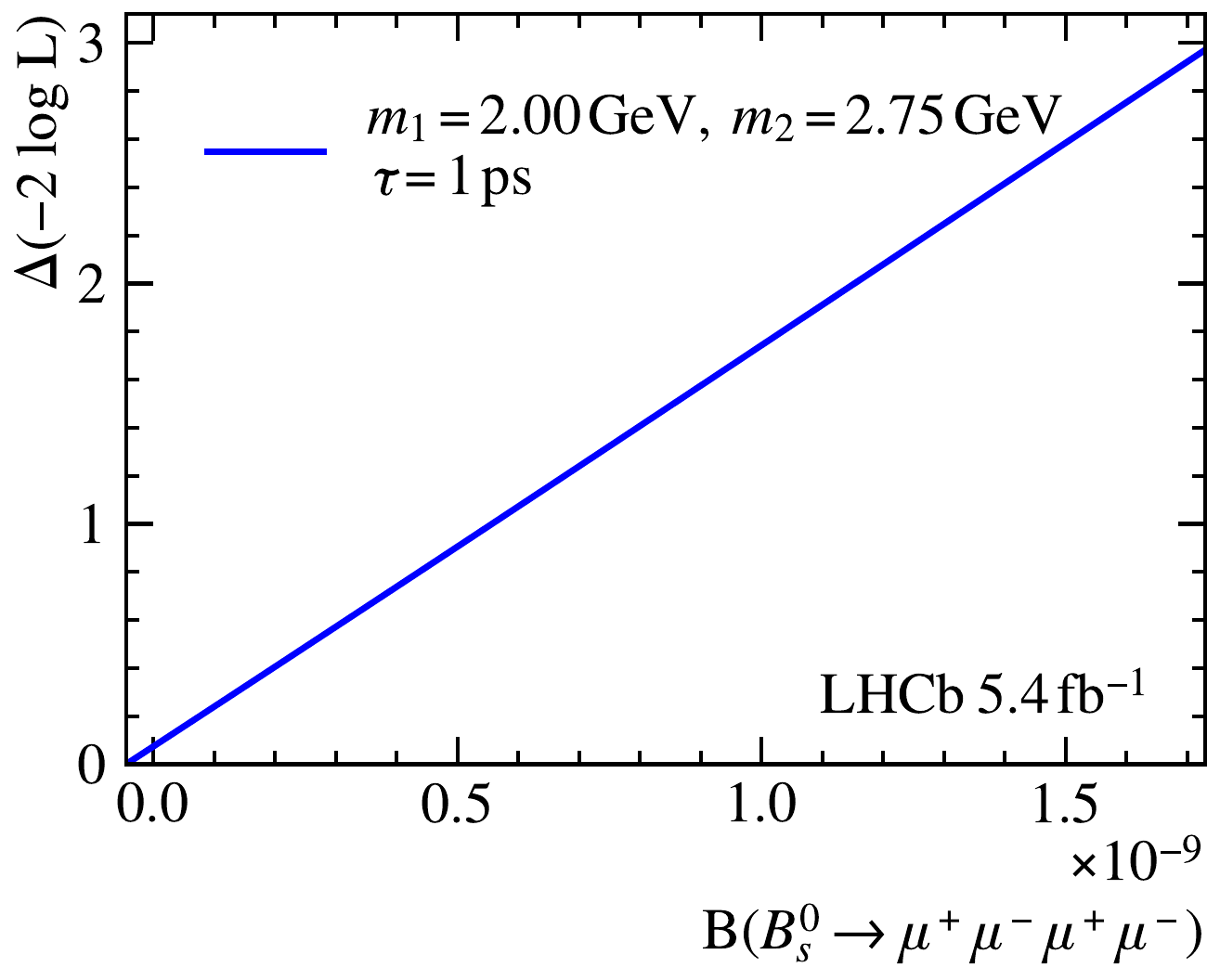}
    \includegraphics[width=0.33\linewidth]{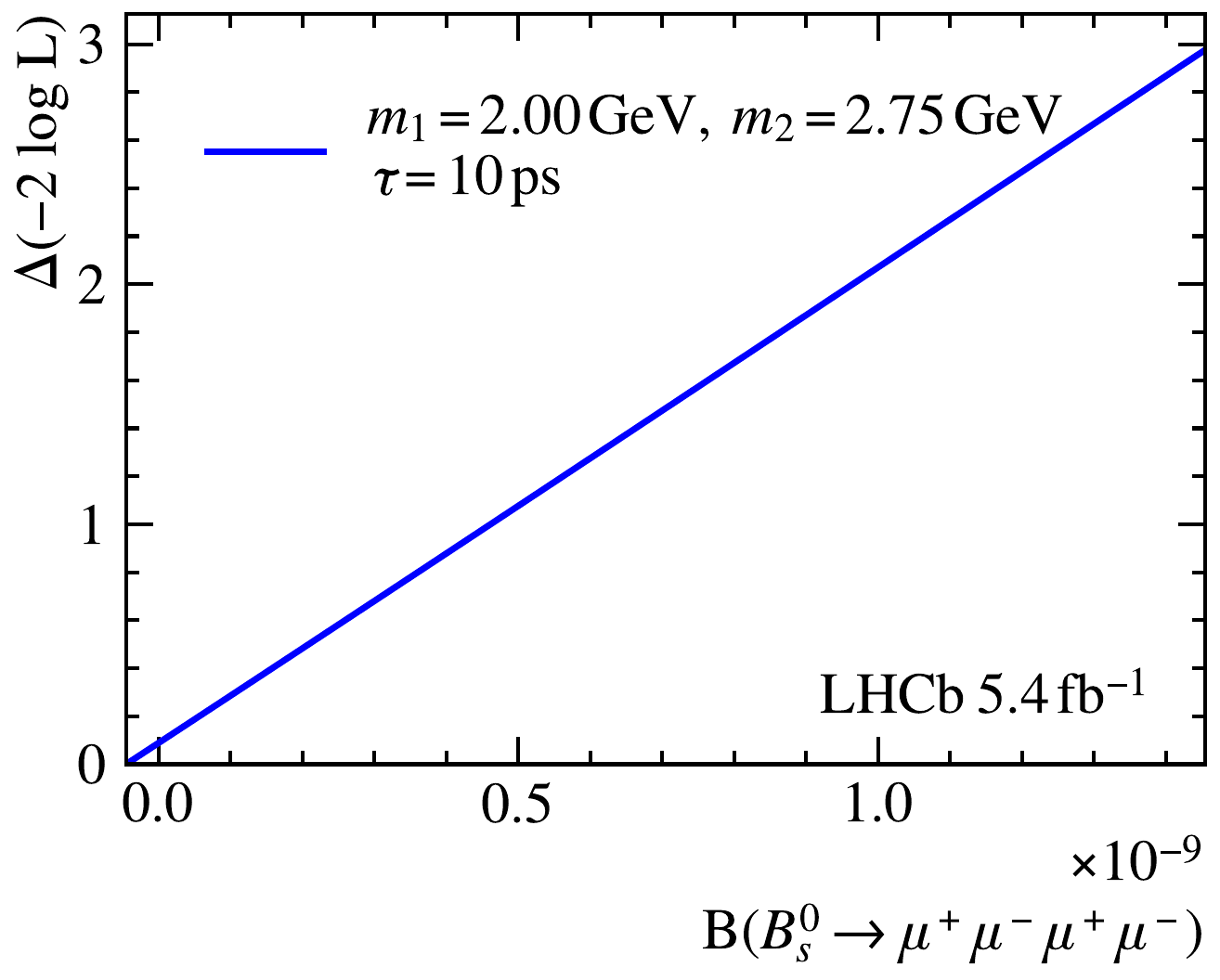}%
    \includegraphics[width=0.33\linewidth]{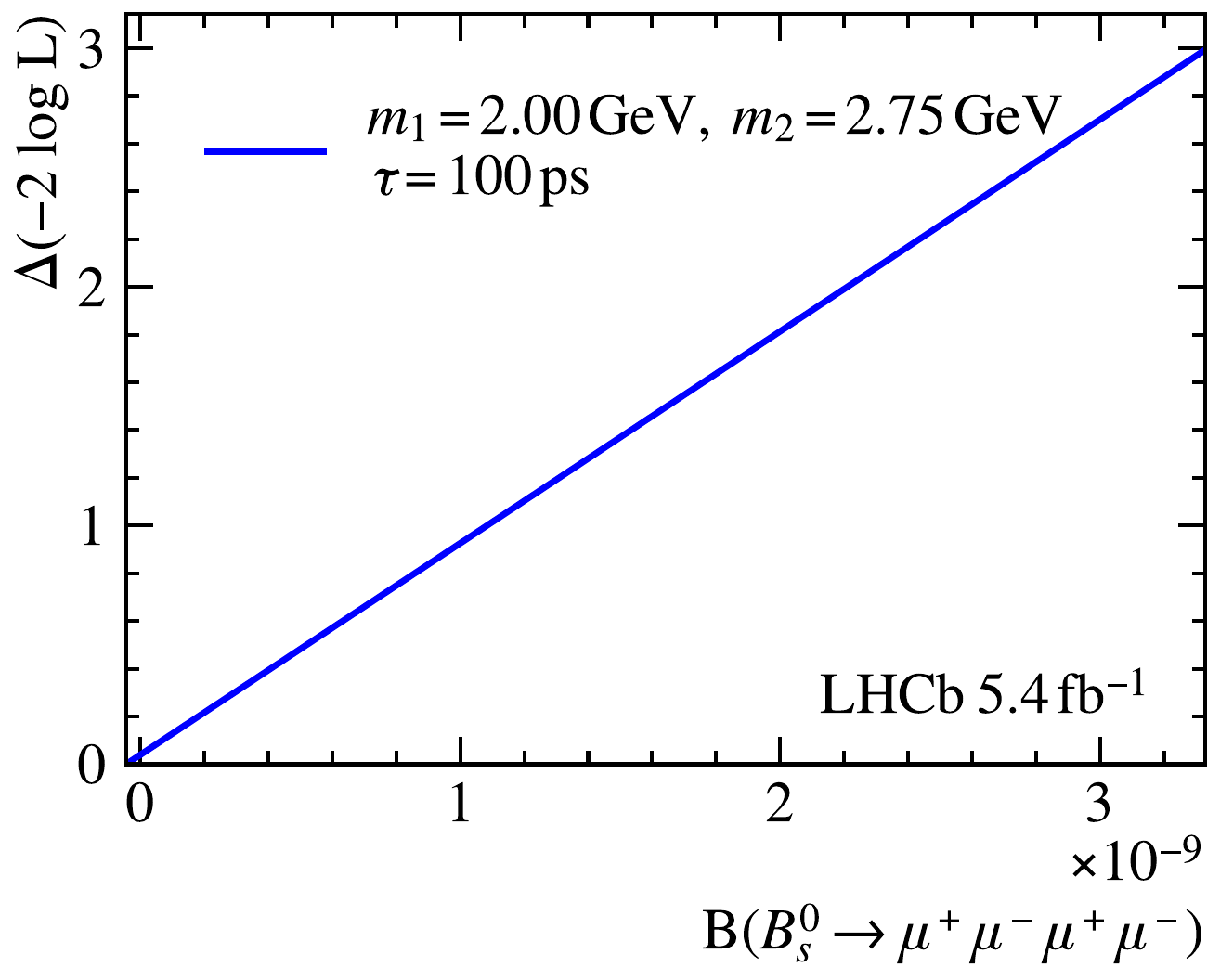}%
    \includegraphics[width=0.33\linewidth]{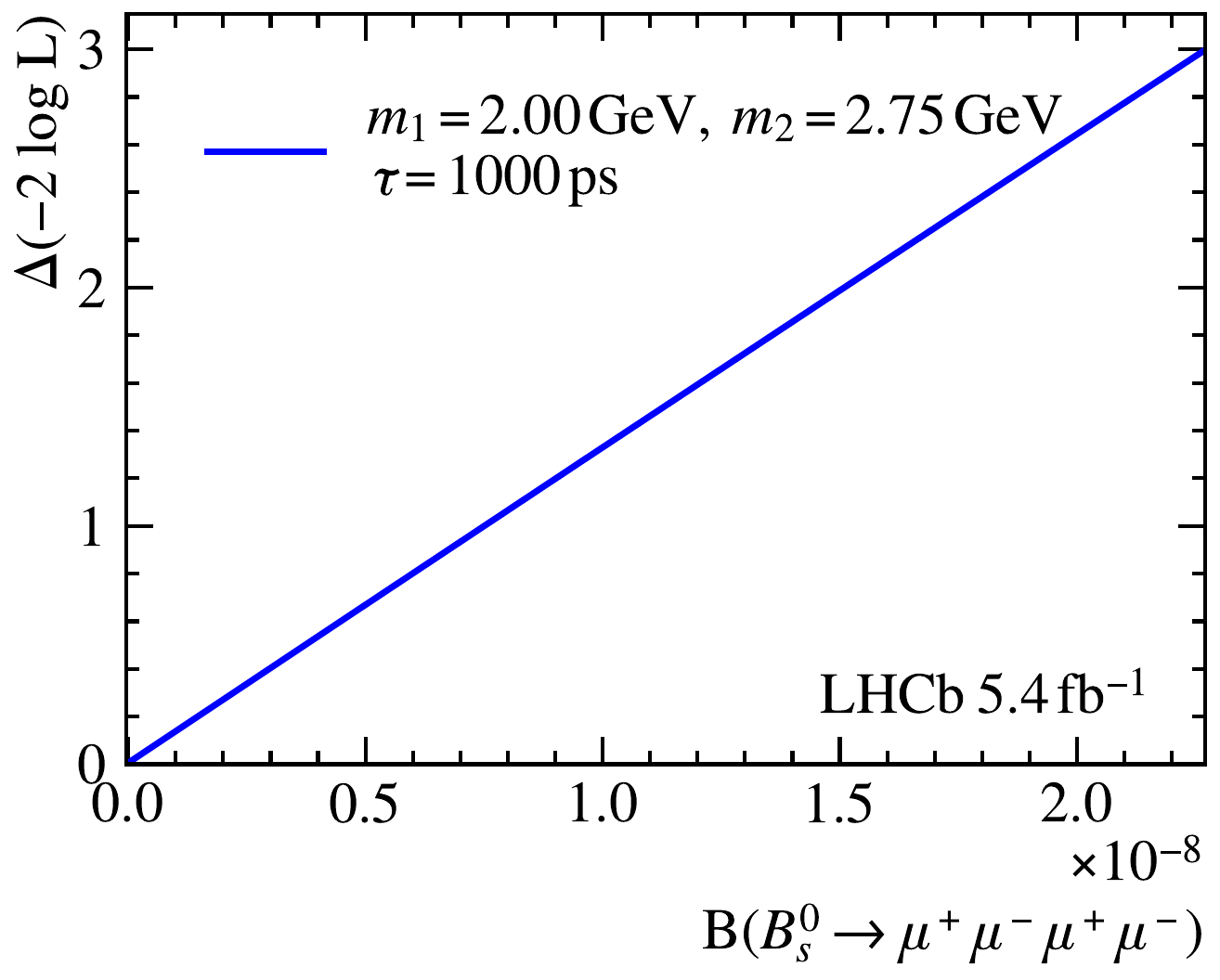}
    \caption{Profile Likelihoods for the fits for \decay{\Bs}{\mu^+ \mu^- \mu^+ \mu^-}. From left to right, top to bottom, the fits correspond to the prompt scenario, the lifetimes $[1,10,100,1000]\ps$ for the mass configuration $(m_{a_1}=0.25\gev,m_{a_2}=0.40\gev)$ and the lifetimes $[1,10,100,1000]\ps$ for the mass configuration $(m_{a_1}=2.0\gev,m_{a_2}=2.75\gev)$. The linear behaviour of the likelihoods for the displaced selections arises because there is no data observed in the signal region.}
    \label{fig:NLL_Bs24mu}
\end{figure}

\begin{figure}
    \centering
    \includegraphics[width=0.33\linewidth]{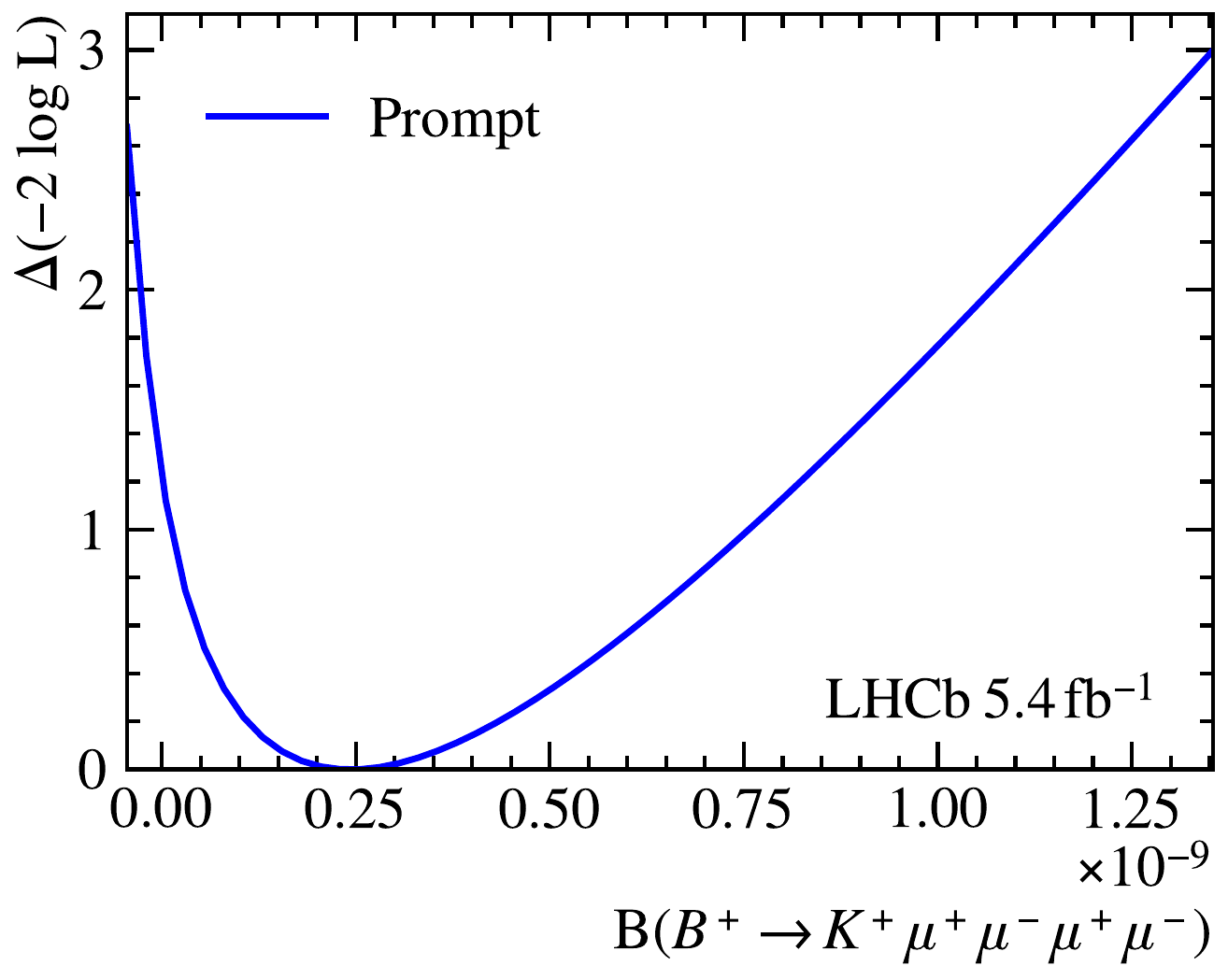}%
    \includegraphics[width=0.33\linewidth]{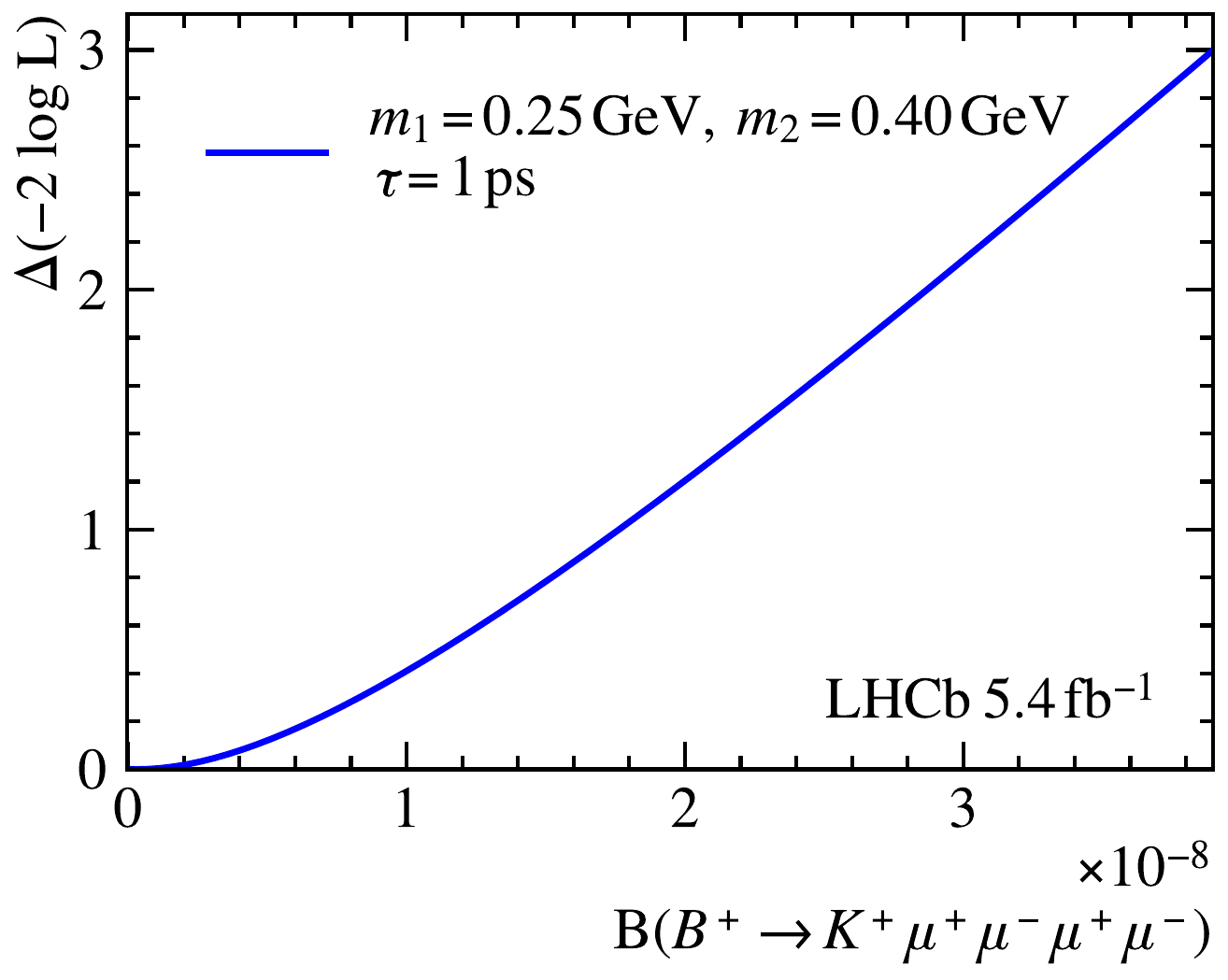}%
    \includegraphics[width=0.33\linewidth]{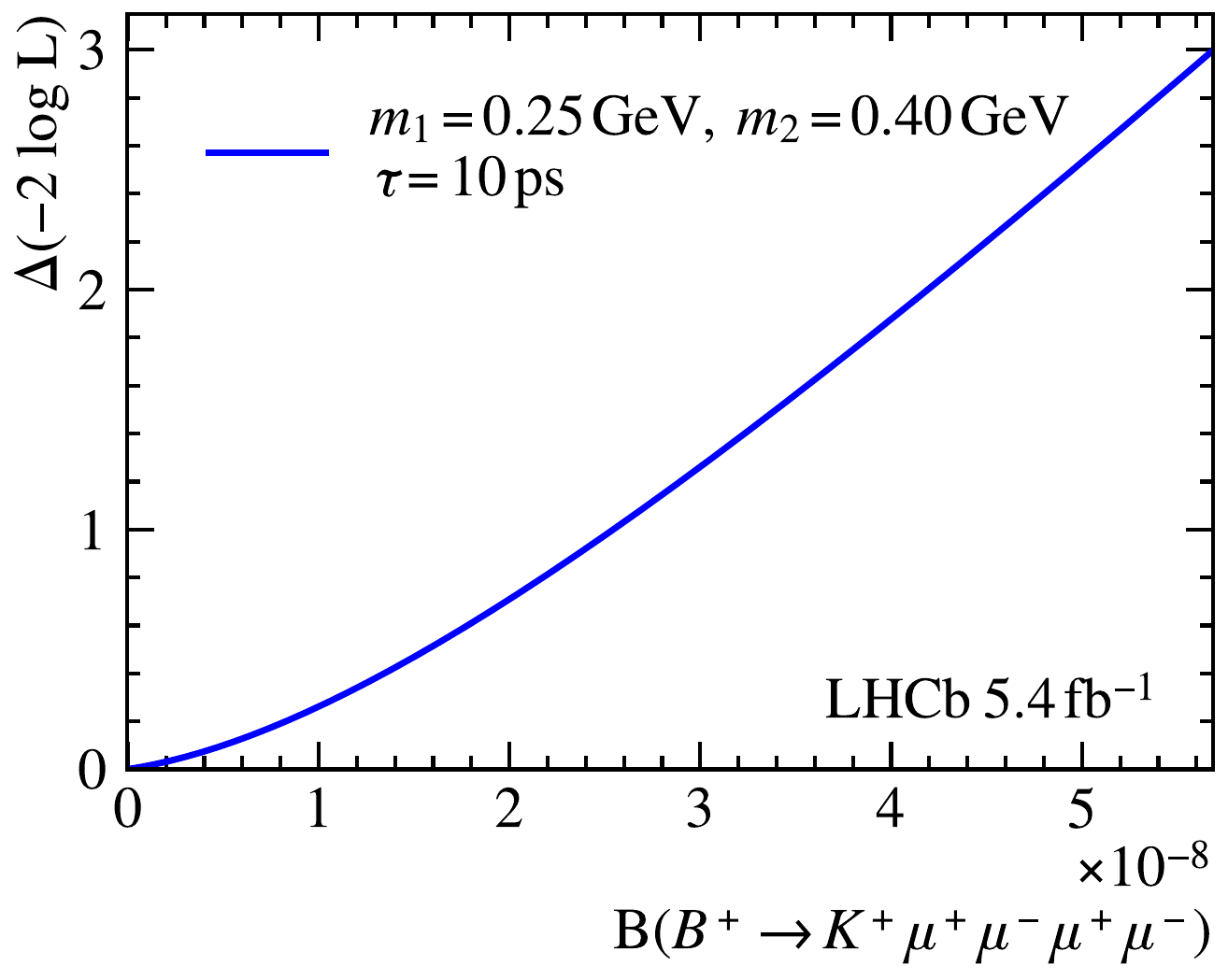}
    \includegraphics[width=0.33\linewidth]{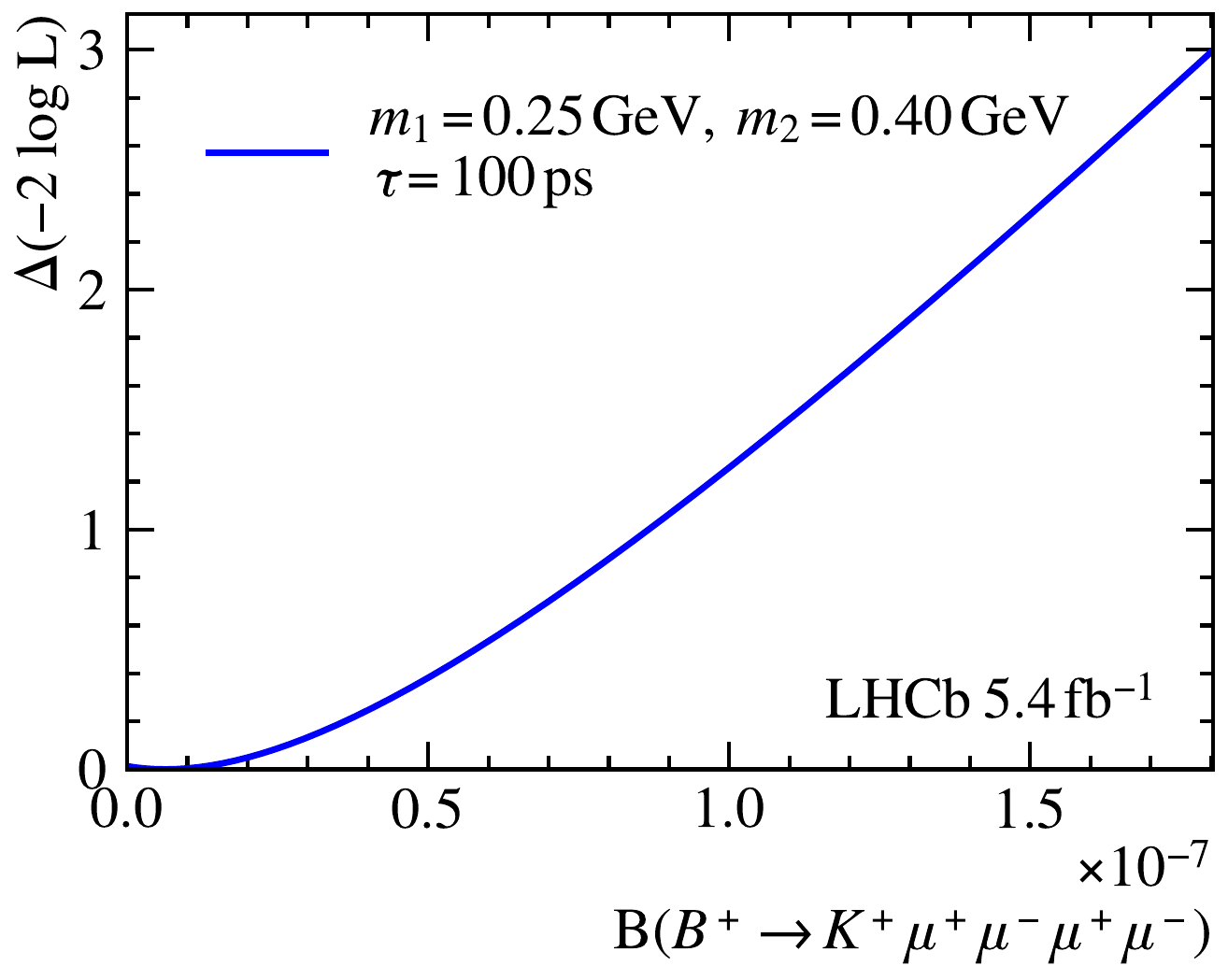}%
    \includegraphics[width=0.33\linewidth]{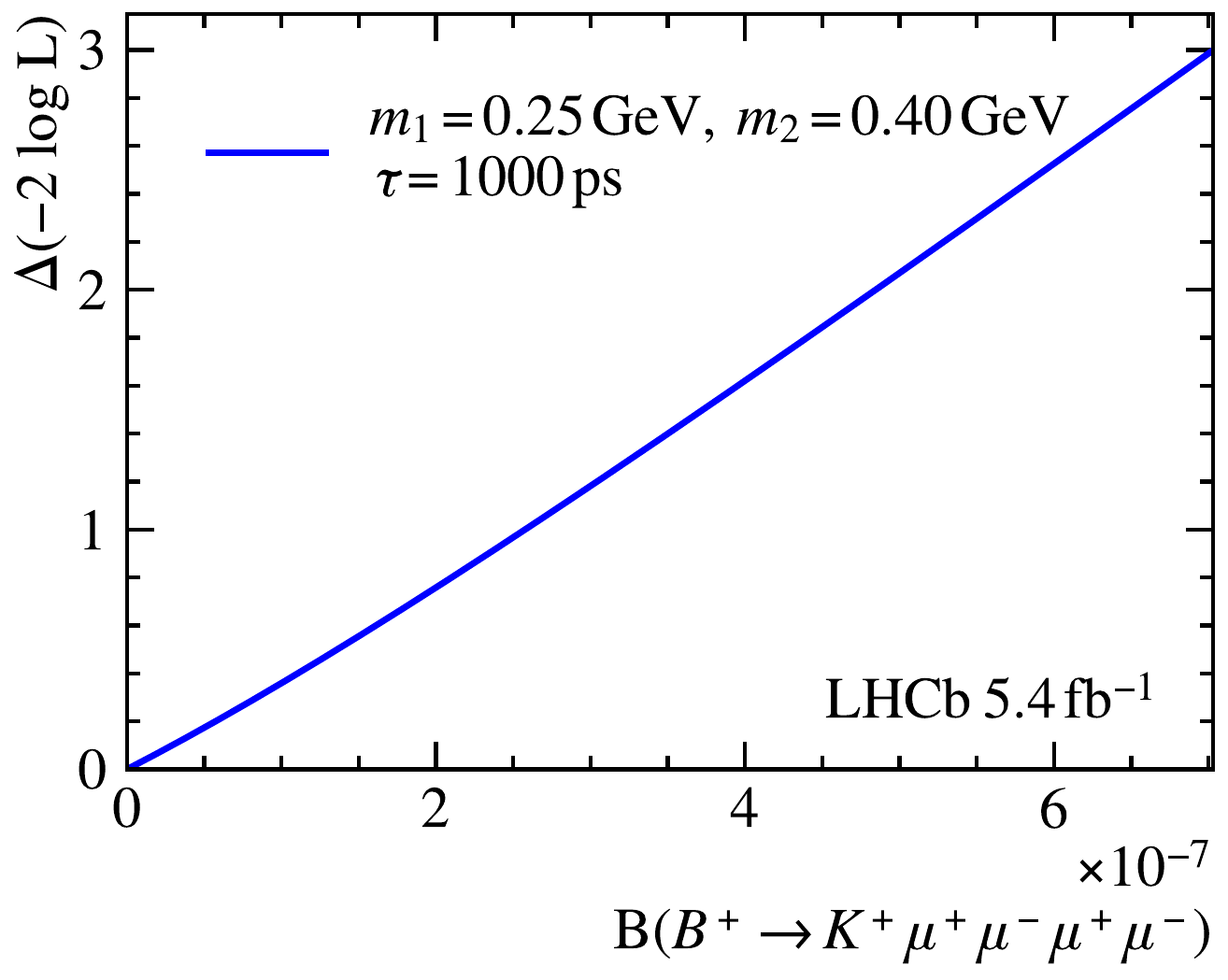}%
    \includegraphics[width=0.33\linewidth]{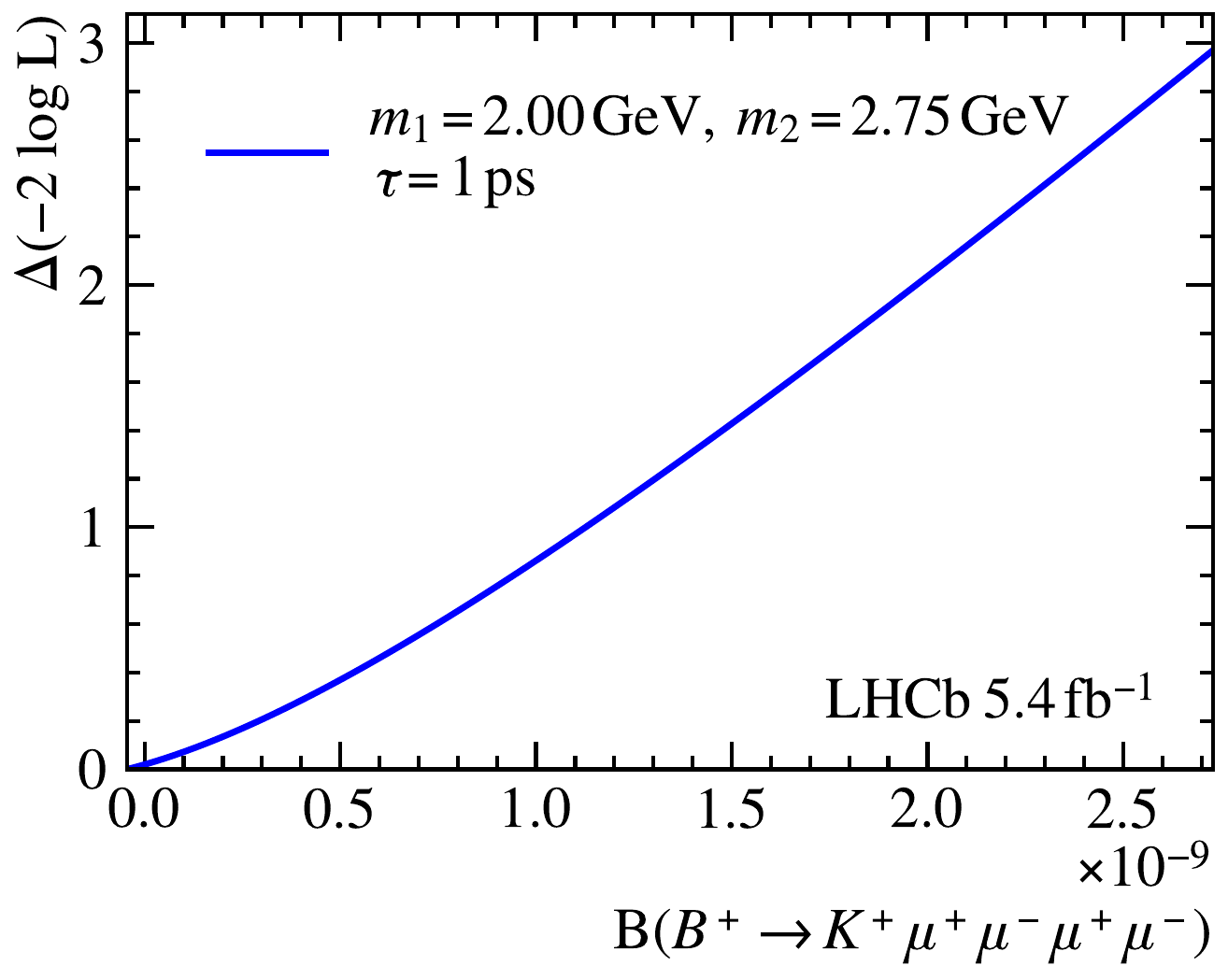}
    \includegraphics[width=0.33\linewidth]{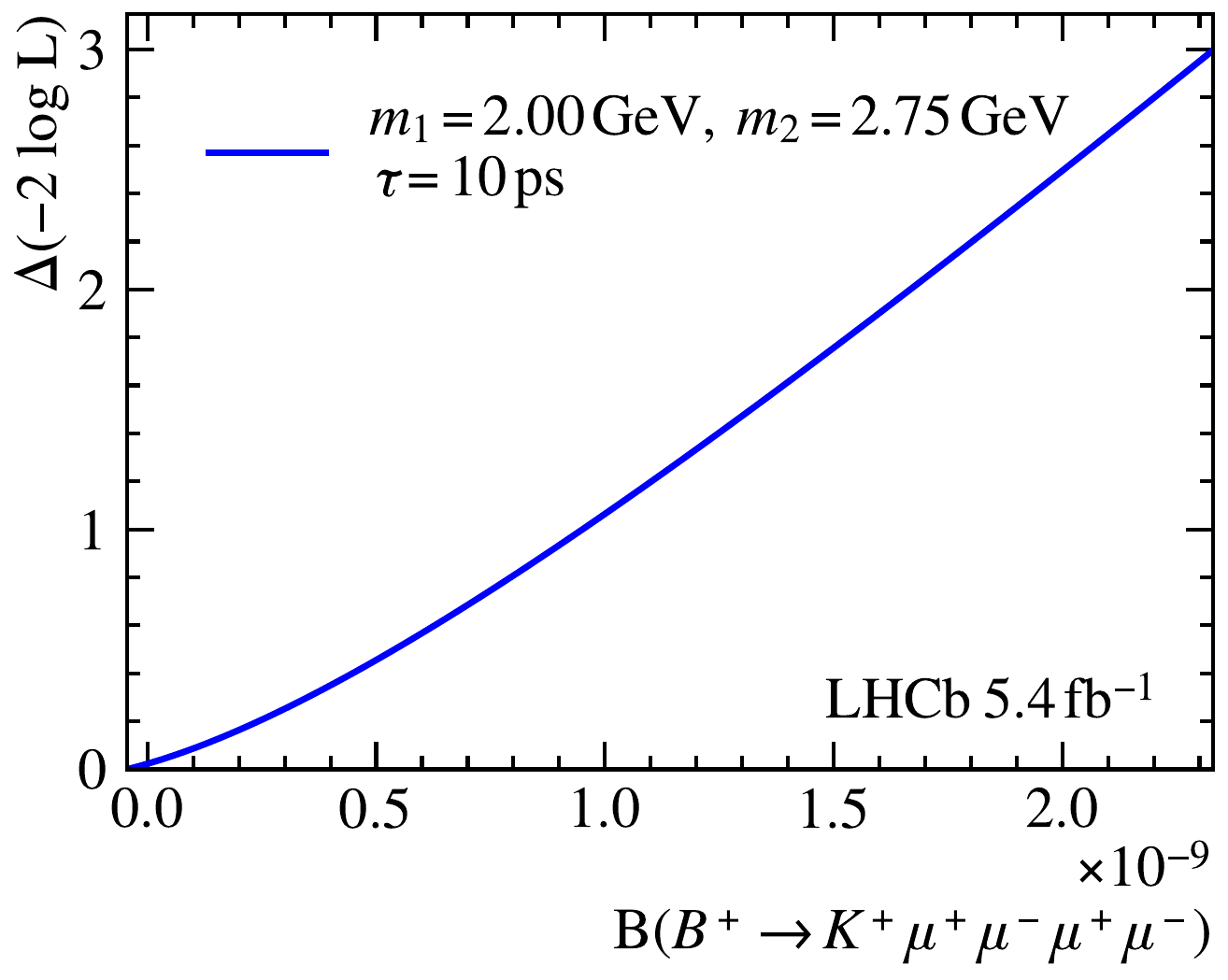}%
    \includegraphics[width=0.33\linewidth]{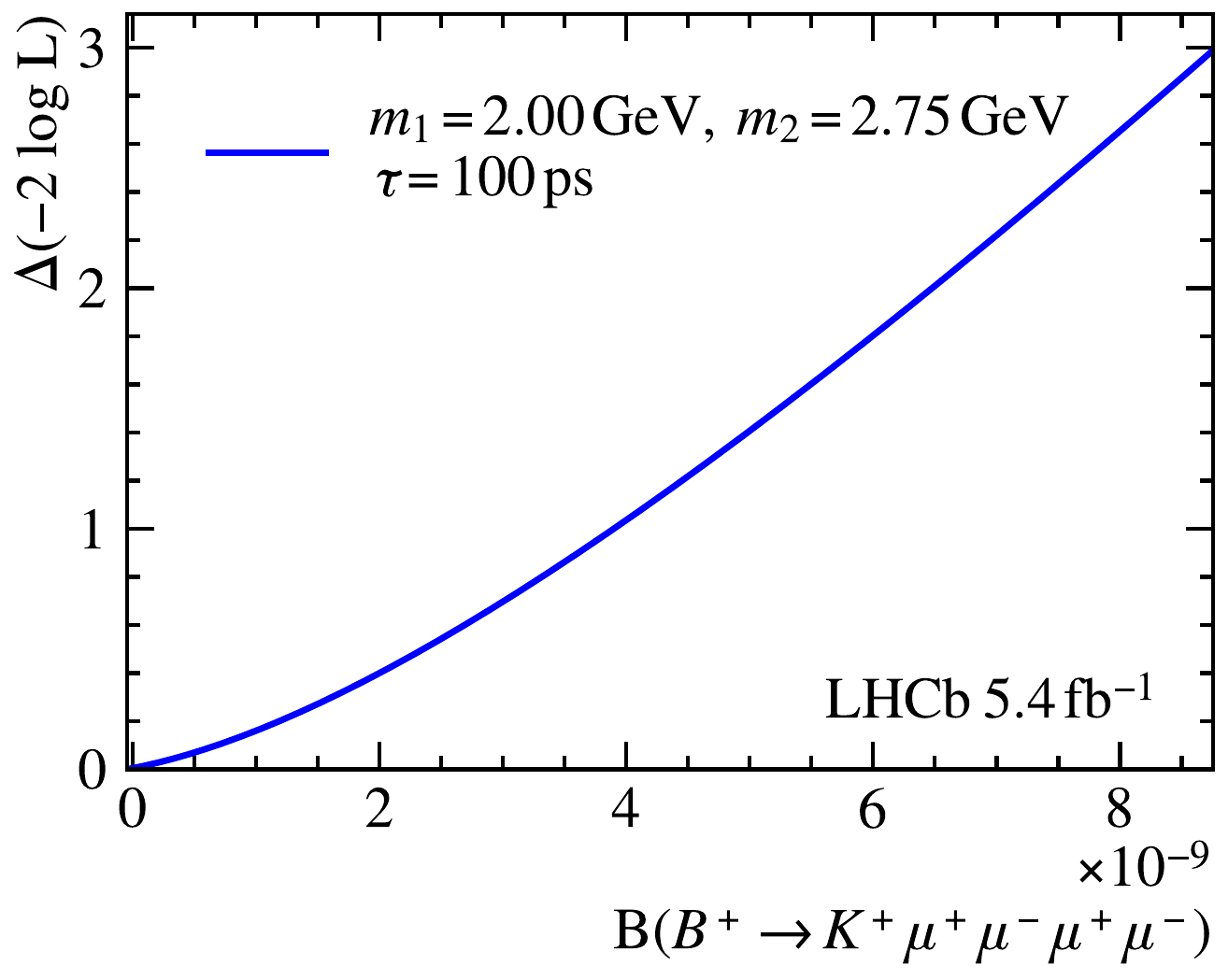}%
    \includegraphics[width=0.33\linewidth]{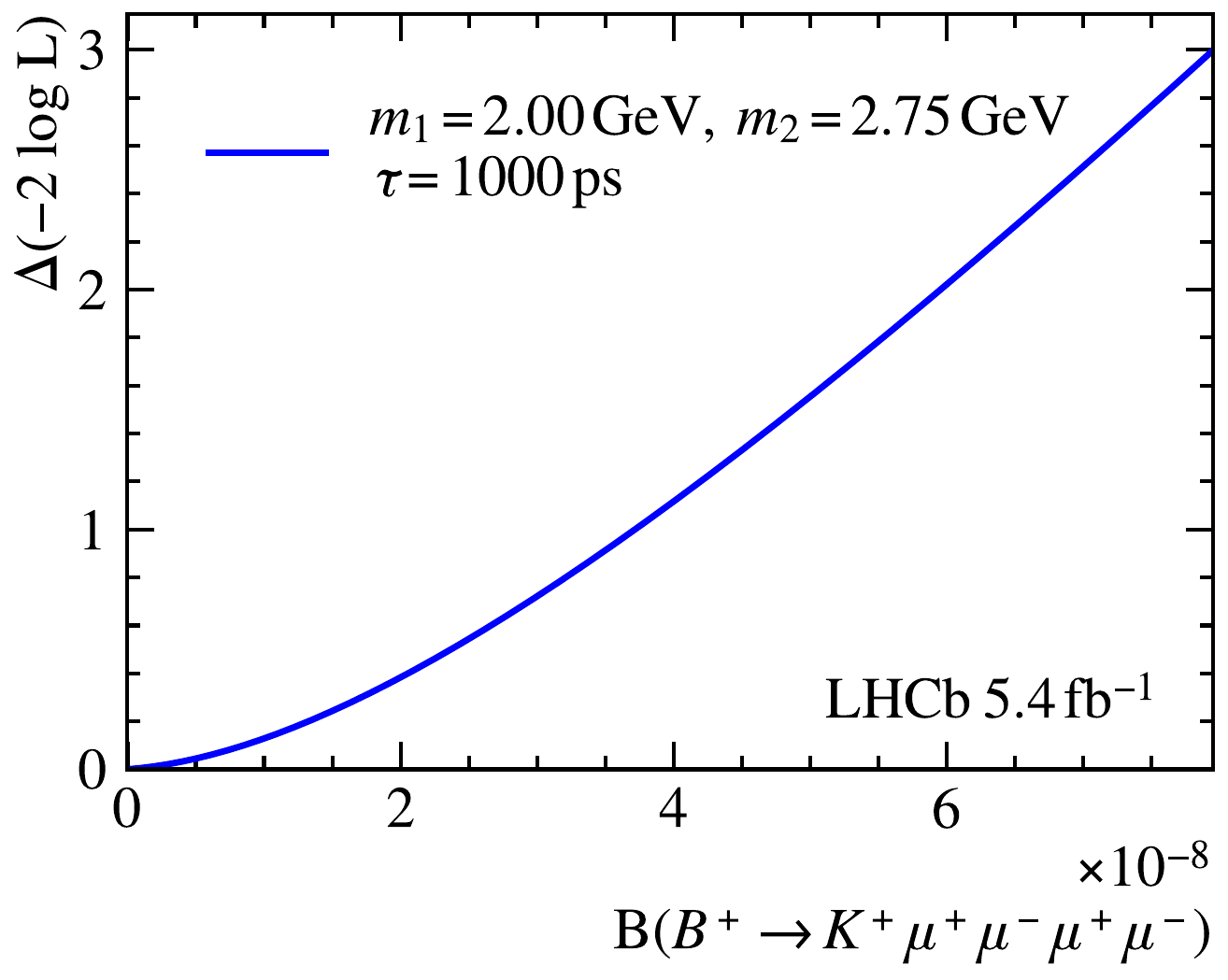}
    \caption{Profile Likelihoods for the fits for \decay{\Bu}{\Kp \mu^+ \mu^- \mu^+ \mu^-}. From left to right the fits correspond to the prompt scenario, the lifetime $[1,10,100,1000]\ps$ for the mass configuration $(m_{a_1}=0.25\gev,m_{a_2}=0.40\gev)$ and the lifetimes $[1,10,100,1000]\ps$ for the mass configuration \mbox{$(m_{a_1}=2.0\gev,m_{a_2}=2.75\gev)$}.}
    \label{fig:NLL_B24muK}
\end{figure}

\begin{figure}
    \centering
    \includegraphics[width=0.33\linewidth]{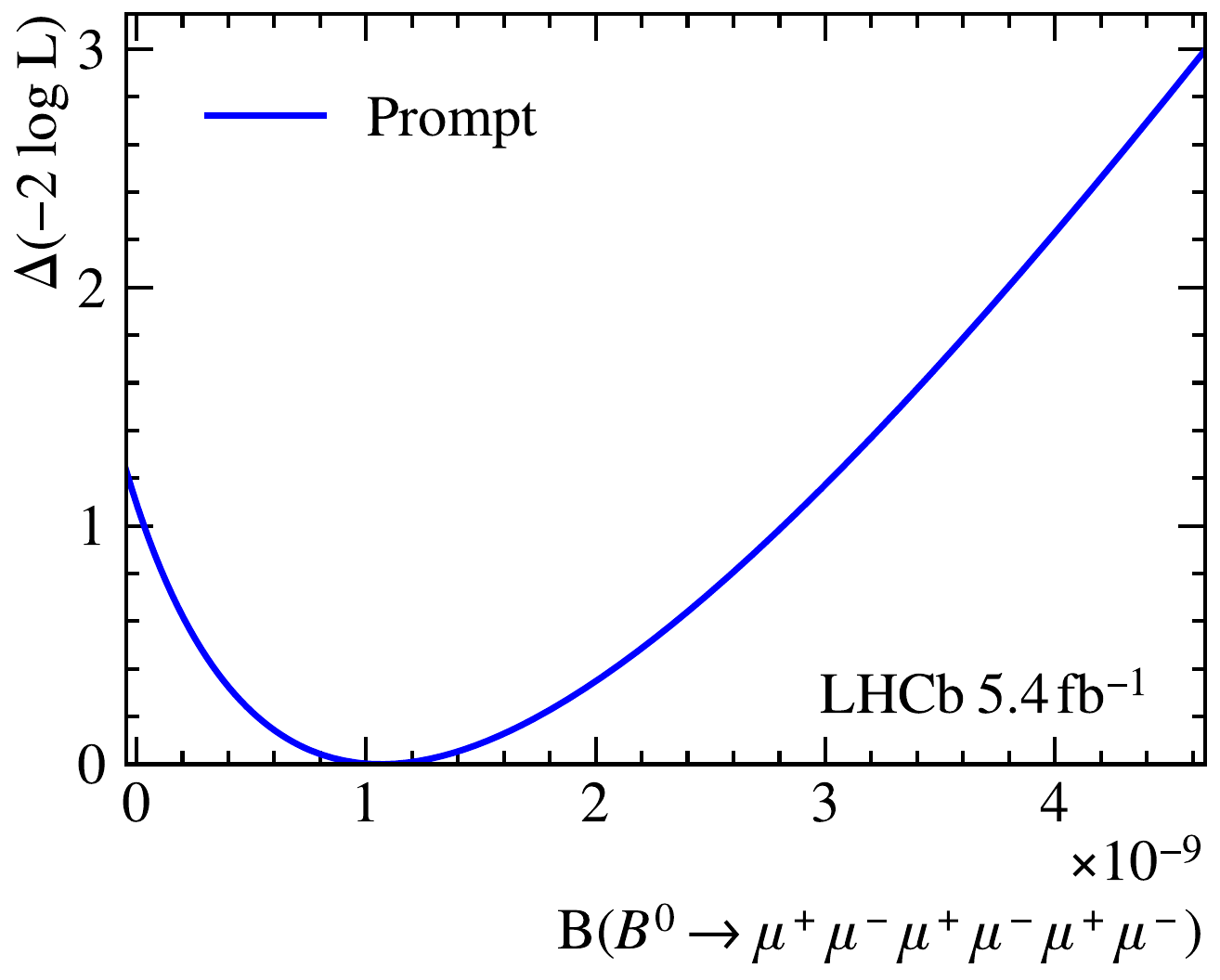}%
    \includegraphics[width=0.33\linewidth]{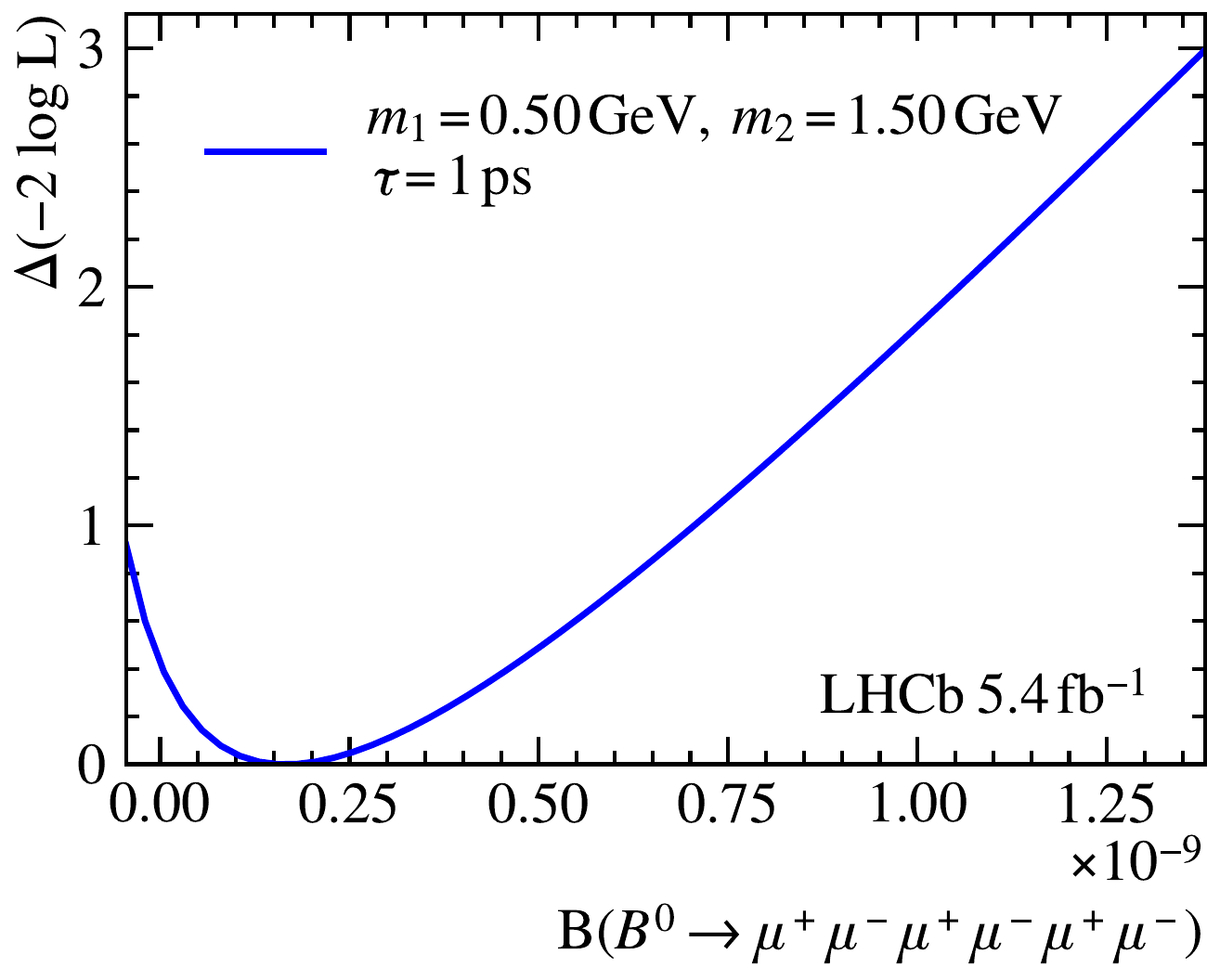}%
    \includegraphics[width=0.33\linewidth]{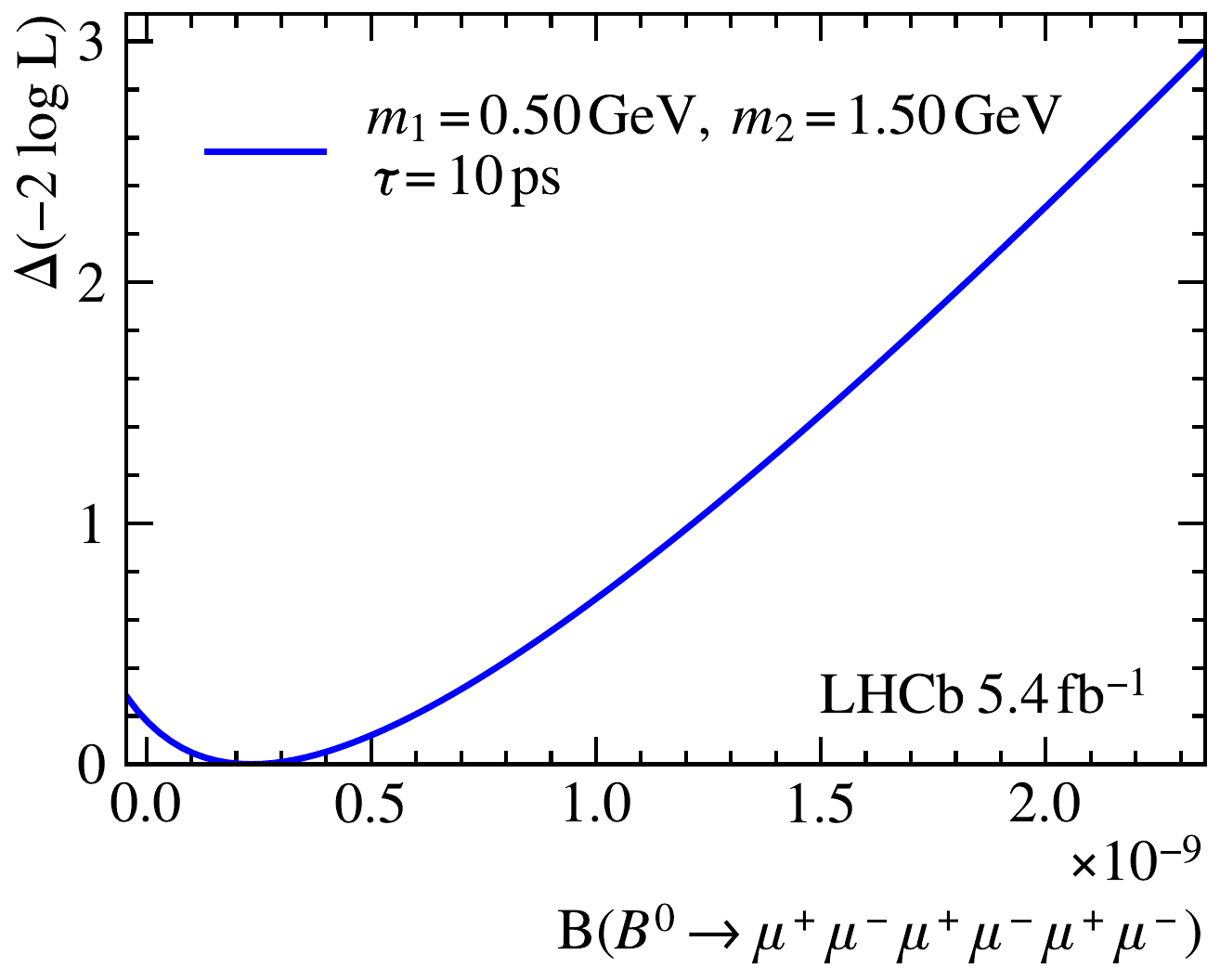}
    \includegraphics[width=0.33\linewidth]{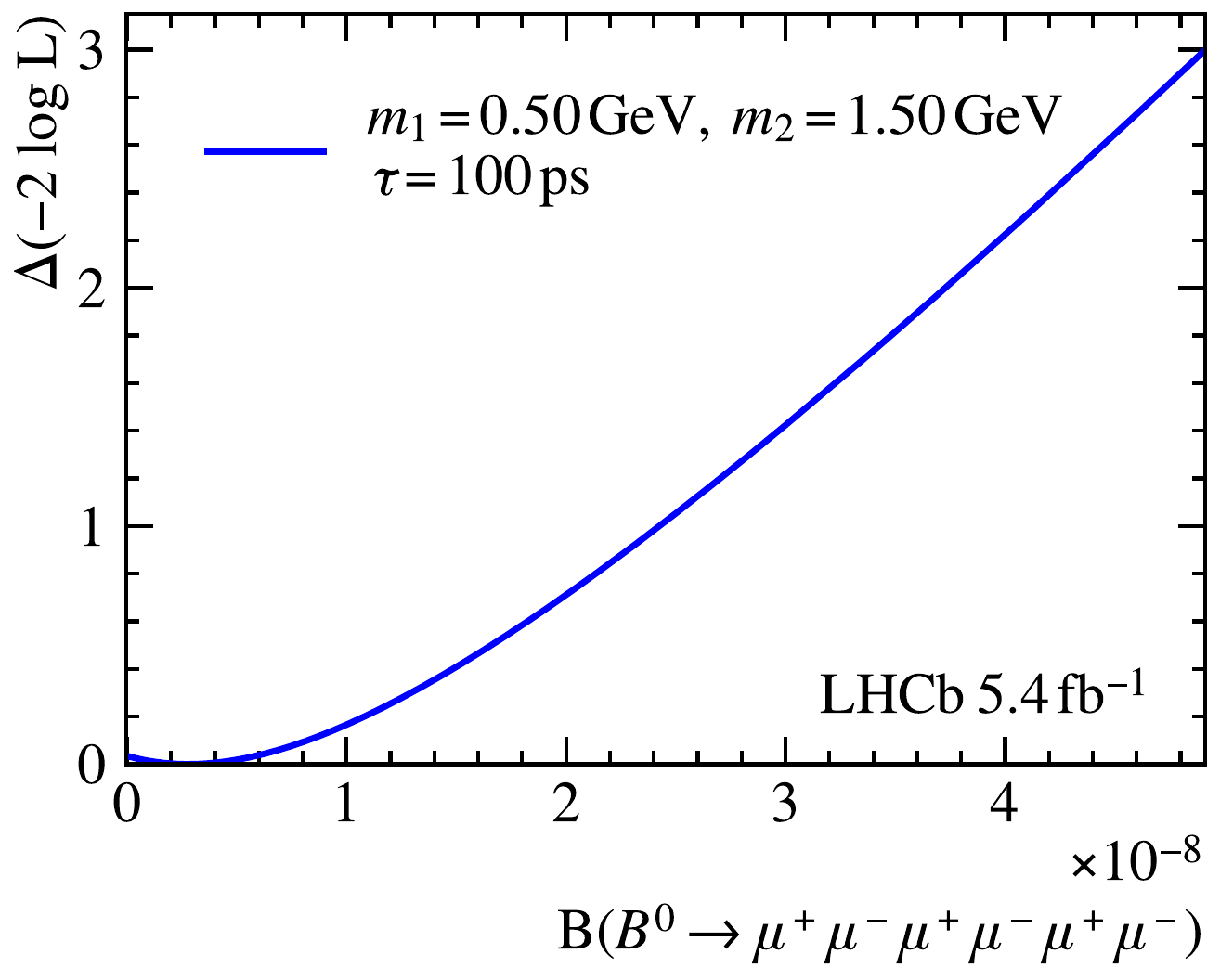}%
    \includegraphics[width=0.33\linewidth]{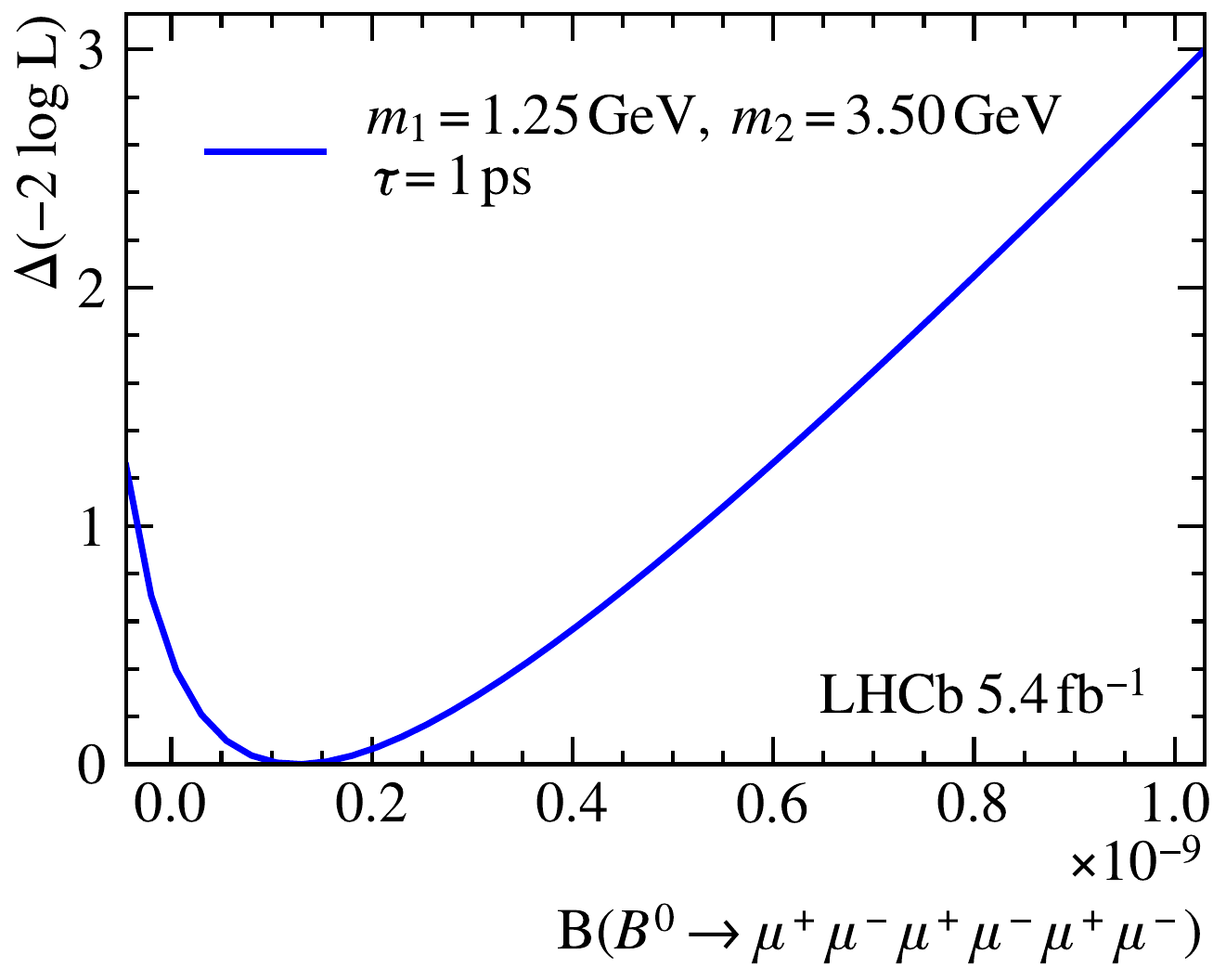}%
    \includegraphics[width=0.33\linewidth]{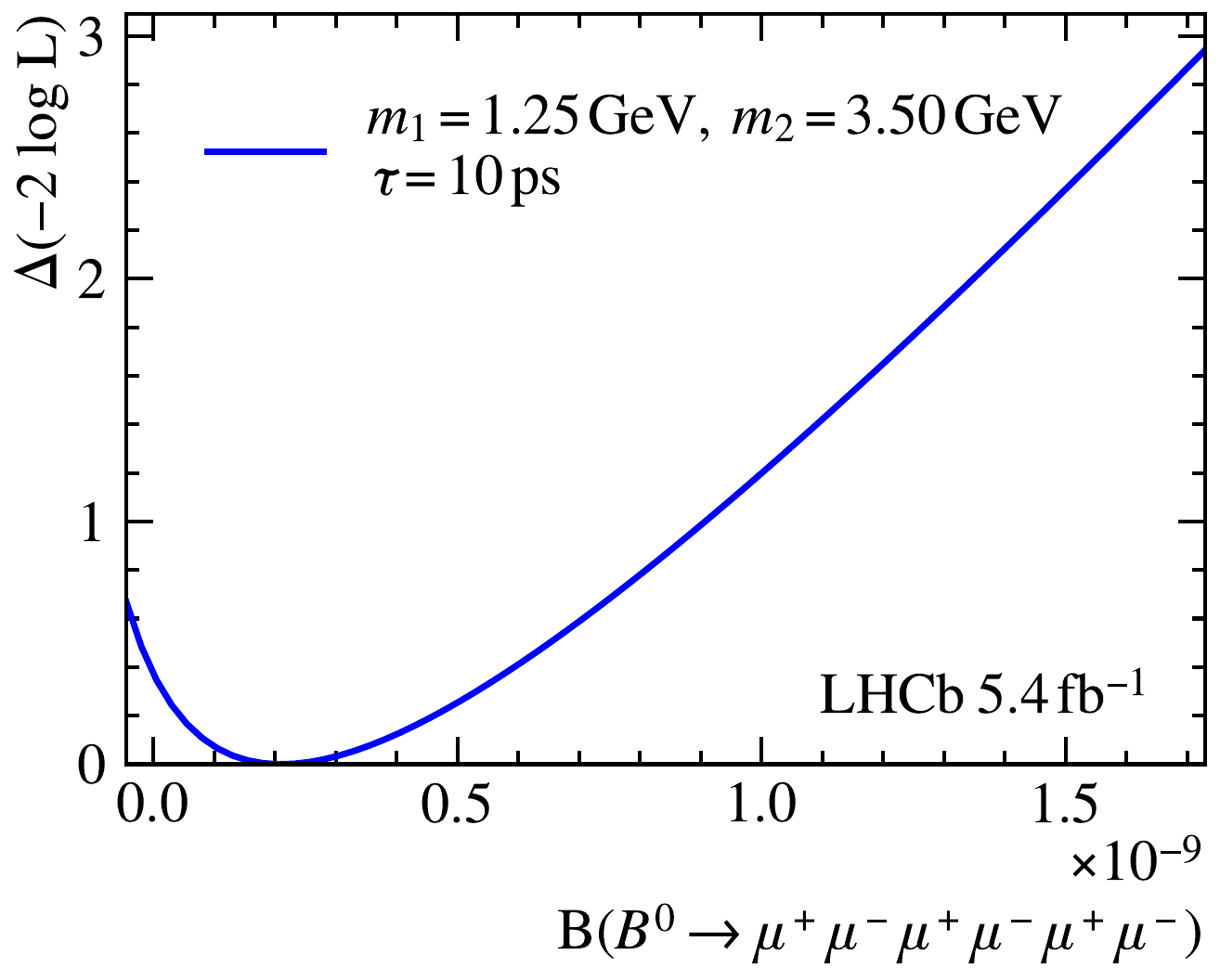}
    \includegraphics[width=0.33\linewidth]{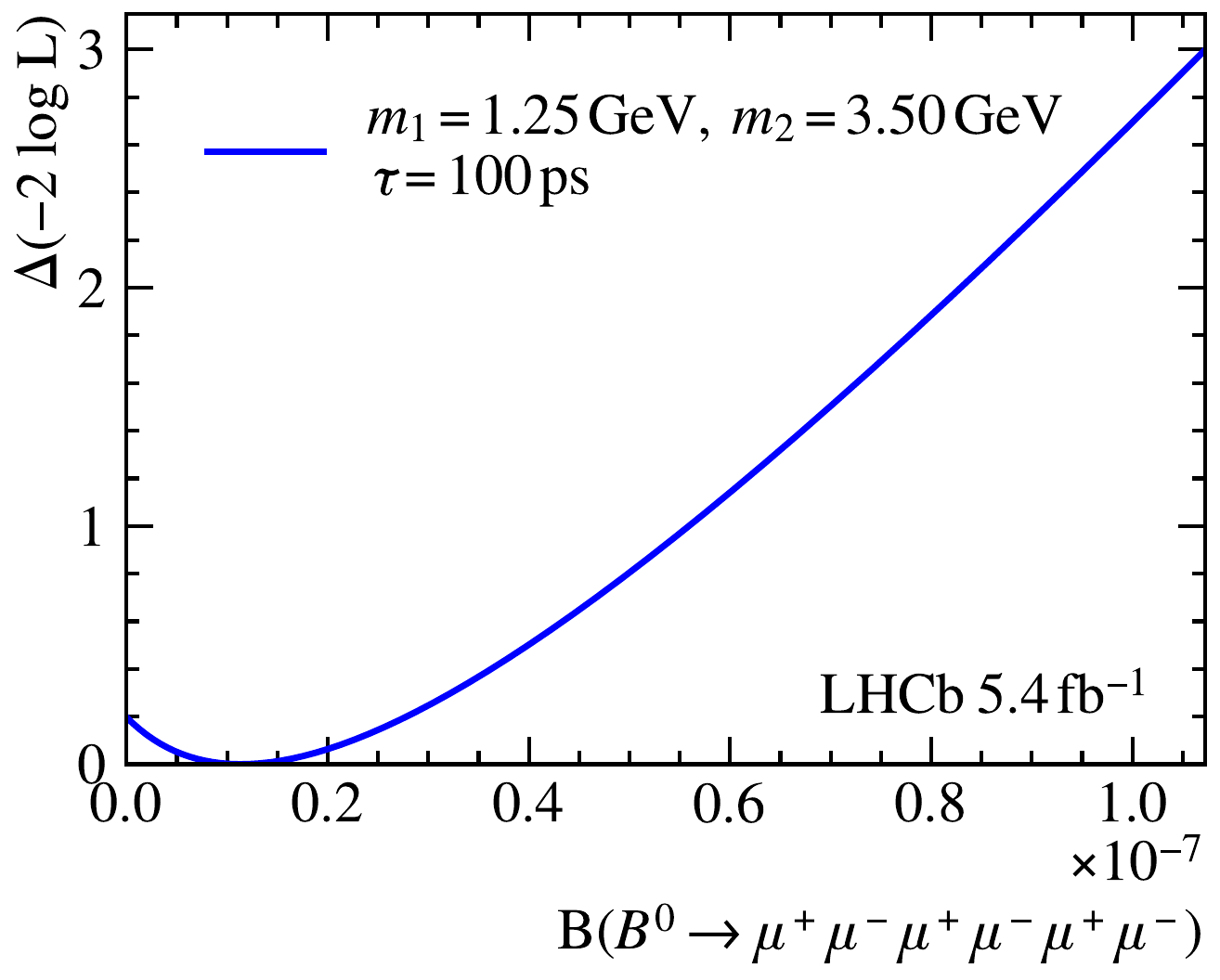}
    \caption{Profile Likelihoods for the fits for \decay{\Bd}{\mu^+ \mu^- \mu^+ \mu^- \mu^+ \mu^-}. From left to right, top to bottom, the fits correspond to the prompt scenario, the lifetimes $[1,10,100]\ps$ for the mass configuration $(m_{a_1}=0.5\gev,m_{a_2}=1.5\gev)$ and the lifetimes $[1,10,100]\ps$ for the mass configuration $(m_{a_1}=1.25\gev,m_{a_2}=3.5\gev)$.}
    \label{fig:NLL_Bd26mu}
\end{figure}

\begin{figure}
    \centering
    \includegraphics[width=0.33\linewidth]{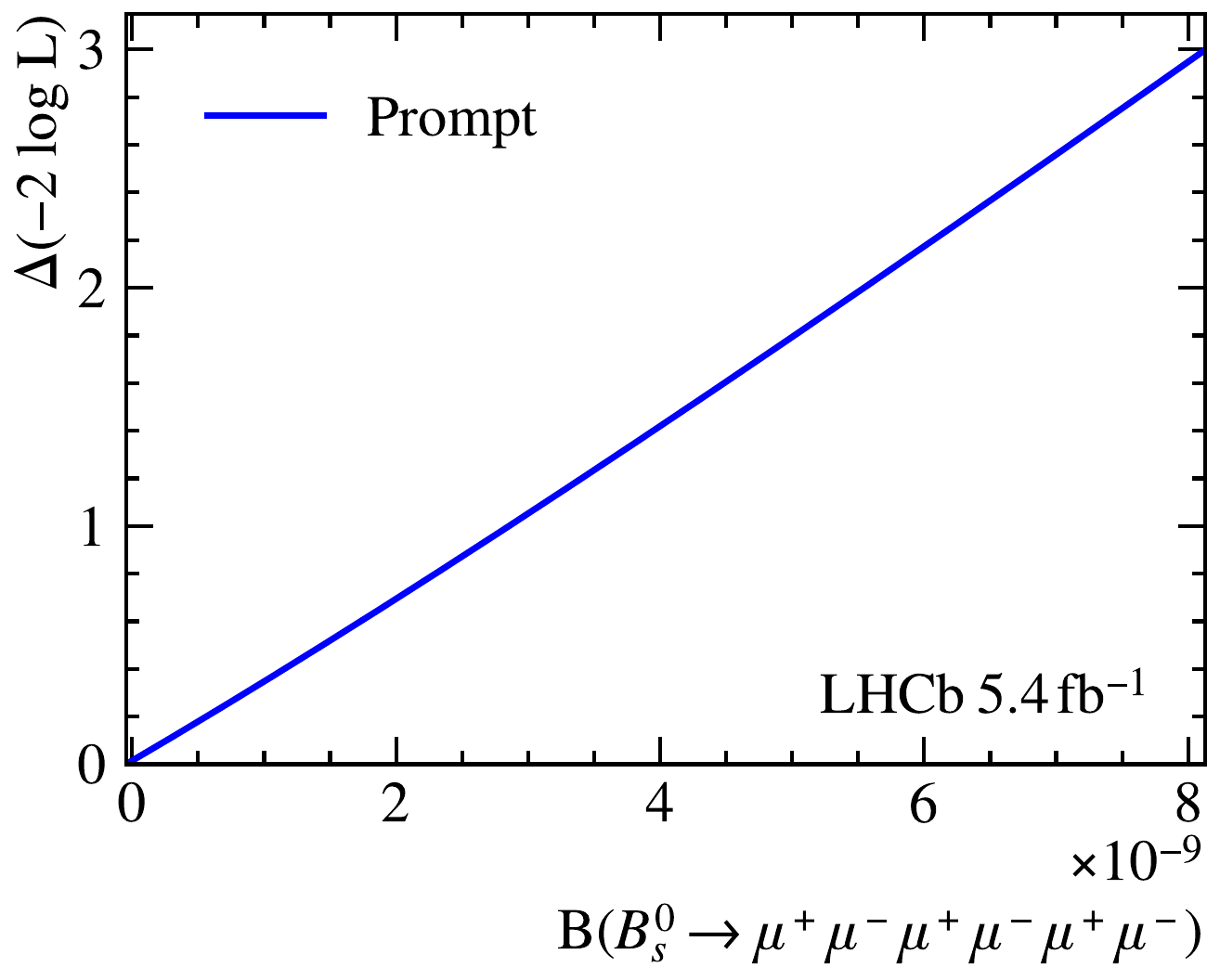}%
    \includegraphics[width=0.33\linewidth]{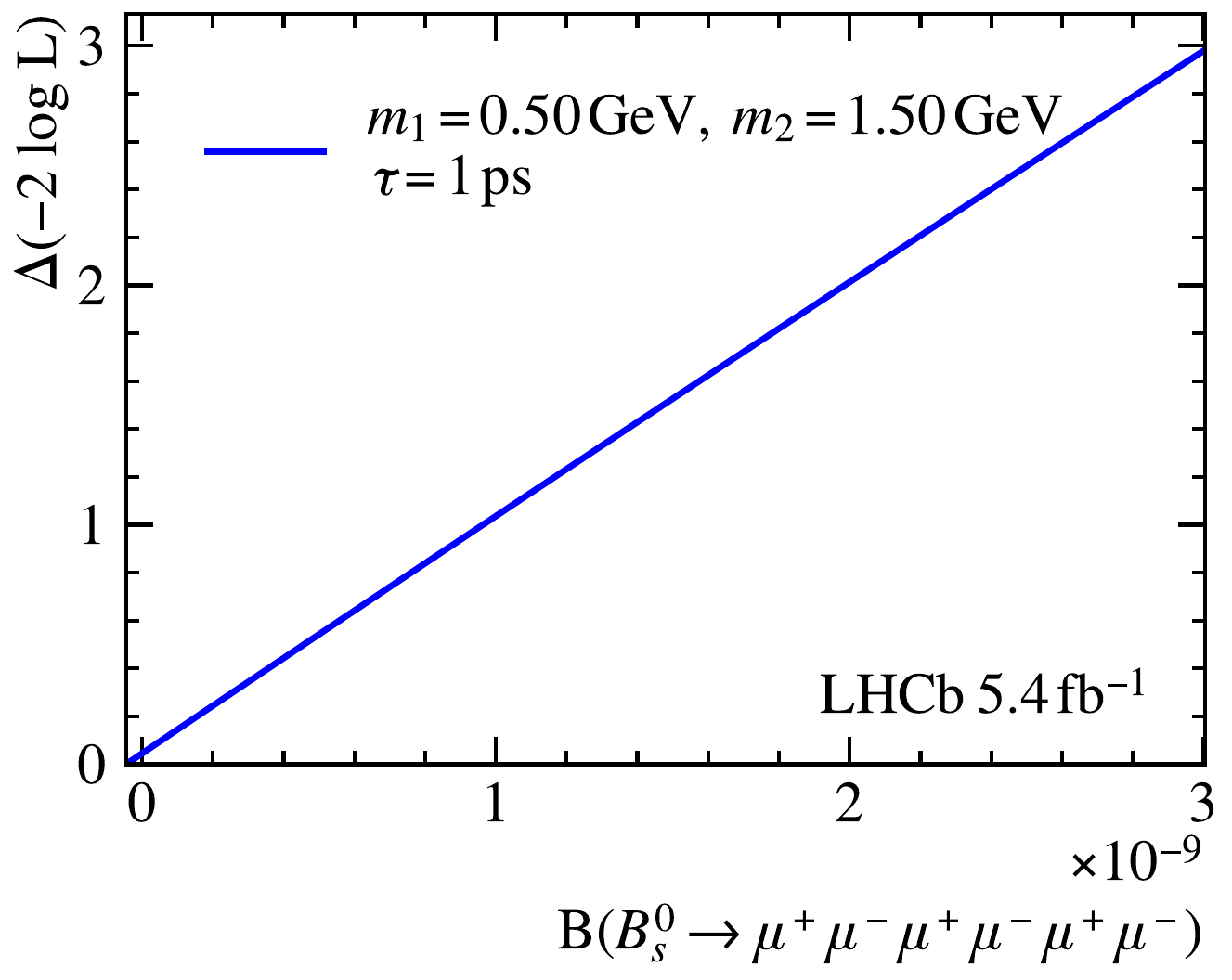}%
    \includegraphics[width=0.33\linewidth]{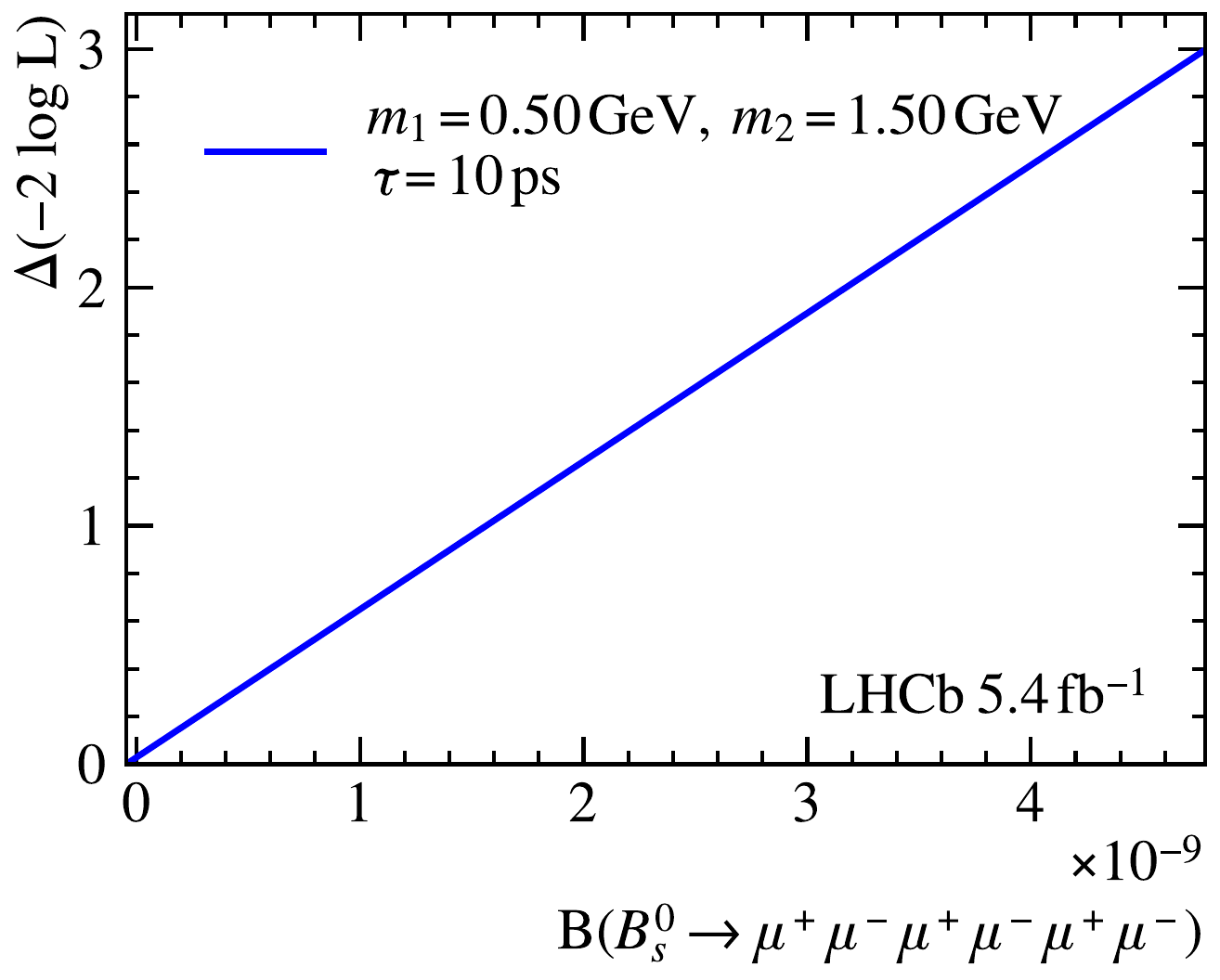}
    \includegraphics[width=0.33\linewidth]{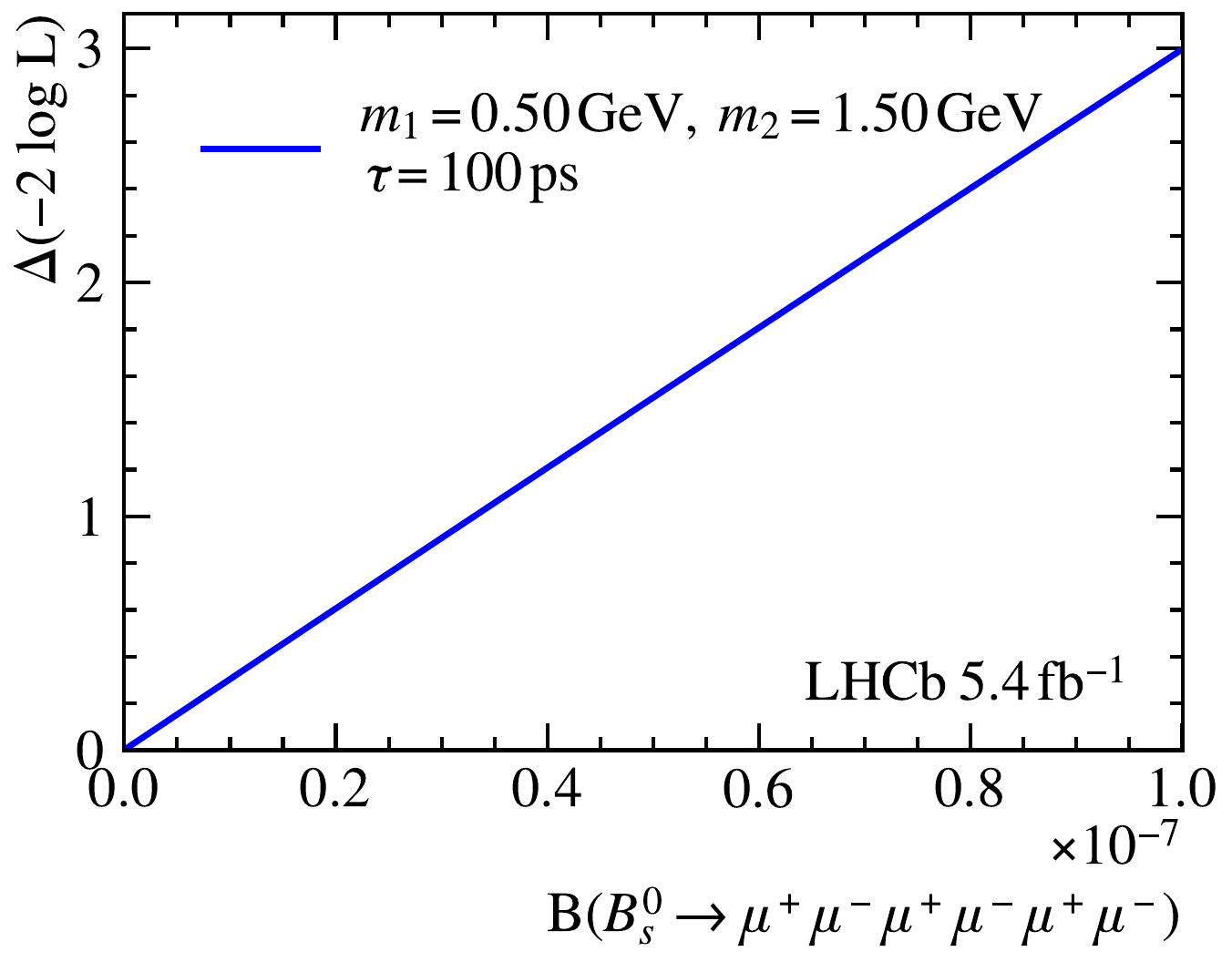}%
    \includegraphics[width=0.33\linewidth]{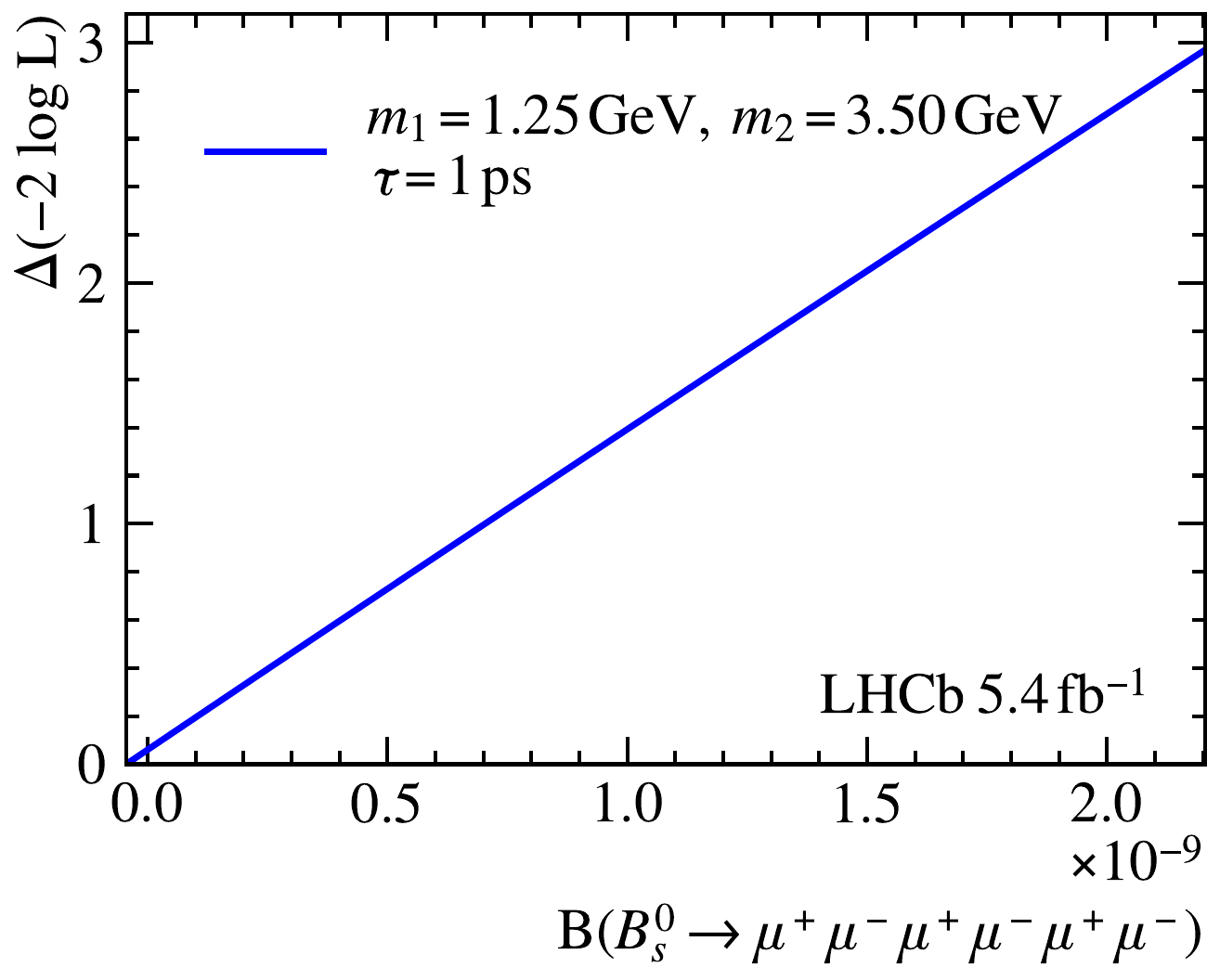}%
    \includegraphics[width=0.33\linewidth]{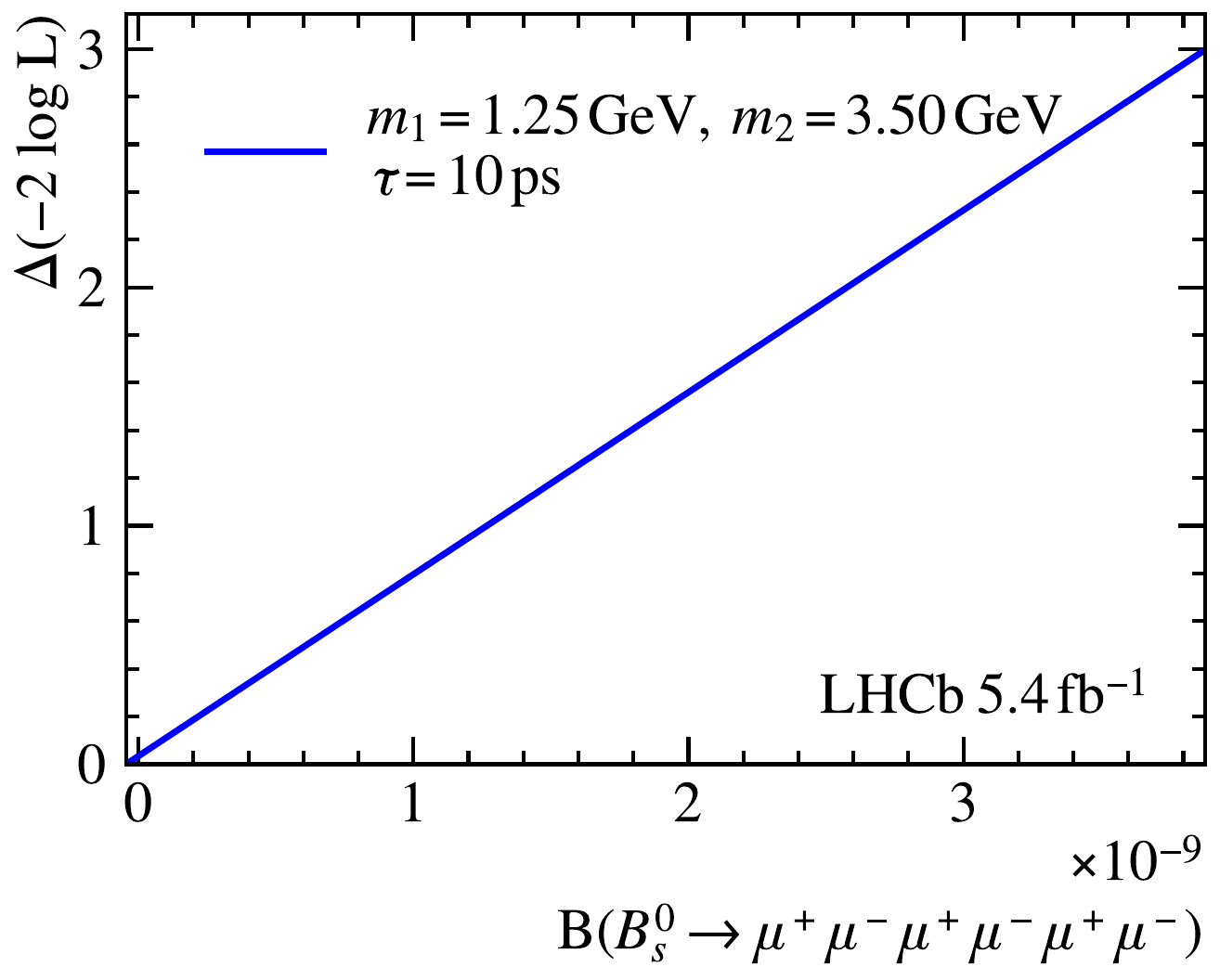}
    \includegraphics[width=0.33\linewidth]{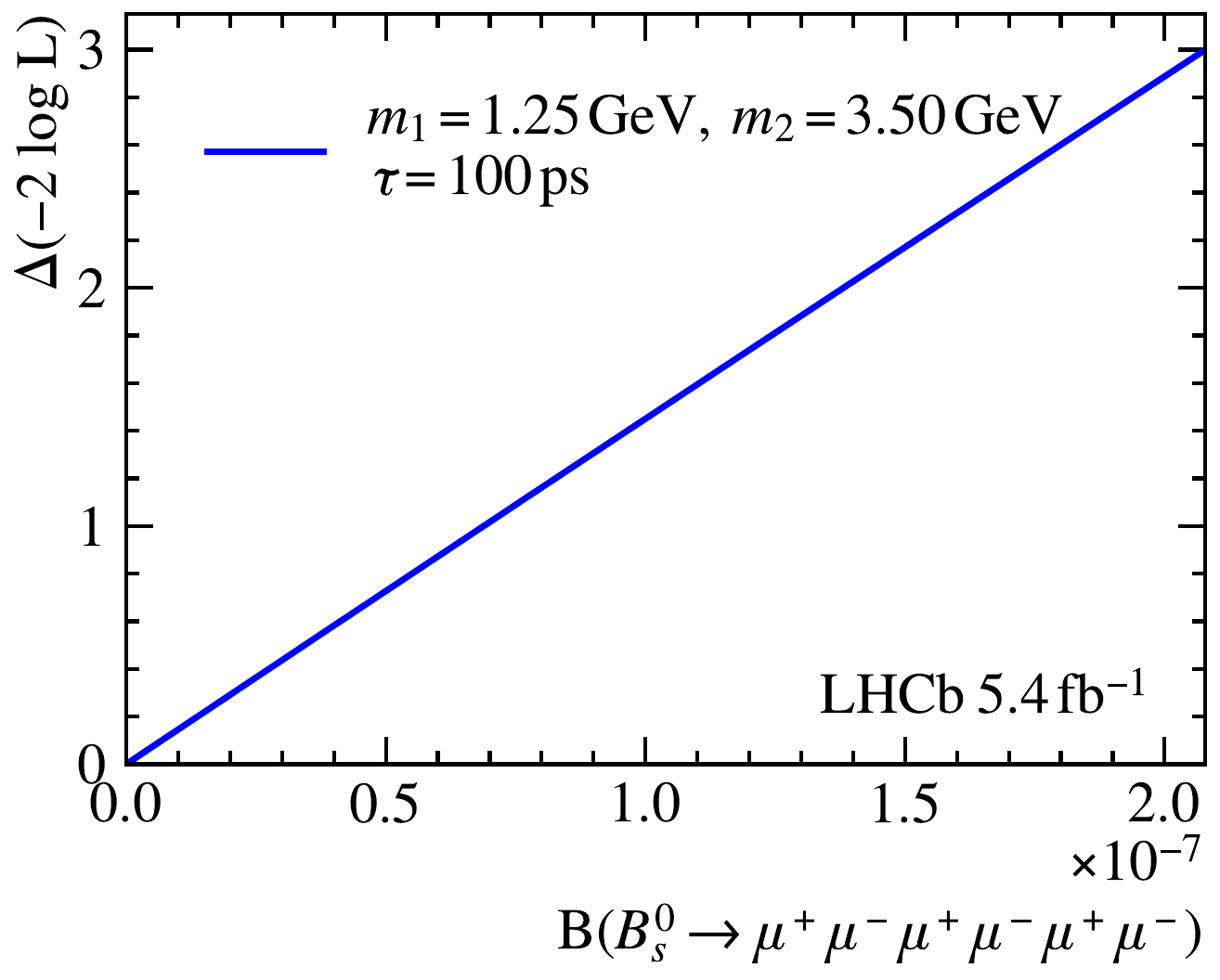}
    \caption{Profile Likelihoods for the fits for \decay{\Bs}{\mu^+ \mu^- \mu^+ \mu^- \mu^+ \mu^-}. From left to right, top to bottom, the fits correspond to the prompt scenario, the lifetimes $[1,10,100]\ps$ for the mass configuration $(m_{a_1}=0.5\gev,m_{a_2}=1.5\gev)$ and the lifetimes $[1,10,100]\ps$ for the mass configuration $(m_{a_1}=1.25\gev,m_{a_2}=3.5\gev)$. The linear behaviour of the likelihoods for the displaced selections arises because there is no data observed in the signal region.}
    \label{fig:NLL_Bs26mu}
\end{figure}

\begin{figure}
    \centering
    \includegraphics[width=0.33\linewidth]{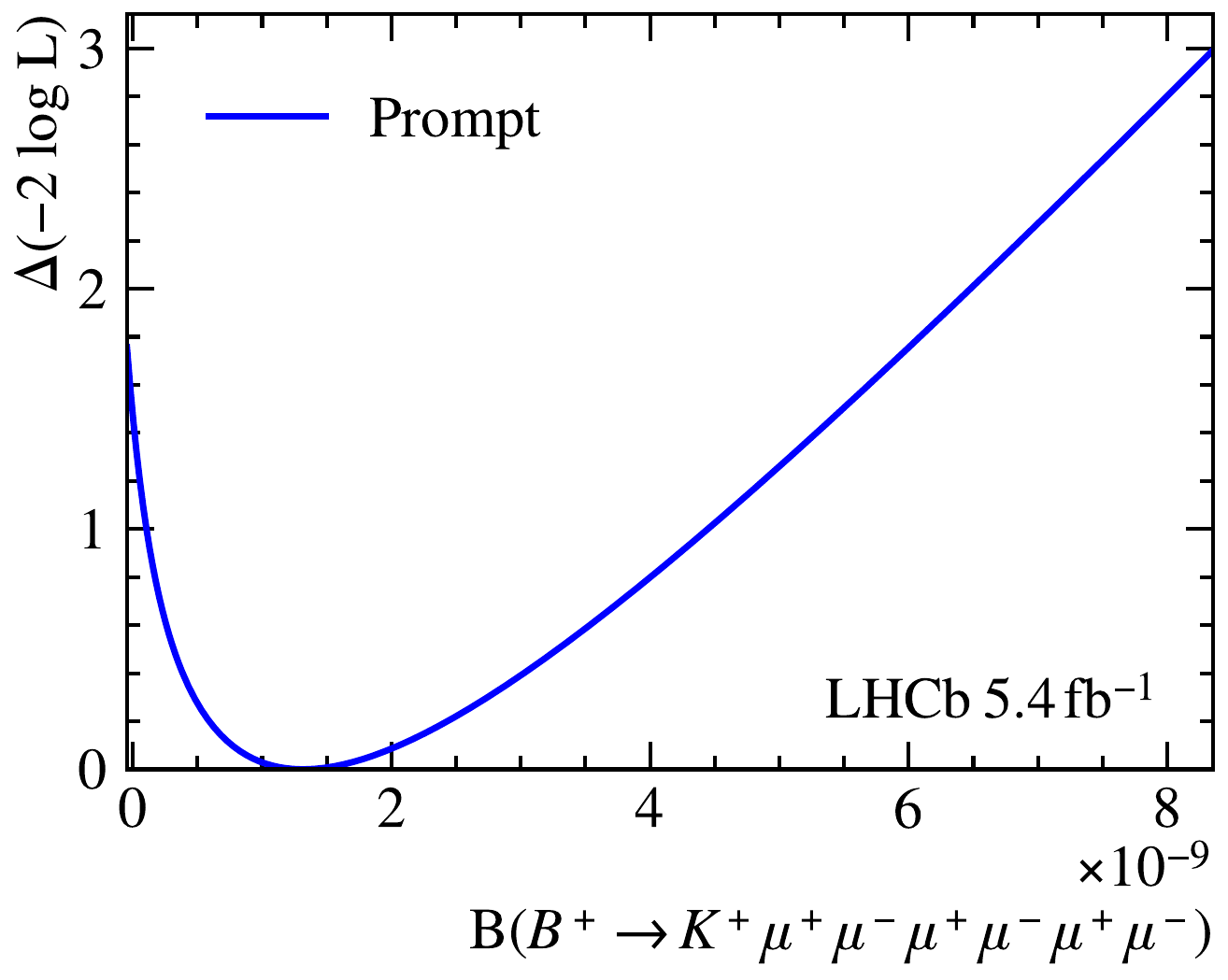}%
    \includegraphics[width=0.33\linewidth]{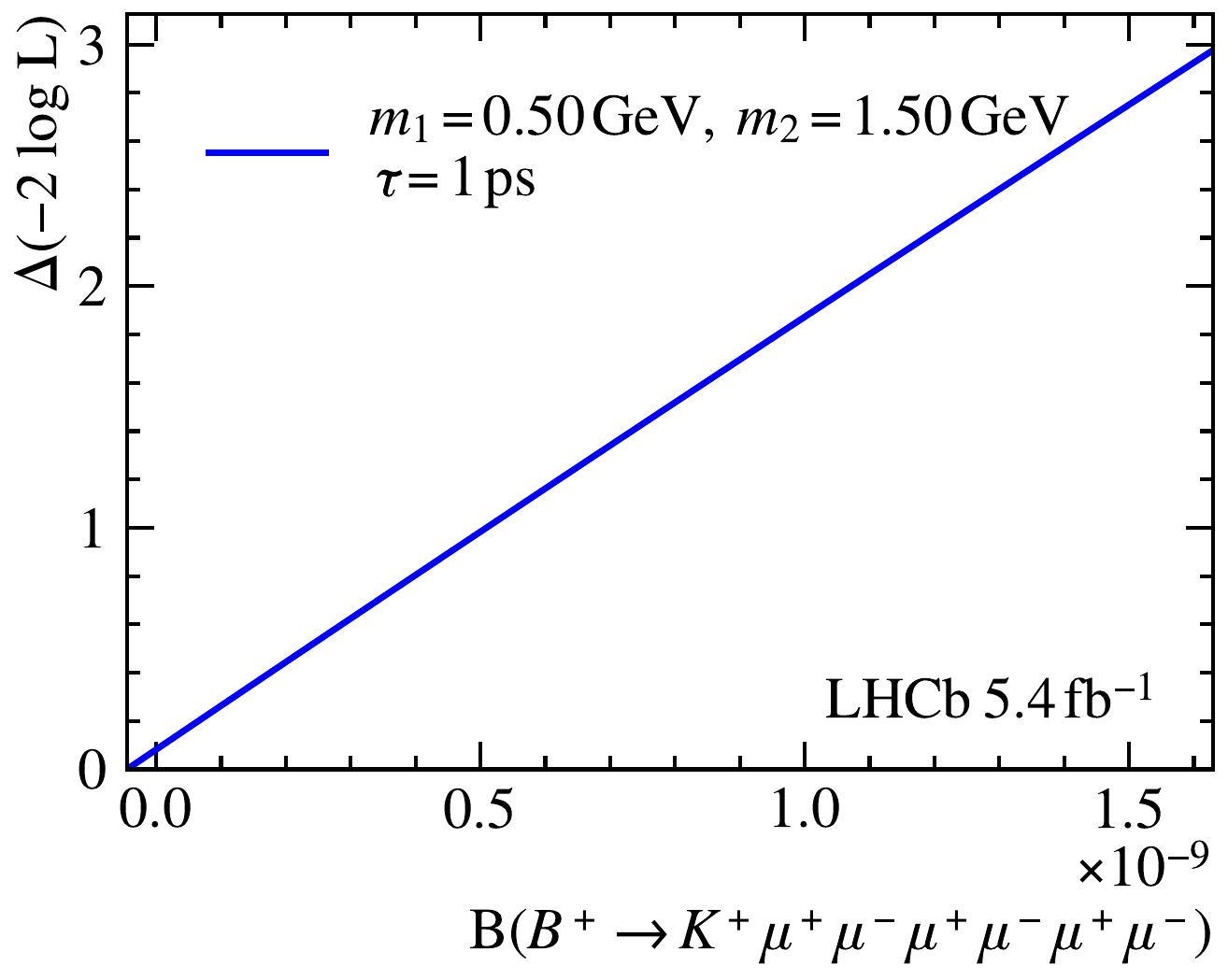}%
    \includegraphics[width=0.33\linewidth]{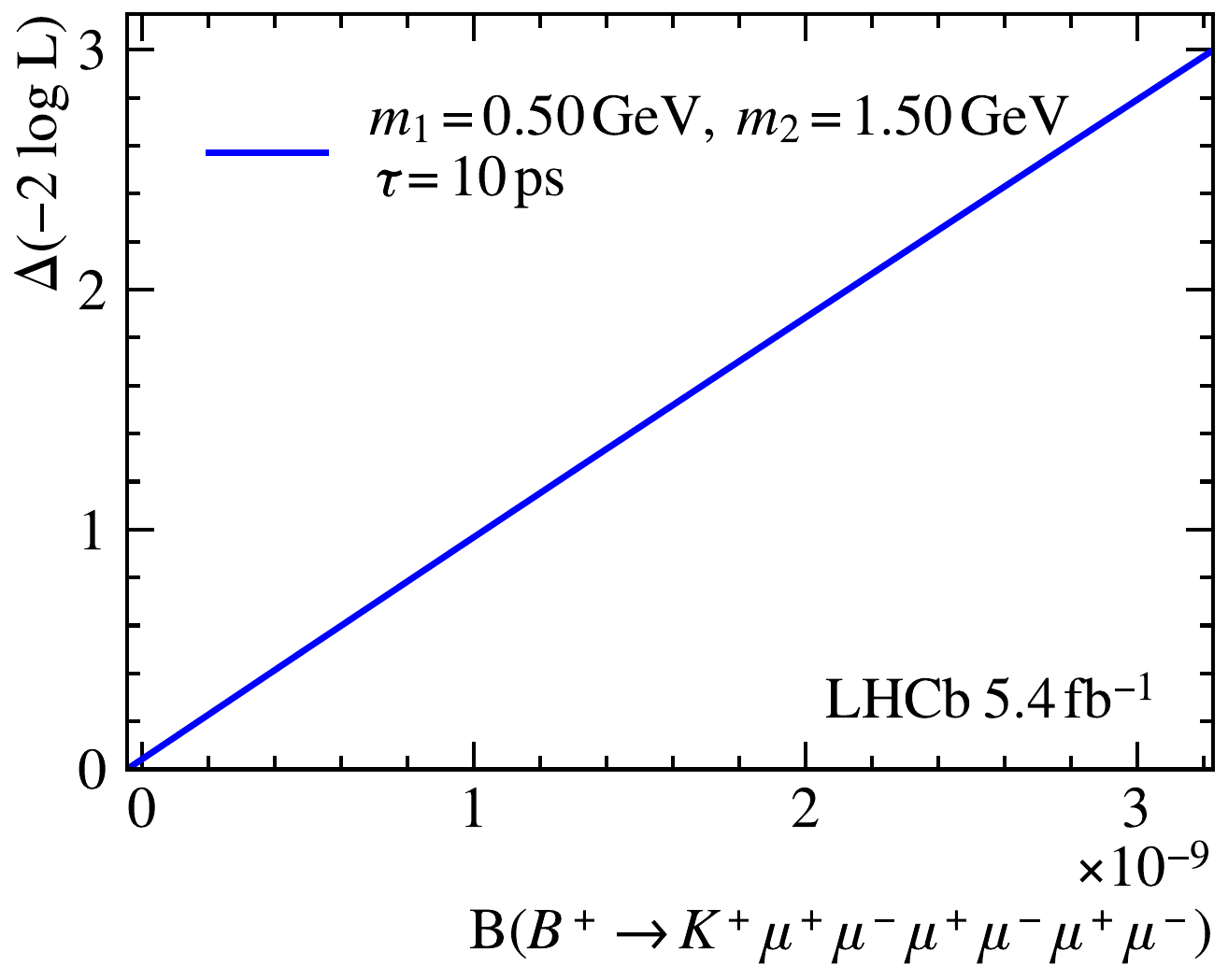}
    \includegraphics[width=0.33\linewidth]{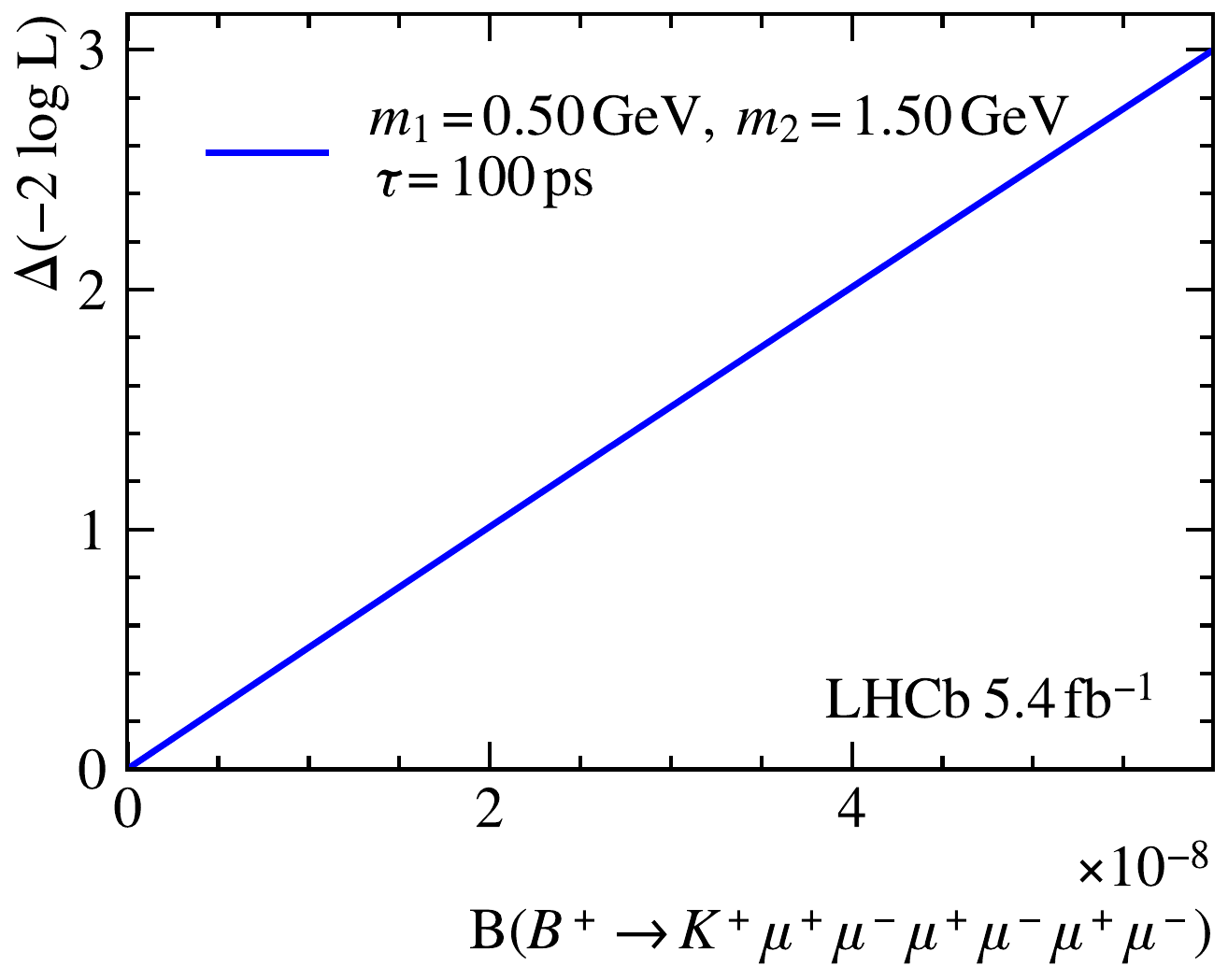}%
    \includegraphics[width=0.33\linewidth]{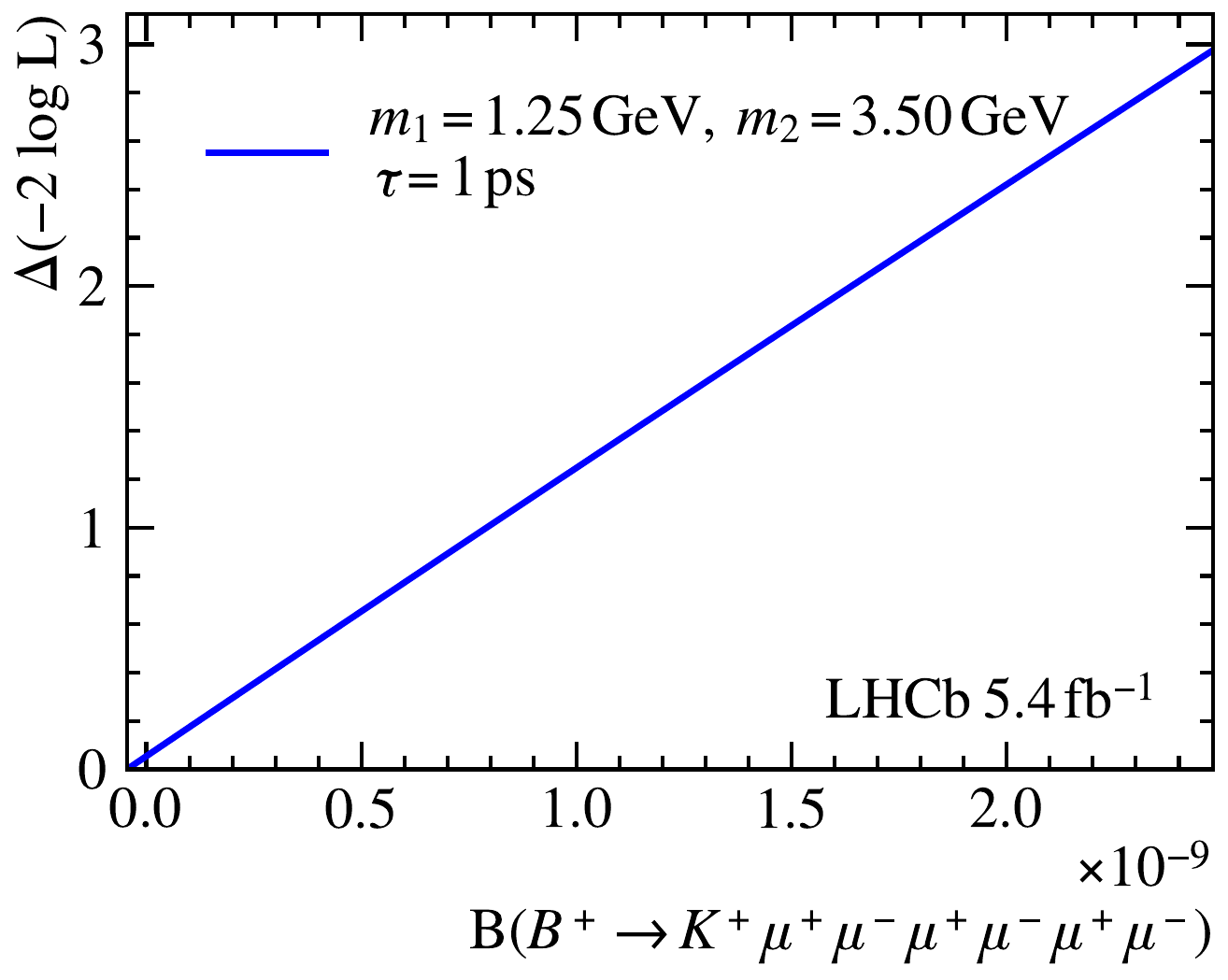}%
    \includegraphics[width=0.33\linewidth]{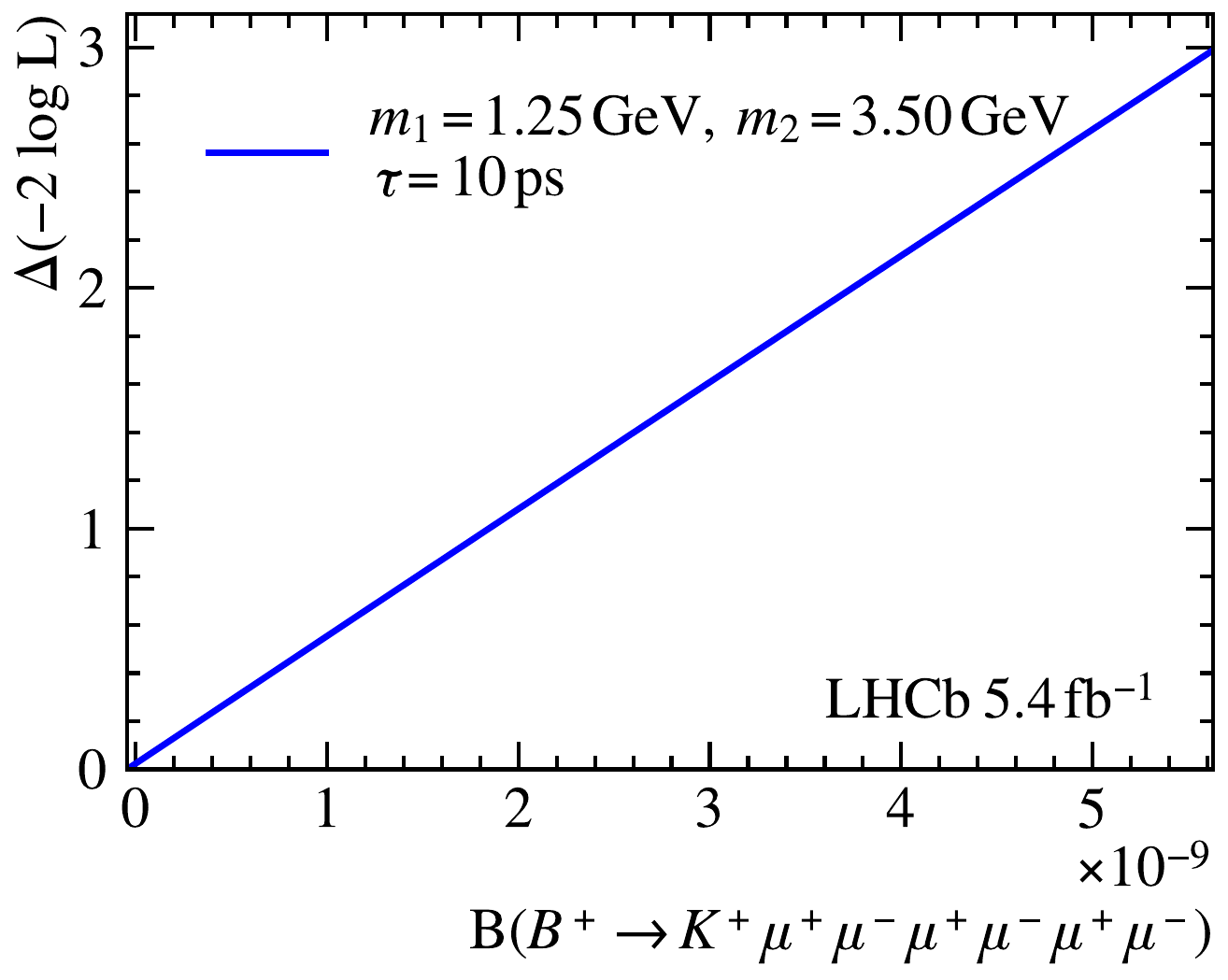}
    \includegraphics[width=0.33\linewidth]{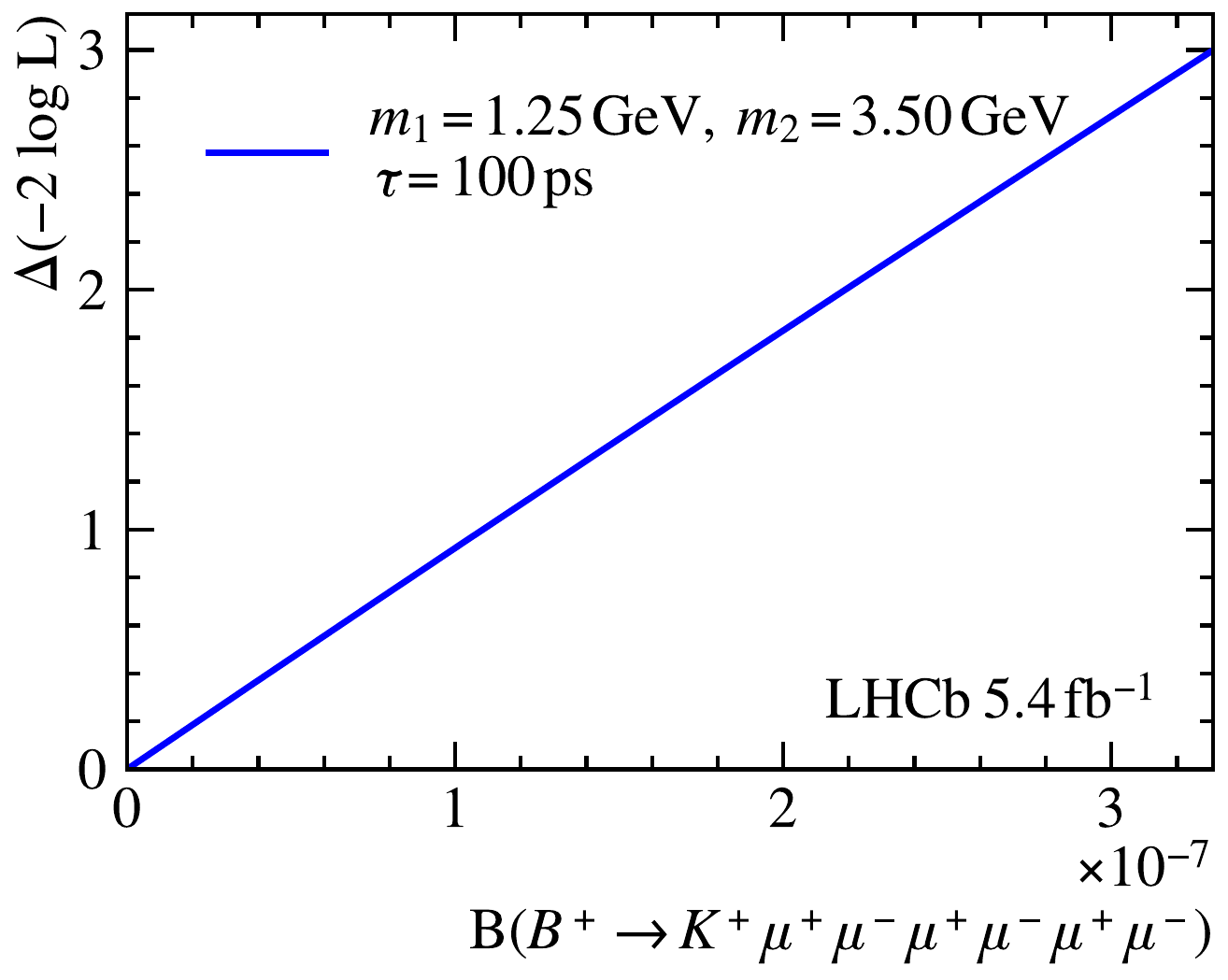}%
    \caption{Profile Likelihoods for the fits for \decay{\Bu}{\Kp \mu^+ \mu^- \mu^+ \mu^- \mu^+ \mu^-}. From left to right, top to bottom, the fits correspond to the prompt scenario, the lifetimes $[10,100]\ps$ for the mass configuration $(m({a_1})=0.5\gev,m({a_2})=1.5\gev)$ and the lifetimes $[10,100]\ps$ for the mass configuration $(m({a_1})=1.25\gev,m({a_2})=3.5\gev)$. The linear behaviour of the likelihoods for the displaced selections arises because there is no data observed in the signal region.}
    \label{fig:NLL_B26muK}
\end{figure}

Figures~\ref{fig:NLL_Bd24mu}, \ref{fig:NLL_Bs24mu}, \ref{fig:NLL_B24muK}, \ref{fig:NLL_Bd26mu}, \ref{fig:NLL_Bs26mu}, and \ref{fig:NLL_B26muK} provide the profile likelihoods (NLL), calculated from a scan of the branching-ratio parameter and systematic uncertainties being included as constraints with a Gaussian prior in the fit model.

\FloatBarrier


\newpage
\addcontentsline{toc}{section}{References}
\bibliographystyle{LHCb}
\bibliography{main,standard,LHCb-PAPER,LHCb-CONF,LHCb-DP,LHCb-TDR}

\clearpage
\centerline
{\large\bf LHCb collaboration}
\begin
{flushleft}
\small
R.~Aaij$^{38}$\lhcborcid{0000-0003-0533-1952},
M. ~Abdelfatah$^{69}$,
A.S.W.~Abdelmotteleb$^{57}$\lhcborcid{0000-0001-7905-0542},
C.~Abellan~Beteta$^{51}$\lhcborcid{0009-0009-0869-6798},
F.~Abudin\'en$^{59}$\lhcborcid{0000-0002-6737-3528},
T.~Ackernley$^{61}$\lhcborcid{0000-0002-5951-3498},
A. A. ~Adefisoye$^{69}$\lhcborcid{0000-0003-2448-1550},
B.~Adeva$^{47}$\lhcborcid{0000-0001-9756-3712},
M.~Adinolfi$^{55}$\lhcborcid{0000-0002-1326-1264},
P.~Adlarson$^{87,42}$\lhcborcid{0000-0001-6280-3851}
% ...

{\footnotesize \it

% ...
$^{38}$Nikhef National Institute for Subatomic Physics, Amsterdam, Netherlands\\
% ...
$^{42}$National Center for Nuclear Research (NCBJ), Warsaw, Poland\\
% ...
$^{47}$Instituto Galego de F{\'\i}sica de Altas Enerx{\'\i}as (IGFAE), Universidade de Santiago de Compostela, Santiago de Compostela, Spain\\
% ...
$^{51}$Physik-Institut, Universit{\"a}t Z{\"u}rich, Z{\"u}rich, Switzerland\\
% ...
$^{55}$H.H. Wills Physics Laboratory, University of Bristol, Bristol, United Kingdom\\
% ...
$^{57}$Department of Physics, University of Warwick, Coventry, United Kingdom\\
% ...
$^{59}$School of Physics and Astronomy, University of Edinburgh, Edinburgh, United Kingdom\\
$^{60}$School of Physics and Astronomy, University of Glasgow, Glasgow, United Kingdom\\
$^{61}$Oliver Lodge Laboratory, University of Liverpool, Liverpool, United Kingdom\\
% ...
$^{69}$Syracuse University, Syracuse, NY, United States\\
% ...
$^{87}$Department of Physics and Astronomy, Uppsala University, Uppsala, Sweden, associated to $^{60}$\\
% ...
}
\end{flushleft}

\end{document}